\documentclass[%
 reprint,
 superscriptaddress,
nofootinbib, % show footnotes on same page instead of in bibliography
 amsmath,amssymb,xfrac,
 aps,
 prc,
]{revtex4-2}

\usepackage{graphicx}% Include figure files
\usepackage{dcolumn}% Align table columns on decimal point
\usepackage{bm}% bold math
\usepackage[hidelinks]{hyperref}% add hypertext capabilities
\usepackage{booktabs} % add horizontal lines over specific table columns
\begin{document}

\preprint{APS/123-QED}

\title{Mass measurements of $^{60 - 63}$Ga reduce x-ray burst model uncertainties and extend the evaluated $\mathbf{T=1}$ isobaric multiplet mass equation} % Force line breaks with \\

\author{S.~F.~Paul}
\email[Corresponding author: ]{stefan.paul@triumf.ca}
\affiliation{TRIUMF, 4004 Wesbrook Mall, Vancouver, British Columbia V6T 2A3, Canada}
\affiliation{Ruprecht-Karls-Universit\"at Heidelberg, Heidelberg D-69117, Germany}

\author{J.~Bergmann}
\affiliation{II. Physikalisches Institut, Justus-Liebig-Universit\"at, Gie\ss en D-35392, Germany}

\author{J.~D.~Cardona}
\affiliation{TRIUMF, 4004 Wesbrook Mall, Vancouver, British Columbia V6T 2A3, Canada}
\affiliation{Department of Physics and Astronomy, University of Manitoba, Winnipeg, Manitoba R3T 2N2, Canada}

\author{K.~A.~Dietrich}
\affiliation{TRIUMF, 4004 Wesbrook Mall, Vancouver, British Columbia V6T 2A3, Canada}
\affiliation{Ruprecht-Karls-Universit\"at Heidelberg, Heidelberg D-69117, Germany}

\author{E.~Dunling}
\affiliation{TRIUMF, 4004 Wesbrook Mall, Vancouver, British Columbia V6T 2A3, Canada}
\affiliation{Department of Physics, University of York, York YO10 5DD, United Kingdom}

\author{Z.~Hockenbery}
\affiliation{TRIUMF, 4004 Wesbrook Mall, Vancouver, British Columbia V6T 2A3, Canada}
\affiliation{Department of Physics, McGill University, Montr\'eal, Qu\'ebec H3A 2T8, Canada}

\author{C.~Hornung}
\affiliation{II. Physikalisches Institut, Justus-Liebig-Universit\"at, Gie\ss en D-35392, Germany}
\affiliation{GSI Helmholtzzentrum f\"ur Schwerionenforschung GmbH, Darmstadt D-64291, Germany}

\author{C.~Izzo}
\affiliation{TRIUMF, 4004 Wesbrook Mall, Vancouver, British Columbia V6T 2A3, Canada}

\author{A.~Jacobs}
\affiliation{TRIUMF, 4004 Wesbrook Mall, Vancouver, British Columbia V6T 2A3, Canada}
\affiliation{Department of Physics and Astronomy, University of British Columbia, Vancouver, British Columbia V6T 1Z1, Canada}

\author{A.~Javaji}
\affiliation{TRIUMF, 4004 Wesbrook Mall, Vancouver, British Columbia V6T 2A3, Canada}
\affiliation{Department of Physics and Astronomy, University of British Columbia, Vancouver, British Columbia V6T 1Z1, Canada}

\author{B.~Kootte}
\affiliation{TRIUMF, 4004 Wesbrook Mall, Vancouver, British Columbia V6T 2A3, Canada}
\affiliation{Department of Physics and Astronomy, University of Manitoba, Winnipeg, Manitoba R3T 2N2, Canada}

\author{Y.~Lan}
\affiliation{TRIUMF, 4004 Wesbrook Mall, Vancouver, British Columbia V6T 2A3, Canada}
\affiliation{Department of Physics and Astronomy, University of British Columbia, Vancouver, British Columbia V6T 1Z1, Canada}

\author{E.~Leistenschneider}
%\affiliation{CERN, 1211 Geneva 23, Switzerland}
\altaffiliation{Present address: CERN, 1211 Geneva 23, Switzerland}
\affiliation{TRIUMF, 4004 Wesbrook Mall, Vancouver, British Columbia V6T 2A3, Canada}
\affiliation{Department of Physics and Astronomy, University of British Columbia, Vancouver, British Columbia V6T 1Z1, Canada}

\author{E.~M.~Lykiardopoulou}
\affiliation{TRIUMF, 4004 Wesbrook Mall, Vancouver, British Columbia V6T 2A3, Canada}
\affiliation{Department of Physics and Astronomy, University of British Columbia, Vancouver, British Columbia V6T 1Z1, Canada}

\author{I.~Mukul}
\affiliation{TRIUMF, 4004 Wesbrook Mall, Vancouver, British Columbia V6T 2A3, Canada}

\author{T.~Murb\"ock}
\affiliation{TRIUMF, 4004 Wesbrook Mall, Vancouver, British Columbia V6T 2A3, Canada}

\author{W.~S.~Porter}
\affiliation{TRIUMF, 4004 Wesbrook Mall, Vancouver, British Columbia V6T 2A3, Canada}
\affiliation{Department of Physics and Astronomy, University of British Columbia, Vancouver, British Columbia V6T 1Z1, Canada}

\author{R.~Silwal}
\altaffiliation[Present address: ]{Department of Physics and Astronomy, Appalachian State University, Boone, North Carolina, 28608, USA}
\affiliation{TRIUMF, 4004 Wesbrook Mall, Vancouver, British Columbia V6T 2A3, Canada}

\author{M.~B.~Smith}
\affiliation{TRIUMF, 4004 Wesbrook Mall, Vancouver, British Columbia V6T 2A3, Canada}

\author{J.~Ringuette}
\affiliation{TRIUMF, 4004 Wesbrook Mall, Vancouver, British Columbia V6T 2A3, Canada}
\affiliation{Department of Physics, Colorado School of Mines, Golden, Colorado 80401, USA}

\author{T.~Brunner}
\affiliation{Department of Physics, McGill University, Montr\'eal, Qu\'ebec H3A 2T8, Canada}

\author{T.~Dickel}
\affiliation{II. Physikalisches Institut, Justus-Liebig-Universit\"at, Gie\ss en D-35392, Germany}
\affiliation{GSI Helmholtzzentrum f\"ur Schwerionenforschung GmbH, Darmstadt D-64291, Germany}

\author{I.~Dillmann}
\affiliation{TRIUMF, 4004 Wesbrook Mall, Vancouver, British Columbia V6T 2A3, Canada}
\affiliation{Department of Physics and Astronomy, University of Victoria, Victoria, British Columbia V8P 5C2, Canada}

\author{G.~Gwinner}
\affiliation{Department of Physics and Astronomy, University of Manitoba, Winnipeg, Manitoba R3T 2N2, Canada}

\author{M.~MacCormick}
\affiliation{Université Paris-Saclay, CNRS/IN2P3, IJCLab, 91405 Orsay, France}

\author{M.~P.~Reiter}
\affiliation{Institute for Particle and Nuclear Physics, The University of Edinburgh, EH9 3FD Edinburgh, United Kingdom}

\author{H.~Schatz}
\affiliation{National Superconducting Cyclotron Laboratory, Michigan State University, East Lansing, Michigan 48824, USA}
\affiliation{Department of Physics and Astronomy, Michigan State University, East Lansing, Michigan 48824, USA}
\affiliation{Joint Institute for Nuclear Astrophysics Center for the Evolution of the Elements (JINA-CEE), East Lansing, Michigan 48854, USA}

\author{N.~A.~Smirnova}
\affiliation{CENBG (CNRS/IN2P3 - Université de Bordeaux), 33175 Gradignan Cedex, France}

\author{J.~Dilling}
\affiliation{TRIUMF, 4004 Wesbrook Mall, Vancouver, British Columbia V6T 2A3, Canada}
\affiliation{Department of Physics and Astronomy, University of British Columbia, Vancouver, British Columbia V6T 1Z1, Canada}

\author{A.~A.~Kwiatkowski}
\affiliation{TRIUMF, 4004 Wesbrook Mall, Vancouver, British Columbia V6T 2A3, Canada}
\affiliation{Department of Physics and Astronomy, University of Victoria, Victoria, British Columbia V8P 5C2, Canada}

\date{\today}% It is always \today, today,
             %  but any date may be explicitly specified

\begin{abstract}
We report precision mass measurements of neutron-deficient gallium isotopes approaching the proton drip line. The measurements of $^{60-63}$Ga performed with the TITAN multiple-reflection time-of-flight mass spectrometer provide a more than threefold improvement over the current literature mass uncertainty of $^{61}$Ga and mark the first direct mass measurement of $^{60}$Ga.
The improved precision of the $^{61}$Ga mass has important implications for the astrophysical $\textit{rp}$ process, as it constrains essential reaction $\mathrm{Q}$ values near the $^{60}$Zn waiting point. Based on calculations with a one-zone model, we demonstrate the impact of the improved mass data on prediction uncertainties of x-ray burst models.  
The first-time measurement of the $^{60}$Ga ground-state mass establishes the proton-bound nature of this nuclide; thus, constraining the location of the proton drip line along this isotopic chain. Including the measured mass of $^{60}$Ga further enables us to extend the evaluated $T=1$ isobaric multiplet mass equation up to $A=60$. 
\end{abstract}

%\keywords{Suggested keywords}%Use showkeys class option if keyword
                              %display desired
\maketitle

%\tableofcontents

\section{\label{sec:intro}Introduction}
Type I x-ray bursts are recurring thermonuclear explosions occurring on the surface of accreting neutron stars~\cite{Schatz2006b,Parikh2013,Meisel2018,Galloway2021}. 
Comparisons of the observed light curves with model predictions constrain the properties of neutron stars~\cite{Schatz2006,Parikh2013,Meisel2018b,Meisel2019} and the extreme states of neutron-rich matter found in the neutron star crust~\cite{Fisker2008,Chamel2008p139}. 
The composition of burst ashes and the characteristic shape of the burst light curves, with a fast  rise followed by a slow decay of the luminosity, are determined by the underlying nuclear reaction sequences~\cite{Kratz1998,Schatz2001,Parikh2013}.
The primary energy source of x-ray bursts is the rapid proton capture process ($\textit{rp}$ process)~\cite{Wallace1981}, a succession of fast proton captures (p,$\gamma$) and slower $\beta$ decays that can synthesize neutron-deficient elements up to the $A\approx100$ region~\cite{Schatz2001,Koike2004}.

The $\textit{rp}$ process rapidly drives matter towards heavier nuclides along the $N=Z$ line, until the combination of a small or negative proton capture $\mathrm{Q}$ value $Q_{(\text{p},\gamma)}$ and a comparatively long $\beta$-decay half-life causes the mass flow to stall~\cite{Schatz1998} at so-called waiting point nuclei. The reaction flow impedance imposed by the slow $\beta$ decay of a waiting point nucleus $i$ can be bypassed by the sequential \text{two-proton} capture $i(\text{p},\gamma)j(\text{p},\gamma)k$. In this case, a local (p,$\gamma$)-($\gamma$,p)-equilibrium develops between the waiting point nucleus and the intermediate nucleus $j$, and the net forward reaction flow is determined by the Saha equation for the two-proton capture reaction rate~\cite{Schatz1998}: 
\begin{equation}\label{eqn:2p_capture_rate}
    \langle2\text{p},\gamma\rangle_i = \frac{2G_j}{(2J_\text{p}+1)G_i}\left( \frac{2\pi \hbar^2}{\mu k_B T}\right)^\frac{3}{2}\exp{\left(\frac{Q_{i(\text{p},\gamma)}}{k_B T}\right)}\langle \text{p},\gamma \rangle_{j},
\end{equation}
where $G_{i}$ and $G_{j}$ denote the respective partition functions, $J_\text{p}$ is the proton spin, $\mu$ is the reduced mass for a proton and nucleus $i$, $T$ is the temperature, $\langle\text{p},\gamma \rangle_{j}$ is the proton-capture rate on nucleus $j$, and $Q_{i(\text{p},\gamma)}$ is the $\mathrm{Q}$ value for proton capture on nucleus $i$, i.e. the mass difference between the initial and final states. Due to the exponential dependence of the two-proton capture rate on the $\mathrm{Q}$ value, the $\textit{rp}$-process flow is highly sensitive to nuclear masses, and x-ray burst models rely critically on accurate nuclear mass inputs. In addition, most of the relevant proton capture rates $\langle \text{p},\gamma \rangle$ have not been measured and must be calculated~\cite{Cyburt2016}; these reaction rate calculations require well-known $Q$-values to pin down the energies of unmeasured reaction resonances~\cite{Parikh2009,Schatz2017}. In this way, nuclear masses additionally affect burst model predictions through theoretical proton capture rates. Finally, accurate resonance energies are also critical to guide and enable future direct measurements of reaction rates. Despite extensive efforts at rare isotope beam facilities~\cite{Clark2007,Breitenfeldt2009,Kankainen2010,Tu2011b,Valverde2018} some of the relevant input masses are still not known to the required precision of $\approx10$~keV~\cite{Schatz2006,Schatz2017,Duy2020}.

Here, we address nuclear uncertainties near the $^{60}$Zn $\textit{rp}$-process waiting point~\cite{Fisker2008} with precision mass measurements of $^{60}$Ga and $^{61}$Ga. The $^{60}$Zn waiting point ($T_{1/2}=2.38$~min.~\cite{Browne2013}) is of particular importance, as it may produce departures from the standard power-law decay of the burst luminosity~\cite{in'tZand2017}. Two complementary, large-scale sensitivity studies~\cite{Schatz2017,Cyburt2016} identified relevant nuclear masses and reaction rates for x-ray burst models respectively, and both emphasized the urgent need for improved experimental data near $^{60}$Zn. Schatz~\emph{et al.}~\cite{Schatz2017} found the 40~keV literature uncertainty~\cite{Wang2021} of $^{61}$Ga to be one of only a few remaining mass uncertainties that have significant impact on predicted x-ray burst light curves. Based on this result and possible underestimation of systematic uncertainties in earlier measurements, they recommended a more accurate re-determination of this mass value. The analysis by Cyburt~\emph{et al.}~\cite{Cyburt2016} and other reaction rate sensitivity studies~\cite{Parikh2008,Meisel2018} have identified the unmeasured $^{61}$Ga(p,$\gamma)^{62}$Ge reaction rate as a key ingredient for reliable burst models. Accurate mass values for $^{61}$Ga and $^{62}$Ge are also important inputs for reliable calculations of this reaction rate. In principle, it is also possible that the $\textit{rp}$ process at least partially bypasses the $^{60}$Zn waiting point via the $^{59}$Zn(p,$\gamma)^{60}$Ga(p,$\gamma)^{61}$Ge two-proton capture sequence. The strength of this branching depends sensitively on the $^{59}$Zn(p,$\gamma)^{60}$Ga $\mathrm{Q}$ value (see Eqn.~\ref{eqn:2p_capture_rate}) and therefore on the previously unknown $^{60}$Ga mass. 

At $A \approx 60 - 100$, many of the masses of neutron-deficient nuclei along the $\textit{rp}$-process path have not been measured accurately~\cite{Schatz2017}, oftentimes forcing astrophysical models to rely on mass predictions. With typical RMS deviations of several 100~keV~\cite{Sobiczewski2018}, global mass models are usually insufficient for astrophysical applications. Better mass predictions can be obtained with local mass extrapolations. With theoretical Coulomb displacement energies, the mass of a proton-rich nucleus can be deduced from the usually well-known mass of its mirror nucleus, yielding typical mass accuracies of $\approx 100$~keV~\cite{Brown2002,Schatz2017}. Even smaller prediction uncertainties have been demonstrated with mass relations involving more than two nuclei, such as the Garvey-Kelson mass relations~\cite{Garvey1966,Zhang2018} or the isobaric multiplet mass equation~\cite{MacCormick2014b,Ong2017}. 

According to the isobaric multiplet mass equation (IMME), the mass excesses (ME) of an isobaric multiplet, characterized by the total isospin $T$ and a given set of other quantum numbers $\alpha$, follow a quadratic form as function of the isospin projection $T_z=(N-Z)/2$:
\begin{equation}\label{eqn:IMME}
    \mathrm{ME}(\alpha, T,T_z) = a(\alpha ,T) + b(\alpha ,T) \, T_z + c(\alpha ,T) \, T_z^2 \,.
\end{equation}
The coefficients $a$, $b$ and $c$ are either determined by fitting experimental mass excesses~\cite{Lam2013,MacCormick2014b} or from theoretical predictions~\cite{Klochko2021}. The IMME can be used to predict masses of proton-rich nuclei with uncertainties as low as a few 10~keV~\cite{MacCormick2014b}. It further provides a conceptually simple, yet powerful framework to study the charge-dependence of the nuclear interaction~\cite{Lam2013b}. Experimental IMME coefficients are crucial for the construction of isospin non-conserving (INC) shell-model Hamiltonians, since they are used to adjust the charge-dependent terms in the Hamiltonian~\cite{Ormand1989}. The inclusion of phenomenological INC forces has been shown to be important for shell-model calculations in the $sd$ shell~\cite{Lam2013b} and the lower $pf$ shell~\cite{Kaneko2013}. The situation is more diffuse in the upper $pf$ shell~\cite{Kaneko2014,Kaneko2015}, as published compilations~\cite{Lam2013,MacCormick2014b} of experimental IMME coefficients for $T=1$ and higher-order multiplets currently only extend up to $A \approx 60$. More experimental data on isobaric multiplets at higher masses is needed to assess the importance of INC forces in the upper $pf$ shell~\cite{Kaneko2014,Kaneko2015} and, if necessary, to tune their associated strength parameters~\cite{Puentes2020}. Since charge-dependent microscopic forces have so far failed to reproduce the experimental data in the $pf$ shell~\cite{Gadea2006,Ormand2017}, a well-tuned phenomenological Hamiltonian would offer an attractive alternative for accurate predictions for nuclear structure and astrophysics.

In this work, mass measurements of neutron-deficient gallium isotopes near the proton drip line are reported. We present a significantly more precise mass value for the key $\textit{rp}$-process nucleus $^{61}$Ga and demonstrate its impact on the reliability of x-ray burst models. We further report the first direct mass measurement of $^{60}$Ga and discuss its implications for a $^{60}$Zn $\textit{rp}$-process bypass, the location of the proton drip line, and the $T=1$ IMME. The experiment was performed at TRIUMF's Ion Trap for Atomic and Nuclear Science (TITAN)~\cite{Dilling2003}, a series of charged-particle traps for precision mass measurements and in-trap decay spectroscopy of short-lived isotopes. For the presented measurements, TITAN's multiple-reflection time-of-flight mass spectrometer~\cite{Jesch2017} was deployed to leverage its combination of high mass accuracy, high sensitivity and drastic suppression of isobaric contamination. 

\section{\label{sec:experiment}Experiment}

The gallium isotopes of interest were produced at the TRIUMF Isotope Separator and Accelerator (ISAC) facility~\cite{Dilling2014ISAC} by impinging a 480~MeV, 55~$\mu$A proton beam onto a ZrC target. The generated nuclides were element-selectively ionized by TRIUMF's Resonant Laser Ion Source TRILIS~\cite{Lassen2009}. The resulting singly charged ions were then accelerated to a transport energy of 20~keV and sent through ISAC's two-stage mass separator~\cite{Bricault2002} for selection of the mass unit of interest. 

The continuous rare isotope beam was guided into TITAN's helium-gas-filled radiofrequency quadrupole (RFQ) cooler \& buncher~\cite{Brunner2012}, which supplied cold ion bunches at an output rate of 50~Hz. %TODO: Add exp. figure?
Using a pulsed drift tube, the ejected ion packets were then reduced to a transport energy of 1.3~keV and subsequently guided to TITAN's multiple-reflection time-of-flight mass spectrometer (MR-TOF-MS). 

The MR-TOF-MS~\cite{Jesch2017} is comprised of a helium-gas-filled RFQ transport section ($p \approx 2\times10^{-2}$~mbar) and a time-of-flight (TOF) analyzer ($p < 1\times10^{-7}$~mbar) with two opposing ion mirrors~\cite{Yavor2015}. The TOF analyzer is framed by a linear Paul trap for preparation of cold ion bunches on the injection end, and a TOF detector (ETP MagneTOF\texttrademark) on the opposite end. After final cooling and preparation in the injection trap, the ions were injected into the TOF analyzer and underwent gradual mass separation by many reflections between the isochronous ion mirrors. A dynamic time-focus shift~\cite{Dickel2017b} was performed by applying a suitable set of mirror voltages during the first reflection. Independent of the number of ion reflections, this procedure guarantees narrow TOF foci, both at the detector and near the injection trap.  
To perform a mass measurement, after a well-defined number of laps in the TOF analyzer, referred to as isochronous turns (IT), the far-side ion mirror was opened to eject the ions towards the detector. The ion arrival times at the detector were recorded using a time-to-digital converter and the dedicated Mass Acquistion software (MAc)~\cite{BergmannMSc2015,Dickel2019}. More details on technical aspects and the performance of TITAN's MR-TOF-MS can be found in~\cite{Reiter2021}. 

\begin{figure}[htb]
\centering
\includegraphics[width=1.0\linewidth,trim=0 0 0 0]{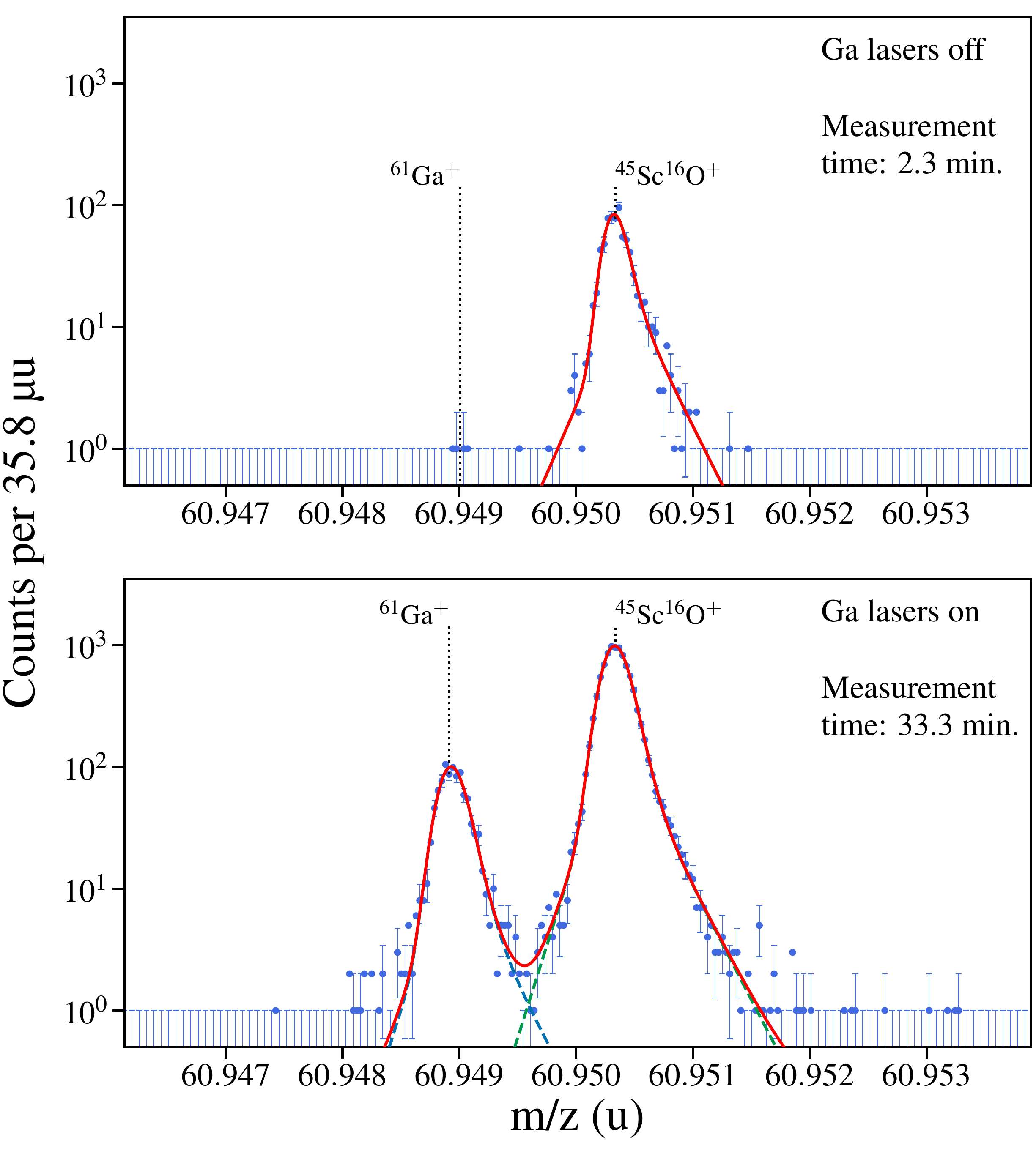}
\caption{\label{fig:Ga61_spec} Mass spectra acquired at $A=61$ with the ionizing laser beam blocked (top) and unblocked (bottom). The laser-induced increase of the relative intensity of the lower-mass peak provided additional verification of the gallium peak assignment. For both spectra mass-selective retrapping was used to suppress contaminant peaks outside the shown mass range. The solid red lines mark maximum-likelihood fits obtained using a hyper-EMG model function~\cite{Purushothaman2017} with two positive and one negative exponential tails. Colored dashed lines indicate the fit components associated with each peak. For clarity, error bars are only shown for every second bin. The total measurement durations are given in the respective mass spectra.}
\end{figure}

In this experiment, the ions of interest were kept in the TOF analyzer for 327--400~IT, corresponding to total flight times of 5.1--6.2~ms. At each mass unit, unambiguous identification of the peak of interest was ensured by verifying the more than fivefold reduction in gallium yield when the ionizing TRILIS laser beam was blocked (see Fig.~\ref{fig:Ga61_spec}).

In the case of overwhelming isobaric contamination, so-called mass-selective retrapping~\cite{Dickel2017} can be used to purify the ion beam within the MR-TOF-MS, thus deploying the mass spectrometer as its own isobar separator. This operation mode is facilitated by splitting each MR-TOF-MS cycle into two subcycles of identical length. The first subcycle is set aside for optional beam purification; the second one is used for the mass measurement. 
The device is then operated in either of the following modes:
\begin{enumerate}
    \item \emph{Regular mass measurement:} Throughout the first subcycle, the ion samples are kept in the injection trap. At the beginning of the second subcycle, they are injected into the TOF analyzer for the mass measurement. 
    \item \emph{Mass measurement with mass-selective retrapping:} At the beginning of the first subcycle, the ion samples are injected into the TOF analyzer for mass separation via time of flight. Once the ions of interest have acquired sufficient separation from undesired isobaric contamination, they are selectively retrapped and recooled in the injection trap. Meanwhile, contaminant ions remaining in the TOF analyzer are removed by electrostatic deflection. In the second subcycle, the purified ion samples are re-injected for the mass measurement, analogous to operation mode (1).
\end{enumerate}
Providing mass separation powers of several $100\,000$ and background suppression factors as high as $10^{4}$~\cite{Reiter2021}, mass-selective retrapping allows one to leverage the characteristics of the isotope-separation on-line (ISOL) method, namely high rare isotope production yields at the expense of heavy isobaric contamination. After the first utilization in~\cite{Beck2021}, the technique has been successfully used in several other on-line measurements at TITAN~\cite{Mukul2021,Izzo2021}.

\begin{figure}[htb]
\centering
\includegraphics[width=1.0\linewidth]{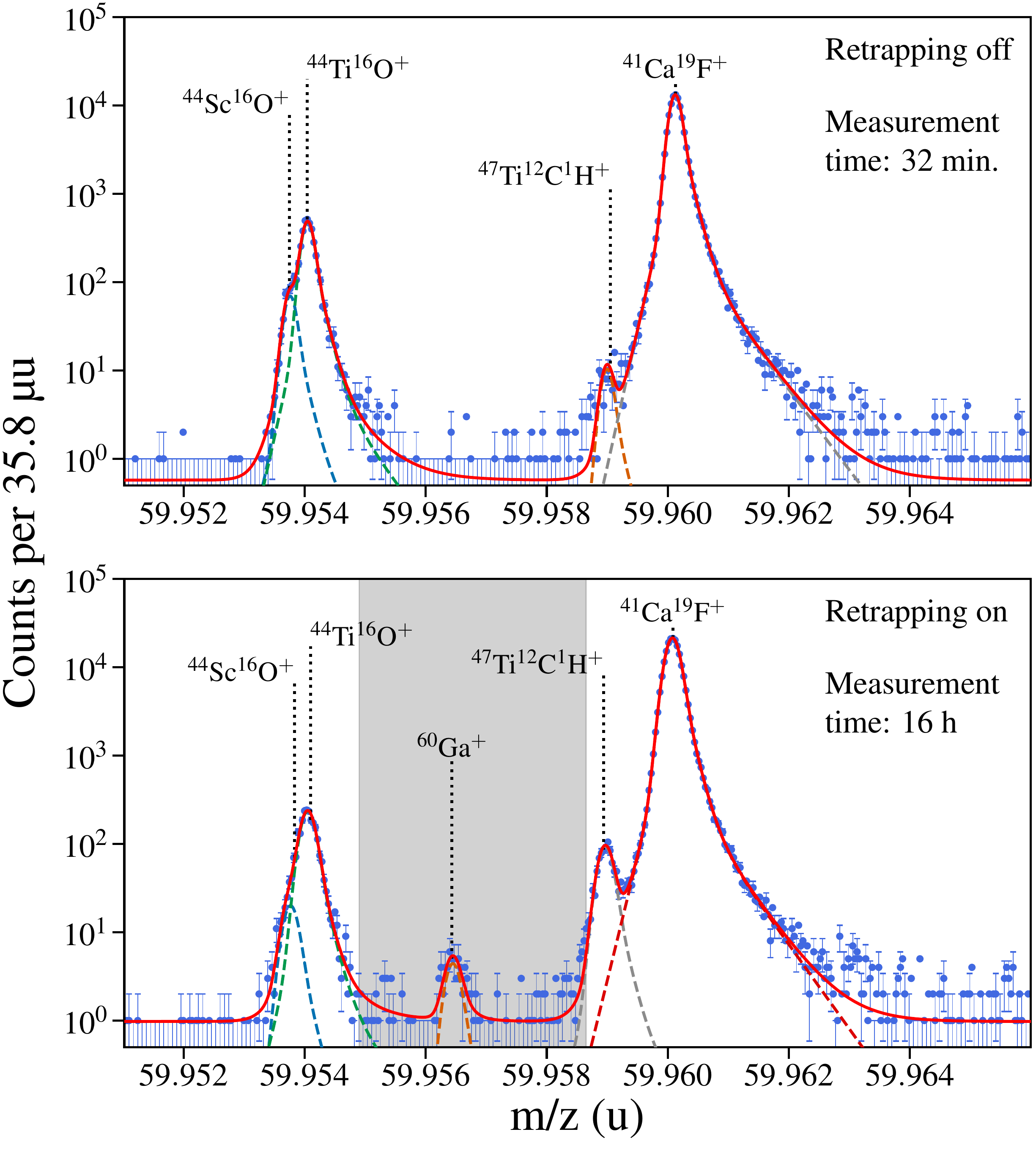}
\caption{\label{fig:Ga60_spec} Mass spectra acquired at $A=60$ with mass-selective retrapping turned off (top) and on (bottom). The intensities of peaks outside the retrapping TOF gate (gray-shaded area) were suppressed by up to two orders of magnitude, enabling a fiftyfold increase of the incoming rare isotope beam intensity and resulting in the observation of the minute $^{60}$Ga signal. The solid red lines mark maximum-likelihood fits obtained using a hyper-EMG model function~\cite{Purushothaman2017} with two positive and two negative exponential tails. Colored dashed lines indicate the fit components associated with each peak. For clarity, only for every second bin is the Poisson counting uncertainty indicated by an error bar. The total measurement durations are given in the respective mass spectra.}
\end{figure}

In this experiment, mass-selective retrapping was used in the mass measurements of $^{60}$Ga and $^{61}$Ga. The suppression of isobaric background allowed the incoming rare isotope beam intensity to be raised by factors of 50 and 5 in the cases of $^{60}$Ga and $^{61}$Ga, respectively, while keeping the total event rate on the detector at $\ll1$ per cycle. This greatly reduced the measurement durations, without adding systematic uncertainties due to ion-ion interactions. The use of mass-selective retrapping proved essential for the first direct mass measurement of $^{60}$Ga. Only the strong suppression of close-lying isobaric contaminants enabled the minute $^{60}$Ga signal (detected rate $\approx 7\times 10^{-4}$ pps) to be elevated above the extended tails of contaminant peaks with orders of magnitude higher intensity (see Fig.~\ref{fig:Ga60_spec}). Out of a total rate of $\approx 10\,000$~pps delivered to the TITAN beam line only $\approx 0.025$~pps corresponded to $^{60}$Ga ions. This highlights the remarkable background-handling capability and sensitivity of \text{TITAN's} MR-TOF-MS. 

\section{\label{sec:analysis} Data analysis}
The data analysis followed the general methodology presented in~\cite{Ayet2019}. In the following, we briefly reiterate the most essential analysis steps and point out noteworthy departures from the procedures described in~\cite{Ayet2019}. Most notably, for the present publication an open-source Python package~\cite{emgfit_v0_3_5} for fitting low-count mass spectra with hyper-exponentially modified Gaussian (hyper-EMG)~\cite{Purushothaman2017} line shapes was developed. Details on the statistical and numerical methods implemented in the fitting package as well as results from extensive tests with synthetic spectra will be reported in a forthcoming publication~\cite{emgfit}.

The measured times of flight $t$ were converted to mass-to-charge ratios ($\frac{m}{z}$) using the following calibration equation: 
\begin{equation}
    \frac{m}{z} = \frac{c\, (t-t_0)^2}{(1+N_\mathrm{IT} \, b)^2},
\end{equation}
where $c$, $b$ and $t_0$ are calibration parameters and $N_\mathrm{IT}$ is the number of completed isochronous turns. The calibration constants $c$ and $t_0$ were determined before the beam time in an off-line measurement with several stable reference ions of well-known mass. Temperature drifts and fluctuations of the isochronous mirror voltages broaden the TOF peaks and degrade the mass resolving power. These effects were compensated with a time-resolved calibration (TRC)~\cite{Ayet2019}, in which the acquired data is split into calibration blocks with lengths on the order of 1--100~s and the calibration parameter $b$ is determined for each block separately by fitting a Gaussian distribution to a well-known isobaric reference peak. This procedure resulted in high mass resolving powers on the order of $R(\mathrm{FWHM})=\frac{m}{\Delta m} = 250\,000$. 

For each spectrum, a high-statistics reference peak was selected as the shape calibrant peak and was individually fitted using Pearson's chi-squared statistic $\chi^2_P$ ~\cite{Mighell1999,Dembinski2019} as the cost function to be minimized. The optimal model function for the description of the mass peaks was automatically determined by fitting the shape calibration peak with hyper-EMG distributions including successively higher orders of exponential tails on either side of the peak. Models for which the best-fit amplitude parameter of any exponential tail agreed with zero within its $1\sigma$-interval were excluded. Amongst the remaining hyper-EMG models, the one resulting in the lowest chi-squared per degree of freedom was selected. Ideally, the shape calibrant peak has high statistics and is baseline-separated from other peaks. For the spectra at $A=60$ and $A=61$ no fully separated peak was available as a shape calibrant (see Fig.~\ref{fig:Ga60_spec} and Fig.~\ref{fig:Ga61_spec}, respectively). Hence, the peak-shape calibration at $A=60$ was performed by fitting the neighboring $^{41}$Ca$^{19}$F$^{+}$ and $^{47}$Ti$^{12}$C$^{1}$H$^{+}$ peaks together, enforcing identical line shapes for both peaks. Similarly, at $A=61$ the peak shape was calibrated by simultaneously fitting the $^{61}$Ga$^{+}$ and $^{45}$Sc$^{16}$O$^{+}$ peaks.

All peaks in the mass range of interest were then simultaneously fitted in a binned maximum likelihood estimation (MLE) using the following Poisson likelihood ratio~\cite{Cash1979, Mighell1999} as cost function: % modified version of Cash's C-statistic 
\begin{equation}
    L = 2\sum_{i=1}^{N} \left[ f(x_i) - y_i + y_i \ln{\left(\frac{y_i}{f(x_i)}\right)} \right],
\end{equation}
where the summation runs over all $N$ bins, $y_i$ denotes the measured number of counts in the i-th bin and $f(x_i)$ is the respective model prediction. In contrast to a regular $\chi^{2}$ fit, this approach properly treats the counting statistics in each bin as a Poisson process and yields unbiased fit results, even in low-count scenarios~\cite{Cash1979}. 
Further, $L$ asymptotically converges to a $\chi^2$ distribution and, hence, provides a convenient goodness-of-fit measure, which has been shown to remain valid in low-statistics situations~\cite{Kaastra2017}.
To account for dark counts and ions reaching the detector after collisions with residual-gas atoms, a uniform background was added to the multi-peak fit model. Since the shape parameters were predetermined in the peak-shape calibration, only the peak amplitudes and positions, and the amplitude of the uniform background were varied in the MLE fits. The fit curves obtained at $A=60$ and $A=61$ are shown in Fig.~\ref{fig:Ga60_spec} and Fig.~\ref{fig:Ga61_spec}, respectively.

The ionic mass values $m_\mathrm{ion}$ and the (atomic) mass excesses ME were calculated as:
\begin{align}
     m_\mathrm{ion} &= \frac{(m/z)_\mathrm{cal,lit}}{(m/z)_\mathrm{cal,MLE}} \, (m/z)_\mathrm{MLE} \, z, \\ 
     \mathrm{ME} &= m_\mathrm{ion}  + z \, m_e - A \, u,
\end{align}
where $(m/z)_\mathrm{MLE}$ and $(m/z)_\mathrm{cal,MLE}$ denote the respective fit centroids of the ion-of-interest and mass-calibrant peak, $(m/z)_\mathrm{cal,lit}$ is the ionic literature mass of the calibrant species, $m_e$ is the electron mass and $z$ denotes the ion charge state. Electron binding energies of the singly charged ions were neglected since they are only relevant at mass precisions $\ll1$~keV. The final mass values were verified in an independent analysis with a different code~\cite{Smith2020} for least-squares fitting with hyper-EMG line shapes.

The final mass uncertainties were derived by adding the following contributions in quadrature: statistical uncertainty of the peak-of-interest centroid, statistical uncertainty of the mass-calibrant centroid, literature uncertainty of the calibrant mass, uncertainty of the TRC, peak-shape uncertainty, uncertainty from ion-ion interactions, and uncertainty from non-ideal ejection from the TOF analyzer. 

The uncertainties of the fitted peak positions were evaluated with Monte Carlo methods. The statistical uncertainties of the peak positions and amplitudes were estimated by performing a parametric bootstrap. Specifically, for each MLE fit 1000 synthetic data sets were created by sampling an identical number of events as in the measured spectrum from the best-fit model. The MLE fit was reperformed on all synthetic spectra and the statistical mass uncertainty was estimated as the sample standard deviation of the obtained peak centroids. Uncertainties in the peak-shape parameters were also propagated into the final mass values using a similar Monte Carlo approach. However, since all ion-of-interest and mass-calibrant peaks were sufficiently separated from other mass peaks, this contribution was found to be on a negligible level of $(\frac{\Delta m}{m})_\mathrm{PS} < 1 \times 10^{-8}$ in all cases.

Particular care was taken to evaluate possible systematic uncertainties arising in the fitting process. The effects of the finite width of the mass bins as well as the impact of the chosen fit ranges in the shape calibration and the final MLE fit were quantified and found to be negligible for all reported cases. 

Systematic mass shifts may arise when the voltages of the exit-side ion mirror are switched to release the ions towards the detector. These shifts are caused if ions of interest and calibrant ions probe different time-varying electric fields. The mass uncertainties related to such non-ideal ejection (NIE) were estimated following the procedure described in~\cite{Ayet2019} and resulted in uncertainty contributions of $(\frac{\Delta m}{m})_\mathrm{NIE}\approx 1.2 - 1.6 \times 10^{-7} $ for the reported cases.

The mass uncertainty due to the time-resolved calibration parameter $b$ was calculated as $(\frac{\Delta m}{m})_\mathrm{TRC}=2\frac{\Delta b}{b}$, where $\Delta b$ denotes the standard error of the mean variation of b between neighboring calibration blocks. This uncertainty contribution typically amounted to $(\frac{\Delta m}{m})_\mathrm{TRC}\approx 1\times10^{-7}$.

The magnitude of possible mass shifts due to ion-ion interactions in the TOF analyzer depends on the specific ion-optical tune, the count rate, the detection efficiency and the composition of the isobaric ion samples under study. Based on test measurements with stable molecular beams from the off-line ion source terminal~\cite{Jayamanna2008} for ISAC, the relative mass uncertainty from ion-ion interactions was estimated by scaling a conservative upper limit of $(\frac{\Delta m}{m})_\mathrm{ion-ion} = 1.7\times10^{-7}$ per detected ion to the respective count rate in each spectrum.
 
\section{\label{sec:results}Results}
The mass values obtained in this work for $^{60 - 63}$Ga are compiled in Table~\ref{tab:ME_vals}, along with literature values from the 2020 Atomic Mass Evaluation (AME2020)~\cite{Wang2021} and results from earlier indirect experimental determinations. 
In the following, we describe our experimental results for each isotope and compare them to previously available data.

\begin{table*}[tb]%The best place to locate the table environment is directly after its first reference in text
\caption{\label{tab:ME_vals}
Overview of the mass excess values obtained in this work (labelled TITAN) in comparison to literature values from AME2020~\cite{Wang2021}, the semiempirical estimate from Mazzocchi et al.~\cite{Mazzocchi2001} and the indirect experimental result from Orrigo et al.~\cite{Orrigo2021}. The last column indicates the deviations between our mass excess values and the ones from AME2020. The respective ion species used for the mass calibration and the number of acquired gallium events $N_\mathrm{events}$ are also listed. Half-lives $T_{1/2}$ are taken from ENSDF~\cite{Browne2013, Zuber2015, Nichols2012, Erjun2001}. 
}
\renewcommand{\arraystretch}{1.25}
\begin{ruledtabular}
\begin{tabular}{ld@{\hspace{-0.3cm}}d@{\hspace{-0.9cm}}cccccc} 
\textrm{Nuclide}&
\textrm{T$_{1/2}$}&
\textrm{N$_\mathrm{events}$}&
\textrm{Calibrant species}&
\multicolumn{5}{c}{Mass excess (keV)} \\
\cmidrule(lr){5-9}
&
\textrm{(ms)}&
&
& 
\textrm{TITAN}&
\textrm{AME2020}&
\textrm{Ref.~\cite{Mazzocchi2001}}&
\textrm{Ref.~\cite{Orrigo2021}}&
\textrm{TITAN - AME2020} \\
\colrule
$^{60}$Ga & 70 & 41 & $^{41}$Ca$^{19}$F$^{+}$ & $-40005(30)$ & $-39590\footnotemark[1](200\footnotemark[1])$ & $-40010(60)$ & $-37045(15)$ & $-415\footnotemark[1](202\footnotemark[1])$ \\
$^{61}$Ga & 167 & 1001  & $^{45}$Sc$^{16}$O$^{+}$ & $-47114(12)$ & $-47130(40)$ & - & - & $16(42)$ \\
$^{62}$Ga & 116 & 954 & $^{46}$Ti$^{16}$O$^{+}$ & $-51992(14)$ & $-51987.0(0.6)$ & - & - & $-5(14)$  \\
$^{63}$Ga & 3240 & 48099 & $^{47}$Ti$^{16}$O$^{+}$ & $-56563(14)$ & $-56547.1(1.3)$ & - & - & $-16(14)$ \\
\end{tabular}
\end{ruledtabular}
\footnotetext[1]{Extrapolated values based on trends from the mass surface~\cite{Huang2021}.}
\end{table*}

\subsection{$^{63}$Ga}
The literature mass of $^{63}$Ga is well-established in AME2020 with an uncertainty of 1.3~keV, based on a Penning trap measurement at ISOLTRAP~\cite{Guenaut2007}. Despite the high precision of the literature value, we re-determined the mass of $^{63}$Ga to (1) provide an independent confirmation of the Penning trap measurement, and (2) diagnose the beam composition. Objective (2) is particularly relevant in the presence of isobaric beam contamination, since tracing isotopic chains through multiple mass units provided additional confirmation for peak assignments. 

Our final $^{63}$Ga mass excess value of $-56563(14)~\mathrm{keV}$ is the variance-weighted mean of three separate measurements at 327 IT, 337 IT \& 347 IT, respectively. %, based on a total of $\approx$ 48100 recorded $^{63}$Ga events. 
The variance weights were derived from the corresponding statistical uncertainties. Since the measurements were performed at total rates of $\approx0.9$ detected particles per cycle, the final mass uncertainty is dominated by the conservative limit placed on mass shifts from  ion-ion interactions, $(\frac{\Delta m}{m})_\mathrm{ion-ion} = 1.7\times 10^{-7}$. The obtained mass value agrees with the more precise AME2020 value within 1.2$\sigma$.

\subsection{$^{62}$Ga}
Due to the superallowed nature of its $\beta$ decay, $^{62}$Ga has been the subject of extensive experimental efforts~\cite{Davids1979,Hyland2006,Finlay2008,MacLean2020}. The evaluated literature mass in AME2020 is known to sub-keV precision based on a Q$_\mathrm{EC}$-value measurement by the JYFLTRAP Penning trap~\cite{Eronen2006}. 

In this work, we obtained a $^{62}$Ga mass excess of $-51992(14)~\mathrm{keV}$, which agrees with the literature value within the quoted uncertainty. The measurement was performed at a total rate of $\approx0.9$ detected ions per cycle; hence, the mass uncertainty is dominated by the estimated limit on shifts due to ion-ion interactions, $(\frac{\Delta m}{m})_\mathrm{ion-ion} = 1.7 \times 10^{-7}$.

\subsection{$^{61}$Ga}
AME2020 lists the mass excess of $^{61}$Ga as $-47130(40)$~keV based on a $\beta$-endpoint measurement~\cite{Weissman2002} and a direct mass measurement at the CSRe storage ring~\cite{Tu2011}. These inputs contributed almost equally to the evaluated value~\cite{Huang2021}. A more precise determination of this mass value has been recommended~\cite{Schatz2017} due to its astrophysical importance and concerns about possible underestimation of systematic uncertainties in the previous measurements.

The experimental mass excess of $-47114(12)$~keV reported here confirms the previous results within the given measurement uncertainties. With a relative mass uncertainty of $\frac{\Delta m}{m} = 2.1 \times 10^{-7}$, our result is three times more precise than the AME2020 value and four times more precise than the storage ring measurement.

\subsection{$^{60}$Ga}
Previous to our study, the ground-state mass of $^{60}$Ga had never been measured directly. Due to the lack of experimental data, multiple theoretical estimates~\cite{Ormand1997,Cole1999,Brown2002,Schatz2017} of the mass excess or one-proton separation energy of $^{60}$Ga have been derived from Coulomb displacement energies (CDEs). 

Until recently, the only detailed experimental investigation of $^{60}$Ga was a decay study performed by Mazzocchi~\emph{et al.}~\cite{Mazzocchi2001} in 2001. This study experimentally established the half-life of $^{60}$Ga and identified the isobaric analog state (IAS) in $^{60}$Zn. Based on the well-known mass of the IAS in $^{60}$Zn and systematics of experimental CDEs~\cite{Antony1997}, they deduced a semi-empirical estimate for the $^{60}$Ga mass excess of $-40010(60)$~keV and a corresponding one-proton separation energy of $40(70)$~keV. 

A recent study of the $\beta$-decay of $^{60}$Ge by Orrigo~\emph{et al.}~\cite{Orrigo2021} determined the mass excess of the IAS in $^{60}$Ga and provided a more precise derivation of the $^{60}$Ga ground-state mass. In that experiment, the $^{60}$Ge decays populated excited states in the daughter nuclide $^{60}$Ga, which then either de-excited via gamma emission or decayed to states in $^{59}$Zn by emission of a proton ($\beta$-delayed proton emission). Combining the measured proton energies with the well-known $^{59}$Zn ground-state mass, Orrigo~\emph{et al.} determined the mass excess of the IAS in $^{60}$Ga as $-37045(15)$~keV. Under the hypothesis that the IAS de-excites through the observed $1775$~keV and $837$~keV gamma transitions, they further deduced a $^{60}$Ga ground-state mass excess of $-40016(15)$~keV and a corresponding proton separation energy of $90(15)$~keV. Their underlying hypothesis about the gamma de-excitation of the IAS is strengthened by mirror symmetry arguments. However, they conceded that the discrepancy between the observed Fermi transition strength ($B(F) = 3.1(1)$) and theoretical expectation ($B(F) = 4$) could be caused by undetected gamma transitions~\cite{Orrigo2021} from the IAS. This raises questions about the reliability of the mass excess deduced in that study. 

The present work marks the first direct measurement of the $^{60}$Ga ground-state mass. The $41(8)$ measured $^{60}$Ga events (see Fig.~\ref{fig:Ga60_spec}) were accumulated over a total measurement duration of $\approx 16$ hours and resulted in a statistically-limited mass uncertainty of $\frac{\Delta m}{m} = 5\times 10^{-7}$. Since our direct mass measurement is essentially free from additional assumptions, our $^{60}$Ga mass value can be considered more reliable than previous results. Our mass excess value of $-40005(30)$~keV lies more than 400~keV below the AME2020~\cite{Wang2021} value of $-39590^{\#}(200^{\#})$~keV extrapolated from trends in the mass surface~($^{\#}$), but is consistent with the AME2003~\cite{Audi2003} value of $-40000^{\#}(110^{\#})$~keV. In this region of the nuclear chart, multiple other works~\cite{DelSanto2014,Orrigo2014,Orrigo2016,Orrigo2021} have found similar discrepancies with extrapolated values from atomic mass evaluations after AME2003. Our measured $^{60}$Ga mass adds an important anchor point to mitigate these issues in future mass evaluations. Additional mass measurements in this region are desirable~\cite{Orrigo2021}. Our result is in excellent agreement with the mass excess of $-40005(100)$~keV obtained from theoretical CDE~\cite{Brown2002} and the measured mass of the $^{60}$Cu mirror nucleus~\cite{Wang2017}. It also shows $1\sigma$ agreement with the respective values deduced by Mazzocchi~\emph{et al.}~\cite{Mazzocchi2001} and Orrigo~\emph{et al.}~\cite{Orrigo2021}. 

\section{\label{sec:discussion}Discussion}
In the following, we discuss the implications of our data in the context of the proton drip line along the gallium chain, give an update on the $T=1$ IMME multiplets in the $pf$ shell, and use the IMME to predict the unmeasured mass excess of $^{61}$Ge. Further, the consequences of the improved mass data for models of type I x-ray bursts are reported. 

\subsection{\label{sec:dripline-position}Location of the proton drip line at $\mathbf{Z=31}$}
The proton drip line is defined as the point along an isotopic chain where the (one-)proton separation energy, 
\begin{equation}
    S_\text{p}(Z,A)=\mathrm{ME}(Z-1,A-1) + \mathrm{ME}(^{1}\mathrm{H}) - \mathrm{ME}(Z,A),
\end{equation}
becomes negative. As a fundamental boundary of nuclear stability, it provides an ideal testing ground for nuclear mass models and local mass predictions. Theoretical mappings of the drip line can further be used to identify candidates for exotic two-proton radioactivity (see for example~\cite{Ormand1997}).

Since the Coulomb barrier allows for some nuclei to exist beyond the proton drip line, its exact localization is an experimental challenge and requires accurately measured $S_\text{p}$ values, ideally obtained from direct mass measurements~\cite{Rauth2008,Thoennessen2004}. The gallium chain is a particularly challenging case, since $^{60}$Ga and $^{61}$Ga lie very close to the expected proton drip line. Combining our mass results with the well-established literature masses for the neighboring zinc isotopes~\cite{Wang2021}, we obtain purely experimental proton separation energies for $^{60-63}$Ga (see Table~\ref{tab:S_p_values}) and can investigate the location of the proton drip line along this isotopic chain. 

\begin{table}[b]
\caption{\label{tab:S_p_values}
Proton separation energies obtained in this work compared to values from AME2020~\cite{Wang2021}, and indirectly deduced results from measurements by Mazzocchi \emph{et al.}~\cite{Mazzocchi2001} and Orrigo \emph{et al.}~\cite{Orrigo2021}.}
\renewcommand{\arraystretch}{1.25}
\begin{ruledtabular}
\begin{tabular}{ccccc}
\textrm{Nuclide}&
\multicolumn{4}{c}{\textrm{S$_\mathrm{p}$ (keV)}} \\
\cmidrule(lr){2-5}
&
\textrm{TITAN}&
\textrm{AME2020}&
\textrm{Ref.~\cite{Mazzocchi2001}}&
\textrm{Ref.~\cite{Orrigo2021}} \\
\colrule
$^{60}$Ga & 78(30)  & $-340\footnotemark[1](200\footnotemark[1])$ & 40(70) & 90(15)\\
$^{61}$Ga & 229(12)  & 250(40) & - & - \\
$^{62}$Ga & 2932(21)  & 2927(16) & - & - \\
$^{63}$Ga & 2684(14)  & 2668.0(1.4) & - & - \\
\end{tabular}
\end{ruledtabular}
\footnotetext[1]{Extrapolated values based on trends from the mass surface~\cite{Huang2021}.}
\end{table}

\begin{figure*}[tb]
\includegraphics[width=0.98\linewidth]{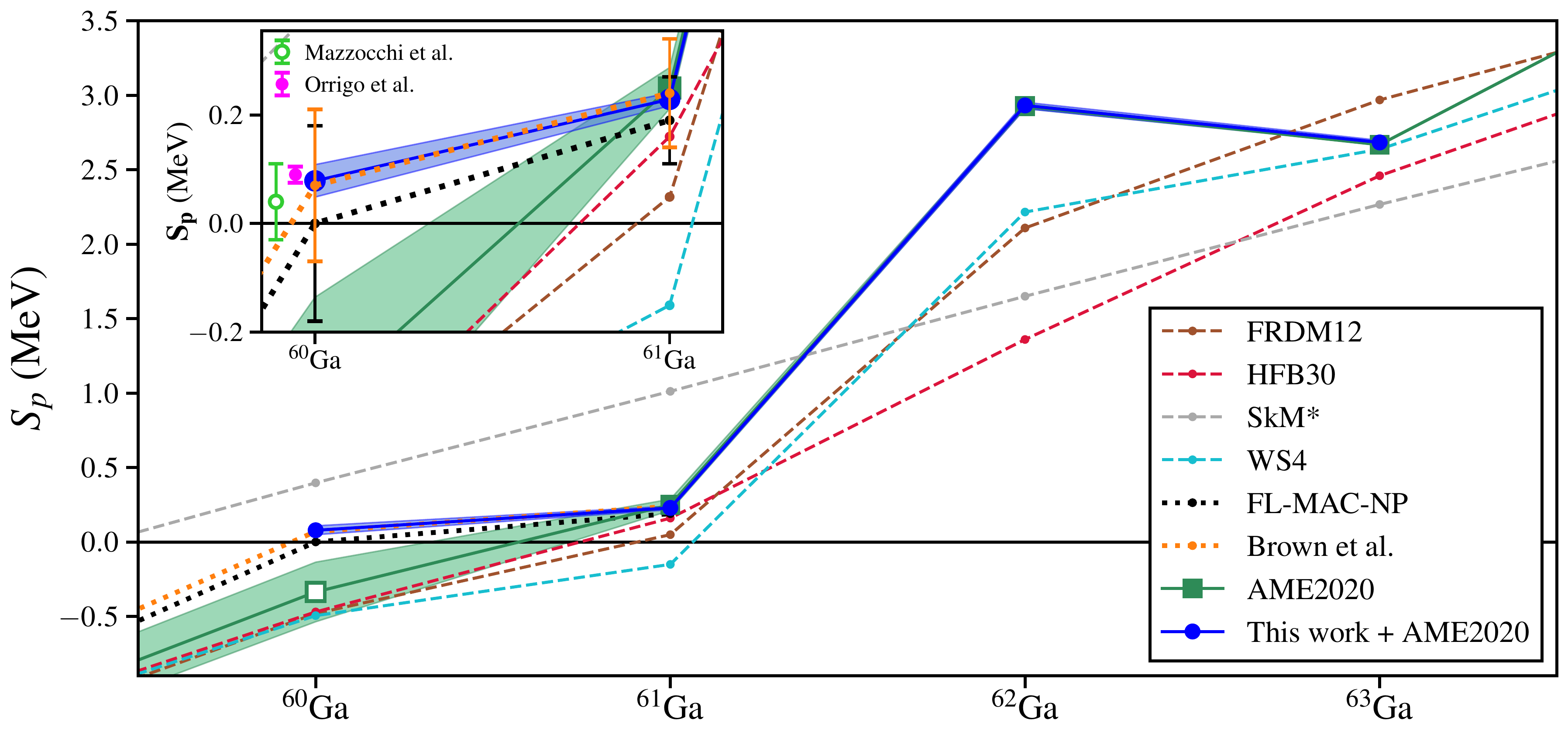}
\caption{\label{fig:S_p_trend} Trend of the proton separation energies along the gallium chain. The inset shows $A=60$ and $A=61$ with expanded ordinate scale. For clarity, the $S_\text{p}(^{60}\mathrm{Ga})$ values of Mazzocchi~\emph{et al.}~\cite{Mazzocchi2001} and Orrigo~\emph{et al.}~\cite{Orrigo2021} are only shown in the inset and were slightly offset horizontally. Due to the small proton separation energies of $^{60}$Ga and $^{61}$Ga, global mass models (dashed lines), and estimates from systematic trends of experimental CDEs (open circle) or the mass surface (open square) cannot reliably predict the location of the proton drip line. Because of the associated uncertainties, drip line predictions from theoretical CDEs (dotted lines) from FL-MAC-NP~\cite{Klochko2021} and Brown~\emph{et al.}~\cite{Brown2002} are likewise insufficient. Only precision measurements (dark blue and magenta dots) can establish $^{60}$Ga to be proton-bound and constrain the location of the drip line along this isotopic chain.}
\end{figure*}

Fig.~\ref{fig:S_p_trend} shows the experimental proton separation energies obtained in this work in comparison to literature values from AME2020 and mass model predictions. The selected global mass models, namely FRDM2012~\cite{Moller2012}, HFB30~\cite{Goriely2016}, SkM*~\cite{Bartel1982}, and WS4~\cite{Wang2014}, are a representative set of well-established models, commonly used for astrophysical calculations. None of these global mass models closely reproduce the experimental results. The spread in the model predictions illustrates the typical accuracy of global mass models, which has been shown to be on the order of $500$~keV~\cite{Sobiczewski2018}. In terms of drip line predictions, this translates into a scatter by circa one isotope. 
Except for SkM*, the global models considered here underpredict the proton separation energy of $^{60}$Ga.

Recent studies~\cite{Niu2018,Neufcourt2020,Shelley2021} have exploited Bayesian inference and machine learning to improve the predictive power of global mass models. The proton-drip line mappings from~\cite{Neufcourt2020b} deployed Bayesian model averaging of various mass models and predicted $^{60}$Ga as a drip line nucleus with $>80\%$ probability for it to be proton-bound. This is in agreement with experiment and indicates the potential of such refinements to global mass models. More experimental tests are needed to evaluate the reliability of mass predictions from these novel approaches.

Decent agreement with the experimental data at $A\leq 61$ is obtained with proton separation energies from theoretical CDEs. A recent mapping of the proton drip line~\cite{Klochko2021} used CDEs deduced from theoretical IMME $b$ coefficients. In this approach, the $b$ coefficients were calculated with FL-MAC-NP, a modified version of the macroscopic part of the finite-range liquid drop model (FRLDM)~\cite{Moller1995}, and then used to derive the masses of proton-rich nuclei from the literature masses of their neutron-rich mirror isotopes. FL-MAC-NP (black dotted line in Fig.~\ref{fig:S_p_trend}) places $^{60}$Ga right on the proton drip line, with $S_\text{p}(^{60}\mathrm{Ga}) = 0(180)$~keV. An earlier mapping of the proton drip line was presented by Brown~\emph{et al.}~\cite{Brown2002} based on CDEs from Skyrme Hartree-Fock calculations. Mass tables derived from these CDEs have been used for one-zone x-ray burst calculations in several studies~\cite{Brown2002,Rodriguez2005,Schatz2017,Cyburt2016}, including the present work. The CDEs from Brown~\emph{et al.} (orange dotted line in Fig.~\ref{fig:S_p_trend}) show impressive agreement with the experimental trend and suggest $^{60}$Ga to be proton-bound with $S_\text{p}(^{60}\mathrm{Ga})=70(140)$~keV. Both theoretical CDE estimates of $S_\text{p}(^{60}\mathrm{Ga})$ are compatible with experimental values, but the prediction uncertainties are too large to pin down the exact location of the proton drip line along the gallium chain.

Based on our direct mass measurement of $^{60}$Ga, we obtain $S_\text{p} = 78(30)$~keV, demonstrating $^{60}$Ga to be proton-bound with $2.6\sigma$ confidence. This is in strong disagreement with the estimated value from trends of the mass surface ($^{\#}$)~\cite{Huang2021} given in AME2020~\cite{Wang2021} ($S_\text{p}=-336^{\#}(200^{\#})$~keV), which suggests $^{60}$Ga to be proton-unbound. Our result shows $1\sigma$ agreement with the semiempirical estimate by Mazzocchi~\emph{et al.}~\cite{Mazzocchi2001} ($S_\text{p}=40(70)$~keV) and confirms the more precise experimental value of $90(15)$~keV indirectly deduced by Orrigo~\emph{et al.}~\cite{Orrigo2021}. We emphasize that the latter, in contrast to our direct mass measurement, relies on assumptions about the gamma de-excitation of the IAS in $^{60}$Ga. The decent agreement with our value supports the assumptions made by Orrigo~\emph{et al.} Combined, these two experimental results yield a variance-weighted mean of $S_\text{p}(^{60}\mathrm{Ga}) = 88(18)$~keV, establishing the proton-bound nature of $^{60}$Ga, and constraining the location of the proton drip line at $Z=31$. 

To firmly establish the location of the proton drip line, it is also necessary to experimentally confirm the first isotope beyond the proposed drip line, in this case $^{59}$Ga, as proton-unbound. 
This question was addressed by Stolz~\emph{et al.}~\cite{Stolz2005,Stolz2005b} in an isotope search with the A1900 fragment separator at the National Superconducting Cyclotron Laboratory.
From the non-observation of $^{59}$Ga in fragmentation reactions induced by a $^{78}$Kr$^{34+}$ beam impinged on a beryllium target, they obtained very strong evidence that $^{59}$Ga is proton-unbound. The aggregated results from Stolz~\emph{et al.}, Orrigo~\emph{et al.} and this work provide very strong evidence that $^{60}$Ga is the last proton-bound gallium isotope and marks the location of the proton drip line. 

\subsection{Status of the $\mathbf{T=1}$ isobaric multiplet mass equation in the \textit{pf} shell}

The most recent comprehensive survey of experimental data on isobaric analog states (IAS) and the associated coefficients in the isobaric multiplet mass equation (IMME) was published  in 2014 by MacCormick~\emph{et al.}~\cite{MacCormick2014b}. In that evaluation, for the first time IAS were included in the mass network, correctly accounting for correlations between nuclei involved in the IAS multiplets. In cases where several observations of the same state had been made, sometimes through different reactions, the relative influence of each measurement was evaluated following the well-established procedures from~\cite{Wang2012}. In their survey, due to missing experimental data, the $T=1$ triplets could only be evaluated up to $A=58$ with gaps at $A=44$,~52~and~56. As demonstrated in~\cite{Puentes2020}, these gaps have all been filled by recent mass measurements of $^{44g,m}$V~\cite{Puentes2020}, $^{52g,m}$Co~\cite{Zhang2018} and $^{56}$Cu~\cite{Valverde2018}, respectively. The addition of our directly measured $^{60}$Ga mass completes the experimental data on the lowest-lying $T=1$ multiplets up to $A=60$.

We performed an updated evaluation of the quadratic IMME for the $T=1$ multiplets from $A=42$ to $A=60$. The selection of input data proceeded according to the scheme detailed in~\cite{MacCormick2014b}. Unless otherwise stated, ground-state masses were taken from AME2020~\cite{Wang2021}, and IAS identifications and the corresponding level energies were adapted from ENSDF, after careful evaluation of the input data sources.
The associated IMME coefficients were then obtained from least-squares fits to the resulting mass excesses of the multiplet members.

The data selection for the $T=1$ multiplet at $A=60$ was complicated by the ambiguous identification of the lowest-lying $T=1$ IAS in zinc, $^{60}$Zn$^{i}$. ENSDF~\cite{Browne2013} lists two candidate levels for $^{60}$Zn$^{i}$, with evaluated excitation energies $E_x=4852.2(7)$~keV and $E_x=4913.3(9)$~keV, respectively. Detailed review of the available data sources showed that neither of the two reaction studies~\cite{Greenfield1972,Schubank1989} which observed the higher-lying candidate level could assign it an unambiguous spin-parity or provide conclusive arguments for its identification as the IAS. The lower-lying candidate level was first observed in $^{58}$Ni($^{16}$O,$^{14}$C)$^{60}$Zn reactions~\cite{Pougheon1972} and suggested as the IAS, based on the agreement of the observed excitation energy with a CDE prediction for E$_x$($^{60}$Zn$^{i}$). The same level was later observed as a cascade of $\beta$-delayed gamma rays in $\beta\gamma$-spectroscopy studies by Mazzocchi~\emph{et.al}~\cite{Mazzocchi2001} and Orrigo~\emph{et al.}~\cite{Orrigo2021}, which both identified it as the $T=1$ IAS. In~\cite{Mazzocchi2001} this identification was supported by the coincident detection of the observed $\beta$-delayed $\gamma$ emission, and the agreement of the observed $\beta$-decay strength with the theoretical expectation for a pure Fermi transition. The latter suggests that potential unobserved $\gamma$ transitions only contributed marginally, if at all, to the $\gamma$ de-excitation of $^{60}$Zn$^{i}$. In agreement with NUBASE2020~\cite{Kondev2021}, we hence identified the lower-lying evaluated level at $E_x=4852.2(7)$~keV as $^{60}$Zn$^{i}$ and used its excitation energy in our experimental evaluation.

To compare the experimental results to shell-model predictions, we derived theoretical IMME coefficients from calculations in the full $pf$ shell. The calculations were performed using the NuShellX@MSU shell-model code~\cite{Brown2014} with the isospin-nonconserving Hamiltonian cdGX1A~\cite{Smirnova2017}, an extension of the effective interaction GXPF1A~\cite{Honma2004} with charge-dependent terms from~\cite{Ormand1989} and~\cite{Ormand1995}. The experimental and theoretical IMME coefficients obtained at $A=60$ and the respective excitation energy of $^{60}$Zn$^{i}$ are shown in Table~\ref{tab:IMME_coeffs_60u}. 

\begin{table}[tb]
\caption{\label{tab:IMME_coeffs_60u}
Experimental and theoretical IMME coefficients obtained for the lowest-lying $T=1$ multiplet at $A=60$.
The obtained $a$ coefficients were omitted as, for isospin triplets, they always equal the mass excess of the $T_z=0$ state.
}
\renewcommand{\arraystretch}{1.25}
\begin{ruledtabular}
\begin{tabular}{lc |cc}
Source of &
$E_{x}(T_z=0)$  &
$b$&
$c$\\
mass data &
(keV) & 
(keV) & 
(keV) \\
\colrule
This work + AME2020 & 4852.2(7)  & $-9170(15)$    & 147(15) \\
\hline
cdGX1A & 4235  &   $-9061.9$  &  103.5
\end{tabular}
\end{ruledtabular}
\end{table}

\begin{figure}[htb]
\includegraphics[width=1.0\linewidth, trim=1cm 0 0 0]{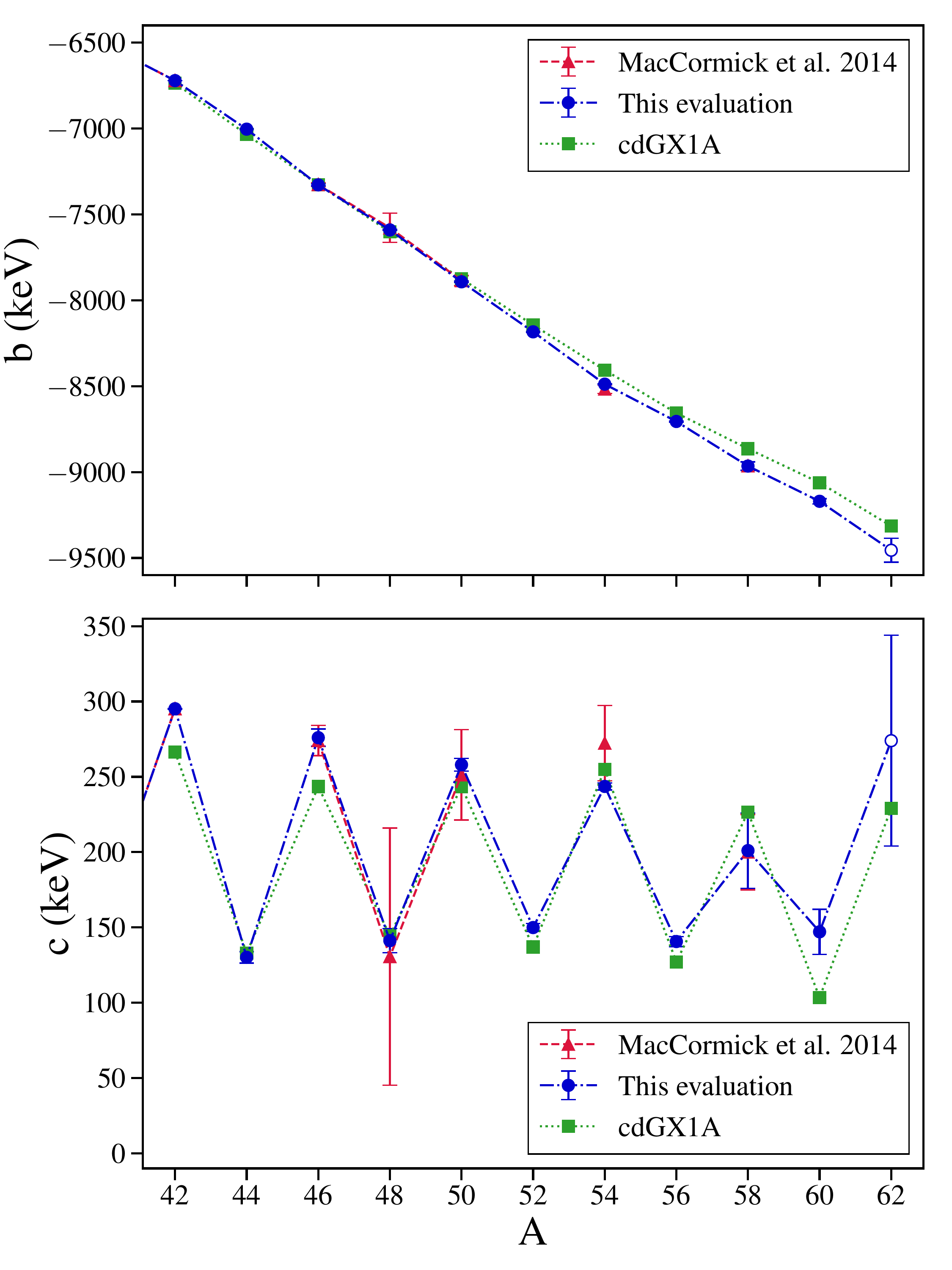}
\caption{\label{fig:IMME_trends} Trends of the IMME $b$ and $c$ coefficients for the lowest-lying $T=1$ multiplets in the $pf$ shell. The directly measured mass value for $^{60}$Ga from this work completes the experimental data on IMME triplets up to $A=60$. The updated experimental coefficients (blue dots) improve on an earlier evaluation~\cite{MacCormick2014b}  (red triangles) and exhibit fair agreement with shell-model predictions obtained with the cdGX1A Hamiltonian (green squares). Unfilled blue circles indicate fit results based on semiempirical input data (see text for details).}
\end{figure}

The resulting trends of the experimental and theoretical IMME coefficients are shown in Fig.~\ref{fig:IMME_trends}. The c coefficients for isospin triplets exhibit a staggering behavior, which has been attributed to Coulomb pairing effects~\cite{Jaenecke1966}, supplemented by smaller contributions from nuclear two-body interactions~\cite{Lam2013}. Comparison of our experimental $b$ and $c$ coefficients with the results from the 2014 survey~\cite{MacCormick2014b} clearly illustrates the reduced uncertainties due to the recent precision mass measurements~\cite{Zhang2018,Valverde2018,Puentes2020}. The coefficients for all lowest-lying triplets up to $A=60$ are now well constrained by experimental data, with the exception of $A=58$, where the current accuracy of the $c$ coefficient is insufficient to rule out a potential reduction in the staggering amplitude. Such a sudden change in the staggering pattern could indicate isospin impurities~\cite{MacCormick2014b} or a misidentified IAS. Currently, the experimental precision of this triplet is limited by the 50~keV uncertainty of the $^{58}$Zn mass deduced from the measured $\mathrm{Q}$ value of $^{58}\mathrm{Ni}(\pi^{+},\pi^{-})^{58}\mathrm{Zn}$~\cite{Seth1986}. To exclude a possible anomaly, a direct mass measurement of $^{58}$Zn with $<10$~keV accuracy is desirable.

The experimental IAS data for the triplet at $A=62$ is incomplete, since the mass of $^{62}$Ge has not been measured. However, Orrigo~\emph{et al.}~\cite{Orrigo2021} recently estimated a $^{62}$Ge mass excess of $-42258(140)$~keV. This estimate was obtained by adjusting the mass value such that the $\beta$-decay strength for their observed Fermi transition between the isobaric analog ground states of $^{60}$Ge and $^{62}$Ga matched the expected transition strength of $B(F)=2$. We used this mass excess to deduce semiempirical IMME coefficients for $A=62$. Within the large uncertainties, the semiempirical coefficients seamlessly extend the experimental trends, continuing both the expected drop in the $b$ coefficients and the staggering in the $c$ coefficients. To verify these tendencies, a ground-state mass measurement of the $T_z=-1$ triplet member $^{62}$Ge is needed.

The shell-model predictions obtained with the cdGX1A framework are overall in reasonable agreement with the experimental IMME trends, as quantified by RMS deviations of 59~keV and 23~keV for the $b$ and $c$ coefficients at $42 \leq A \leq 60$, respectively.  Increased discrepancies between theory and experiment are found at $A \geq 58$, i.e. beyond the $N=Z=28$ shell closures. Re-adjustment of the INC parametrization from~\cite{Ormand1989,Ormand1995} to the much expanded experimental data available today may alleviate these deviations and extend the high predictive power of INC Hamiltonians into the upper $pf$~shell.

\subsection{IMME prediction for the mass of $^{61}$Ge}
The mass of $^{61}$Ge has not yet been measured. We can combine our improved ground-state mass of $^{61}$Ga with experimental data on the other $T=3/2$ quadruplet members ($^{61}$Zn$^{i}$ and $^{61}$Cu) to give an IMME prediction for the mass of $^{61}$Ge. 

With our $^{61}$Ga ground-state mass excess and the literature excitation energy for the IAS~\cite{Zuber2015} ($E_x=3162(30)$~keV), we deduced a mass excess of $-43723(30)$~keV for the lowest $T=3/2$ IAS in $^{61}$Ga. Since the excitation energy listed for $^{61}$Zn$^{i}$ in ENSDF~\cite{Zuber2015} ($E_x=3380$~keV) is based on a reaction $\mathrm{Q}$ value reported without an uncertainty estimate~\cite{Okuma1985}, we deduced $E_x$ from a different set of reaction measurements by Weber~\emph{et al.}~\cite{Weber1979}. Although the IAS identification is not discussed in~\cite{Weber1979}, the observed $^{61}$Zn excitation energies of $3345(20)$~keV and $3370(60)$~keV respectively, match the value reported in~\cite{Okuma1985} and could correspond to the same level. From these two values we obtained a variance-weighted mean of $E_x(^{61}\mathrm{Zn})=3348(19)$~keV. All experimental excitation energies are in tension with the theoretical prediction of 3097~keV deduced with the cdGX1A Hamiltonian. A confirmation of our IAS assignment by a more precise spectroscopy study would be welcomed.

\begin{table}[b]
\caption{\label{tab:ME_Ge61}
Prediction of the unmeasured mass excess of $^{61}$Ge ($T_z=-3/2$) from the $T=3/2$ IMME at $A=61$. All values denote mass excesses in keV. The extrapolated AME2020~\cite{Wang2021} value for the mass excess of $^{61}$Ge and a mass estimate~\cite{Schatz2017} from a theoretical Coulomb displacement energy (CDE)~\cite{Brown2002} are also shown.
}
\renewcommand{\arraystretch}{1.25}
\begin{ruledtabular}
\begin{tabular}{ccc | c}
&
IMME input data &
&
IMME prediction \\
$T_z=+3/2$  &
$T_z=+1/2$&
$T_z=-1/2$&
$T_z=-3/2$\\
\colrule
$-61984(1)$ & $-53001(25)$ & $-43723(30)$ & $-34150(117)$ \\
\colrule
\multicolumn{3}{r}{AME2020:} & $-33790\footnotemark[1](300\footnotemark[1])$ \\
\multicolumn{3}{r}{CDE estimate:} & $-34065(100)$ \\
%\hline
%AME2020   & 4852.2(7)  & $-9378(100)$$   & 355(100) \\
\end{tabular}
\end{ruledtabular}
\footnotetext[1]{Extrapolated values based on trends from the mass surface~\cite{Huang2021}.}
\end{table}

The input mass excesses and the coefficients obtained from the quadratic IMME fit are listed in Table~\ref{tab:ME_Ge61}. To correctly account for correlations between the IMME coefficients, the prediction uncertainties were calculated using the full covariance matrix obtained in the fit. This resulted in a predicted $^{61}$Ge ground-state mass excess of $-34150(117)$~keV. Within the large uncertainties, this value is consistent with the extrapolated value of $-33790^{\#}(300^{\#})$~keV from AME2020. Our mass extrapolation further agrees with a theoretical estimate of $-34065(100)$~keV~\cite{Schatz2017} deduced with a CDE from Brown~\emph{et al.}~\cite{Brown2002}.

\subsection{\label{sec:rp-process_implications}Implications for the $\textit{rp}$ process in x-ray bursts}
We explored the impact of our improved mass values on model predictions of x-ray burst light curves using a self-consistent one-zone model~\cite{Schatz2001,Schatz2017}, which emulates computationally costly multi-zone models closely enough to identify relevant nuclear uncertainties~\cite{Cyburt2016}. We used model A defined in~\cite{Schatz2017}, which reaches a peak temperature of 2~GK and is characterized by a strong $\textit{rp}$-process flow beyond the $A=60$ region. We analyzed the impact of the reduced mass uncertainties of $^{60}$Ga and $^{61}$Ga by varying mass values by $3\sigma$. Proton-capture rates were recalculated for each $\mathrm{Q}$-value variation using the Hauser-Feshbach model code TALYS~\cite{Koning2012}. Reverse photodisintegration rates were determined via detailed balance, again using the new $\mathrm{Q}$ values.

The new masses are relevant for the $^{60}$Zn waiting point in the $\textit{rp}$ process. First we investigated the possibility of the $\textit{rp}$ process bypassing the $^{60}$Zn waiting point via the two-proton capture sequence $^{59}$Zn(p,$\gamma$)$^{60}$Ga(p,$\gamma$)$^{61}$Ge. A possible branching at $^{59}$Zn into this sequence depends critically on the $^{59}$Zn(p,$\gamma$) $\mathrm{Q}$ value, or equivalently the $^{60}$Ga proton separation energy, $S_\text{p}(^{60}$Ga). As (p,$\gamma$)-($\gamma$,p) equilibrium is established between $^{59}$Zn and $^{60}$Ga, the reaction flow and thus the branching depends exponentially on $S_\text{p}(^{60}$Ga) (see Eqn.~\ref{eqn:2p_capture_rate}). With the first direct measurement of the $^{60}$Ga mass in this work, resulting in $S_\text{p}(^{60}$Ga)=78(30)~keV, the branching can now be determined with certainty. Despite $^{60}$Ga being proton-bound, we found a negligible branching of the order of 10$^{-4}$, which has no impact on burst observables.

\begin{figure}[tb] 
\includegraphics[width=1.1\linewidth,trim=0.55cm 0 0 0]{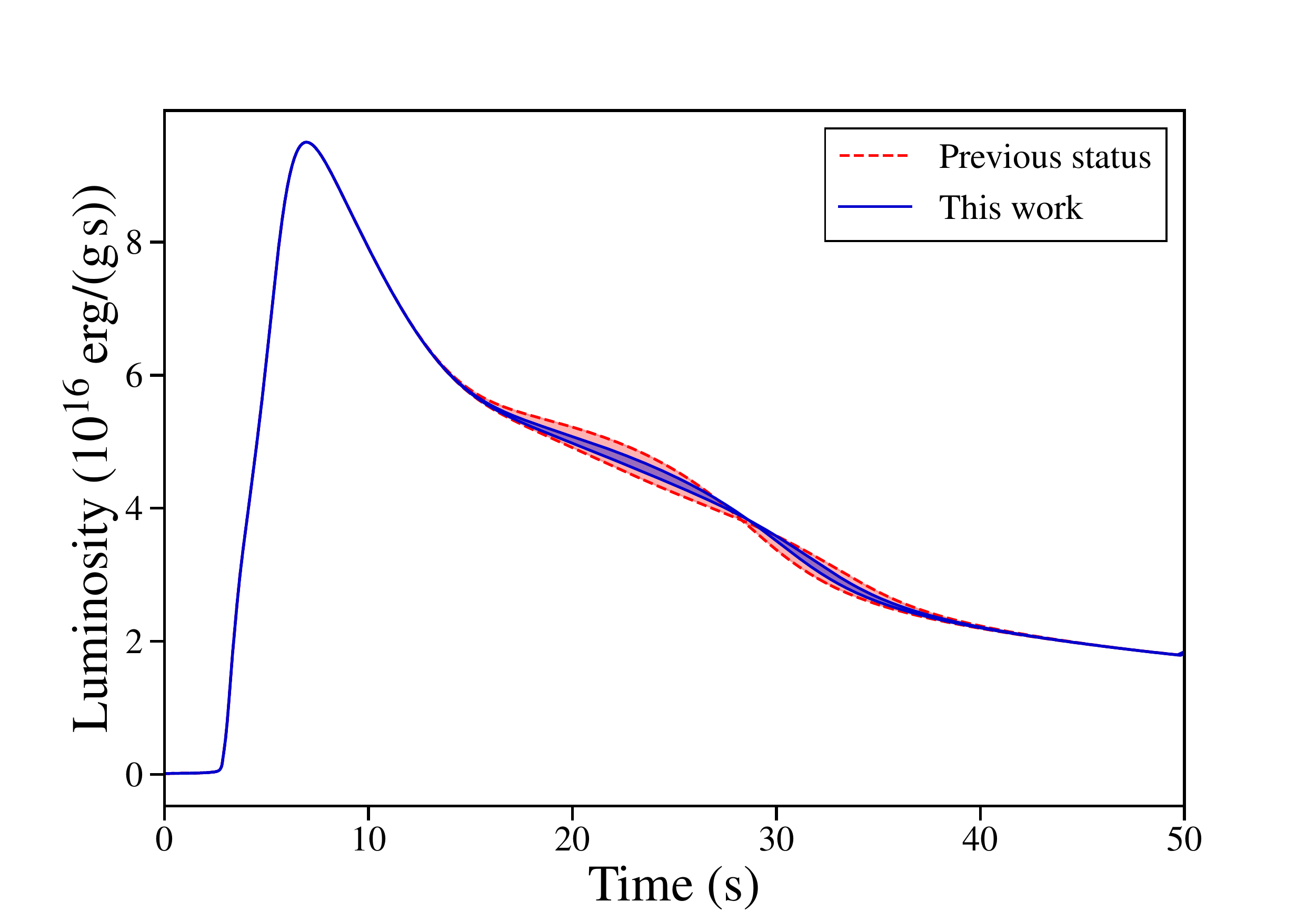}
\caption{\label{fig:lightcurves} Dispersion of predicted x-ray burst light curves resulting from $3\sigma$-variation of the improved $^{60-61}$Ga mass data from this work (blue solid lines) in comparison to AME2020 (red dashed lines). The areas between the obtained light curves were filled for clarity. The more precise $^{61}$Ga mass value significantly constrains the predicted light curve.}
\end{figure}
\begin{figure}[tb]
\includegraphics[width=.99\linewidth]{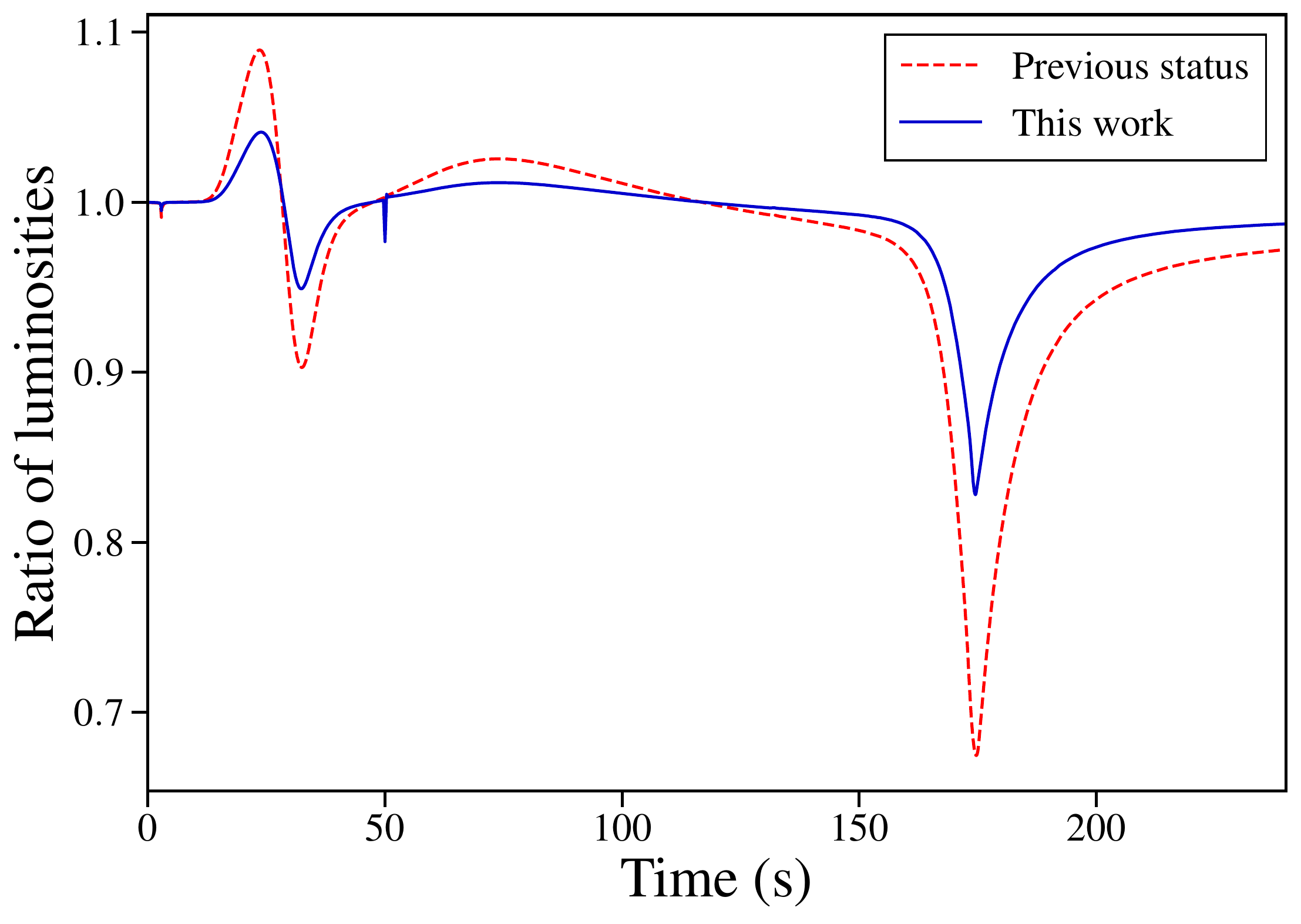}
\caption{\label{fig:ratio} Ratio of the two light curves obtained from $3\sigma$-variation of the $^{60-61}$Ga mass data (ratio of the envelopes in Fig.~\ref{fig:lightcurves}) from this work (blue solid line) and from AME2020 (red dashed line), respectively. The narrow peaks at 5~s and 50~s arise from small discontinuities in the light curves when the model switches between different opacity descriptions. The new mass data reduces the mass-induced prediction uncertainties in the x-ray burst light curve by more than a factor of two.}
\end{figure}

The delay the $^{60}$Zn waiting point imposes on the $\textit{rp}$ process is determined by the two-proton capture sequence $^{60}$Zn(p,$\gamma$)$^{61}$Ga(p,$\gamma$)$^{62}$Ge. The $^{61}$Ga mass determines the efficiency of the first step, which again depends exponentially on $S_\text{p}(^{61}$Ga) due to the (p,$\gamma$)-($\gamma$,p) equilibrium between $^{60}$Zn and $^{61}$Ga. Compared to AME2020, our measurements yield a three times more precise $^{61}$Ga proton separation energy, $S_\text{p}(^{61}\mathrm{Ga})=229(12)$~keV (see Table~\ref{tab:S_p_values}). To investigate the impact of the improved precision of $S_\text{p}(^{61}$Ga), we recalculated the $^{60}$Zn(p,$\gamma$) and the $^{61}$Ga($\gamma$,p) rate using 3$\sigma$ mass variations for $^{60}$Zn and $^{61}$Ga from AME2020 and from this work. The resulting burst light curve variations are shown in Fig.~\ref{fig:lightcurves} and Fig.~\ref{fig:ratio}. They show the characteristic deviations from standard power-law cooling that the $^{60}$Zn waiting point can induce. The new $^{61}$Ga mass value reduces the $S_\text{p}(^{61}$Ga)-induced uncertainties in the burst light curve by more than a factor of two. The remaining 12~keV uncertainty of the $^{61}$Ga mass still produces a small effect. This indicates that, for the chosen astrophysical conditions, even higher mass accuracies on the order of 1~keV would be needed to fully eliminate mass-induced model uncertainties in this particular case. Such precision levels have recently been demonstrated with an MR-TOF mass spectrometer~\cite{Mador2021}, encouraging a future re-measurement of the $^{61}$Ga mass. However, other nuclear uncertainties in this region are now of higher astrophysical relevance.

The most critical uncertainty related to the $^{60}$Zn waiting point is now the unknown $^{61}\mathrm{Ga}(\text{p},\gamma)^{62}\mathrm{Ge}$ reaction rate. We note that our $^{61}$Ga mass value also affects the $\mathrm{Q}$ value of this reaction. However, at present, this does not noticeably improve burst model predictions, since the $\mathrm{Q}$-value uncertainty remains dominated by the large error of the unmeasured mass of $^{62}$Ge. Nevertheless, our measurement marks an important step towards a reliable determination of this key reaction rate and provides additional motivation for an accurate mass measurement of $^{62}$Ge. 
 
\section{Summary and outlook}
We have performed precision MR-TOF mass measurements of neutron-deficient gallium isotopes in the direct vicinity of the proton drip line. The obtained mass values confirm the literature mass values for $^{62-63}$Ga, improve the mass uncertainty for $^{61}$Ga by more than a factor of 3 and mark the first direct measurement of the $^{60}$Ga ground-state mass.

The new $^{60}$Ga mass enabled an accurate determination of the one-proton separation energy of this nuclide, providing very strong evidence that $^{60}$Ga is the last proton-bound nuclide in its isotopic chain. Using the new $^{60}$Ga mass, we completed the experimentally evaluated IMME triplets up to $A=60$ and resolved literature ambiguities about the assignment of the lowest-lying $T=1$ isobaric analog state in $^{60}$Zn. The obtained IMME coefficients continue the expected trends and were found to be in reasonable agreement with shell-model predictions derived from the isospin-nonconserving cdGX1A Hamiltonian. Earlier studies of nuclides in this region have pointed out issues with masses predicted from trends of the mass surface. Our $^{60}$Ga measurement provides an additional fixpoint for future atomic mass evaluations and can be expected to mitigate the reported discrepancies.

Our $^{60,61}$Ga mass measurements have important implications for the $\textit{rp}$ process in x-ray bursts. We find that the new $^{60}$Ga mass excludes the possibility of a bypass of the $^{60}$Zn waiting point via two-proton capture on $^{59}$Zn. The new, more precise $^{61}$Ga mass significantly reduces the uncertainty in x-ray burst light curve predictions due to the uncertainty of $S_\text{p}(^{61}$Ga)~\cite{Schatz2017}. The remaining 12~keV uncertainty still produces a noticeable effect. However, the largest remaining uncertainty related to the $^{60}$Zn waiting point is now the $^{61}$Ga(p,$\gamma$) reaction rate. Our improved $^{61}$Ga mass value is an important step towards an accurate $\mathrm{Q}$ value, and thus reliable calculations of this key reaction rate. Currently, rate calculations remain limited by the large uncertainty of the experimentally unkown mass of $^{62}$Ge. We therefore recommend that future experimental efforts focus on a precision mass measurement of this nuclide. With tighter constraints on the $^{61}$Ga(p,$\gamma$) rate, the sensitivity of x-ray burst models to nuclear uncertainties near $^{60}$Zn could be reassessed.\\

\begin{acknowledgments}
We would like to thank the TRIUMF Beam Delivery and Resonant Ionization Laser Ion Source groups for their excellent support in the realization of this experiment. This work was supported by the Natural Sciences and Engineering Research Council of Canada (NSERC), the Canada Foundation for Innovation (CFI), the Canada-UK Foundation, the Brazilian Conselho Nacional de Desenvolvimento Cient\'{ı}fico e Technol\'{o}gico (CNPq) under grant 249121/2013-1, the Helmholtz Association of German Research Centers through the Nuclear Astrophysics Virtual Institute (VH-VI-417), German Research Foundation (DFG), grant No. SCHE 1969/2-1, the German Federal Ministry for Education and Research (BMBF), grant No. 05P19RGFN1 and 05P21RGFN1, the Hessian Ministry for Science and Art through the LOEWE Center HICforFAIR, the JLU and GSI under the JLU-GSI strategic Helmholtz partnership agreement, the UKRI Science and Technology Facilities Council (STFC) grant No. ST/P004008/1, the US National Science Foundation under PHY-1419765, PHY 14-30152 (Joint Institute for Nuclear Astrophysics JINA-CEE), PHY-1913554, IN2P3/CNRS (France) via the ENFIA Master Project, the US Department of Energy (DOE) Office of Science under Grant DE-SC0017649, the Deutsche Forschungsgemeinschaft (DFG) under Grant FR 601/3-1. TRIUMF receives federal funding via the National Research Council of Canada (NRC).

\end{acknowledgments}

\bibliographystyle{apsrev4-1}

% Include precompiled .bbl file 

%\bibliography{Literature_n-deficient_Ga}% Produces the bibliography via BibTeX.

\begin{thebibliography}{119}
\expandafter\ifx\csname natexlab\endcsname\relax\def\natexlab#1{#1}\fi
\expandafter\ifx\csname bibnamefont\endcsname\relax
  \def\bibnamefont#1{#1}\fi
\expandafter\ifx\csname bibfnamefont\endcsname\relax
  \def\bibfnamefont#1{#1}\fi
\expandafter\ifx\csname citenamefont\endcsname\relax
  \def\citenamefont#1{#1}\fi
\expandafter\ifx\csname url\endcsname\relax
  \def\url#1{\texttt{#1}}\fi
\expandafter\ifx\csname urlprefix\endcsname\relax\def\urlprefix{URL }\fi
\providecommand{\bibinfo}[2]{#2}
\providecommand{\eprint}[2][]{\url{#2}}

\bibitem[{\citenamefont{Schatz and Rehm}(2006)}]{Schatz2006b}
\bibinfo{author}{\bibfnamefont{H.}~\bibnamefont{Schatz}} \bibnamefont{and}
  \bibinfo{author}{\bibfnamefont{K.}~\bibnamefont{Rehm}},
  \bibinfo{journal}{Nuclear Physics A} \textbf{\bibinfo{volume}{777}},
  \bibinfo{pages}{601 } (\bibinfo{year}{2006}), ISSN \bibinfo{issn}{0375-9474},
  \bibinfo{note}{special Issue on Nuclear Astrophysics},
  \urlprefix\url{https://www.sciencedirect.com/science/article/pii/S0375947405008791}.

\bibitem[{\citenamefont{Parikh et~al.}(2013)\citenamefont{Parikh, José, Sala,
  and Iliadis}}]{Parikh2013}
\bibinfo{author}{\bibfnamefont{A.}~\bibnamefont{Parikh}},
  \bibinfo{author}{\bibfnamefont{J.}~\bibnamefont{José}},
  \bibinfo{author}{\bibfnamefont{G.}~\bibnamefont{Sala}}, \bibnamefont{and}
  \bibinfo{author}{\bibfnamefont{C.}~\bibnamefont{Iliadis}},
  \bibinfo{journal}{Progress in Particle and Nuclear Physics}
  \textbf{\bibinfo{volume}{69}}, \bibinfo{pages}{225 } (\bibinfo{year}{2013}),
  ISSN \bibinfo{issn}{0146-6410},
  \urlprefix\url{https://www.sciencedirect.com/science/article/pii/S0146641012001354}.

\bibitem[{\citenamefont{Meisel}(2018)}]{Meisel2018}
\bibinfo{author}{\bibfnamefont{Z.}~\bibnamefont{Meisel}}, \bibinfo{journal}{The
  Astrophysical Journal} \textbf{\bibinfo{volume}{860}}, \bibinfo{pages}{147}
  (\bibinfo{year}{2018}),
  \urlprefix\url{https://doi.org/10.3847%2F1538-4357%2Faac3d3}.

\bibitem[{\citenamefont{Galloway and Keek}(2021)}]{Galloway2021}
\bibinfo{author}{\bibfnamefont{D.~K.} \bibnamefont{Galloway}} \bibnamefont{and}
  \bibinfo{author}{\bibfnamefont{L.}~\bibnamefont{Keek}},
  \emph{\bibinfo{title}{Thermonuclear {X}-ray Bursts}}
  (\bibinfo{publisher}{Springer Berlin Heidelberg}, \bibinfo{address}{Berlin,
  Heidelberg}, \bibinfo{year}{2021}), pp. \bibinfo{pages}{209--262}, ISBN
  \bibinfo{isbn}{978-3-662-62110-3},
  \urlprefix\url{https://doi.org/10.1007/978-3-662-62110-3_5}.

\bibitem[{\citenamefont{Schatz}(2006)}]{Schatz2006}
\bibinfo{author}{\bibfnamefont{H.}~\bibnamefont{Schatz}},
  \bibinfo{journal}{International Journal of Mass Spectrometry}
  \textbf{\bibinfo{volume}{251}}, \bibinfo{pages}{293 } (\bibinfo{year}{2006}).

\bibitem[{\citenamefont{Meisel et~al.}(2018)\citenamefont{Meisel, Deibel, Keek,
  Shternin, and Elfritz}}]{Meisel2018b}
\bibinfo{author}{\bibfnamefont{Z.}~\bibnamefont{Meisel}},
  \bibinfo{author}{\bibfnamefont{A.}~\bibnamefont{Deibel}},
  \bibinfo{author}{\bibfnamefont{L.}~\bibnamefont{Keek}},
  \bibinfo{author}{\bibfnamefont{P.}~\bibnamefont{Shternin}}, \bibnamefont{and}
  \bibinfo{author}{\bibfnamefont{J.}~\bibnamefont{Elfritz}},
  \bibinfo{journal}{Journal of Physics G: Nuclear and Particle Physics}
  \textbf{\bibinfo{volume}{45}}, \bibinfo{pages}{093001}
  (\bibinfo{year}{2018}),
  \urlprefix\url{https://doi.org/10.1088/1361-6471/aad171}.

\bibitem[{\citenamefont{Meisel et~al.}(2019)\citenamefont{Meisel, Merz, and
  Medvid}}]{Meisel2019}
\bibinfo{author}{\bibfnamefont{Z.}~\bibnamefont{Meisel}},
  \bibinfo{author}{\bibfnamefont{G.}~\bibnamefont{Merz}}, \bibnamefont{and}
  \bibinfo{author}{\bibfnamefont{S.}~\bibnamefont{Medvid}},
  \bibinfo{journal}{The Astrophysical Journal} \textbf{\bibinfo{volume}{872}},
  \bibinfo{pages}{84} (\bibinfo{year}{2019}),
  \urlprefix\url{https://doi.org/10.3847%2F1538-4357%2Faafede}.

\bibitem[{\citenamefont{Fisker et~al.}(2008)\citenamefont{Fisker, Schatz, and
  Thielemann}}]{Fisker2008}
\bibinfo{author}{\bibfnamefont{J.~L.} \bibnamefont{Fisker}},
  \bibinfo{author}{\bibfnamefont{H.}~\bibnamefont{Schatz}}, \bibnamefont{and}
  \bibinfo{author}{\bibfnamefont{F.-K.} \bibnamefont{Thielemann}},
  \bibinfo{journal}{The Astrophysical Journal Supplement Series}
  \textbf{\bibinfo{volume}{174}}, \bibinfo{pages}{261 } (\bibinfo{year}{2008}),
  \urlprefix\url{https://doi.org/10.1086/521104}.

\bibitem[{\citenamefont{Chamel and Haensel}(2008)}]{Chamel2008p139}
\bibinfo{author}{\bibfnamefont{N.}~\bibnamefont{Chamel}} \bibnamefont{and}
  \bibinfo{author}{\bibfnamefont{P.}~\bibnamefont{Haensel}}, in
  \emph{\bibinfo{booktitle}{Living Reviews in Relativity}}
  (\bibinfo{publisher}{Springer}, \bibinfo{year}{2008}),
  vol.~\bibinfo{volume}{11}, pp. \bibinfo{pages}{139 -- 141}.

\bibitem[{\citenamefont{Kratz et~al.}(1998)\citenamefont{Kratz, Pfeiffer,
  Hannawald, Thielemann, G{\"o}rres, Schatz, and Wiescher}}]{Kratz1998}
\bibinfo{author}{\bibfnamefont{K.-L.} \bibnamefont{Kratz}},
  \bibinfo{author}{\bibfnamefont{B.}~\bibnamefont{Pfeiffer}},
  \bibinfo{author}{\bibfnamefont{M.}~\bibnamefont{Hannawald}},
  \bibinfo{author}{\bibfnamefont{F.-K.} \bibnamefont{Thielemann}},
  \bibinfo{author}{\bibfnamefont{J.}~\bibnamefont{G{\"o}rres}},
  \bibinfo{author}{\bibfnamefont{H.}~\bibnamefont{Schatz}}, \bibnamefont{and}
  \bibinfo{author}{\bibfnamefont{M.}~\bibnamefont{Wiescher}},
  \bibinfo{journal}{Il Nuovo Cimento A (1971-1996)}
  \textbf{\bibinfo{volume}{111}}, \bibinfo{pages}{1043 }
  (\bibinfo{year}{1998}),
  \urlprefix\url{https://link.springer.com/article/10.1007/BF03035991}.

\bibitem[{\citenamefont{Schatz et~al.}(2001)\citenamefont{Schatz, Aprahamian,
  Barnard, Bildsten, Cumming, Ouellette, Rauscher, Thielemann, and
  Wiescher}}]{Schatz2001}
\bibinfo{author}{\bibfnamefont{H.}~\bibnamefont{Schatz}},
  \bibinfo{author}{\bibfnamefont{A.}~\bibnamefont{Aprahamian}},
  \bibinfo{author}{\bibfnamefont{V.}~\bibnamefont{Barnard}},
  \bibinfo{author}{\bibfnamefont{L.}~\bibnamefont{Bildsten}},
  \bibinfo{author}{\bibfnamefont{A.}~\bibnamefont{Cumming}},
  \bibinfo{author}{\bibfnamefont{M.}~\bibnamefont{Ouellette}},
  \bibinfo{author}{\bibfnamefont{T.}~\bibnamefont{Rauscher}},
  \bibinfo{author}{\bibfnamefont{F.-K.} \bibnamefont{Thielemann}},
  \bibnamefont{and} \bibinfo{author}{\bibfnamefont{M.}~\bibnamefont{Wiescher}},
  \bibinfo{journal}{Phys. Rev. Lett.} \textbf{\bibinfo{volume}{86}},
  \bibinfo{pages}{3471 } (\bibinfo{year}{2001}),
  \urlprefix\url{https://link.aps.org/doi/10.1103/PhysRevLett.86.3471}.

\bibitem[{\citenamefont{Wallace and Woosley}(1981)}]{Wallace1981}
\bibinfo{author}{\bibfnamefont{R.}~\bibnamefont{Wallace}} \bibnamefont{and}
  \bibinfo{author}{\bibfnamefont{S.~E.} \bibnamefont{Woosley}},
  \bibinfo{journal}{The Astrophysical Journal Supplement Series}
  \textbf{\bibinfo{volume}{45}}, \bibinfo{pages}{389 } (\bibinfo{year}{1981}).

\bibitem[{\citenamefont{Koike et~al.}(2004)\citenamefont{Koike, {A}ki
  Hashimoto, Kuromizu, and {I}chirou Fujimoto}}]{Koike2004}
\bibinfo{author}{\bibfnamefont{O.}~\bibnamefont{Koike}},
  \bibinfo{author}{\bibfnamefont{M.}~\bibnamefont{{A}ki Hashimoto}},
  \bibinfo{author}{\bibfnamefont{R.}~\bibnamefont{Kuromizu}}, \bibnamefont{and}
  \bibinfo{author}{\bibfnamefont{S.}~\bibnamefont{{I}chirou Fujimoto}},
  \bibinfo{journal}{The Astrophysical Journal} \textbf{\bibinfo{volume}{603}},
  \bibinfo{pages}{242 } (\bibinfo{year}{2004}),
  \urlprefix\url{https://doi.org/10.1086/381354}.

\bibitem[{\citenamefont{Schatz et~al.}(1998)\citenamefont{Schatz, Aprahamian,
  Görres, Wiescher, Rauscher, Rembges, Thielemann, Pfeiffer, Möller, Kratz
  et~al.}}]{Schatz1998}
\bibinfo{author}{\bibfnamefont{H.}~\bibnamefont{Schatz}},
  \bibinfo{author}{\bibfnamefont{A.}~\bibnamefont{Aprahamian}},
  \bibinfo{author}{\bibfnamefont{J.}~\bibnamefont{Görres}},
  \bibinfo{author}{\bibfnamefont{M.}~\bibnamefont{Wiescher}},
  \bibinfo{author}{\bibfnamefont{T.}~\bibnamefont{Rauscher}},
  \bibinfo{author}{\bibfnamefont{J.}~\bibnamefont{Rembges}},
  \bibinfo{author}{\bibfnamefont{F.-K.} \bibnamefont{Thielemann}},
  \bibinfo{author}{\bibfnamefont{B.}~\bibnamefont{Pfeiffer}},
  \bibinfo{author}{\bibfnamefont{P.}~\bibnamefont{Möller}},
  \bibinfo{author}{\bibfnamefont{K.-L.} \bibnamefont{Kratz}},
  \bibnamefont{et~al.}, \bibinfo{journal}{Physics Reports}
  \textbf{\bibinfo{volume}{294}}, \bibinfo{pages}{167 } (\bibinfo{year}{1998}),
  ISSN \bibinfo{issn}{0370-1573},
  \urlprefix\url{https://www.sciencedirect.com/science/article/pii/S0370157397000483}.

\bibitem[{\citenamefont{Cyburt et~al.}(2016)\citenamefont{Cyburt, Amthor,
  Heger, Johnson, Keek, Meisel, Schatz, and Smith}}]{Cyburt2016}
\bibinfo{author}{\bibfnamefont{R.~H.} \bibnamefont{Cyburt}},
  \bibinfo{author}{\bibfnamefont{A.~M.} \bibnamefont{Amthor}},
  \bibinfo{author}{\bibfnamefont{A.}~\bibnamefont{Heger}},
  \bibinfo{author}{\bibfnamefont{E.}~\bibnamefont{Johnson}},
  \bibinfo{author}{\bibfnamefont{L.}~\bibnamefont{Keek}},
  \bibinfo{author}{\bibfnamefont{Z.}~\bibnamefont{Meisel}},
  \bibinfo{author}{\bibfnamefont{H.}~\bibnamefont{Schatz}}, \bibnamefont{and}
  \bibinfo{author}{\bibfnamefont{K.}~\bibnamefont{Smith}},
  \bibinfo{journal}{The Astrophysical Journal} \textbf{\bibinfo{volume}{830}},
  \bibinfo{pages}{55} (\bibinfo{year}{2016}),
  \urlprefix\url{https://doi.org/10.3847%2F0004-637x%2F830%2F2%2F55}.

\bibitem[{\citenamefont{Parikh et~al.}(2009)\citenamefont{Parikh, Jos\'e,
  Iliadis, Moreno, and Rauscher}}]{Parikh2009}
\bibinfo{author}{\bibfnamefont{A.}~\bibnamefont{Parikh}},
  \bibinfo{author}{\bibfnamefont{J.}~\bibnamefont{Jos\'e}},
  \bibinfo{author}{\bibfnamefont{C.}~\bibnamefont{Iliadis}},
  \bibinfo{author}{\bibfnamefont{F.}~\bibnamefont{Moreno}}, \bibnamefont{and}
  \bibinfo{author}{\bibfnamefont{T.}~\bibnamefont{Rauscher}},
  \bibinfo{journal}{Phys. Rev. C} \textbf{\bibinfo{volume}{79}},
  \bibinfo{pages}{045802} (\bibinfo{year}{2009}),
  \urlprefix\url{https://link.aps.org/doi/10.1103/PhysRevC.79.045802}.

\bibitem[{\citenamefont{Schatz and Ong}(2017)}]{Schatz2017}
\bibinfo{author}{\bibfnamefont{H.}~\bibnamefont{Schatz}} \bibnamefont{and}
  \bibinfo{author}{\bibfnamefont{W.-J.} \bibnamefont{Ong}},
  \bibinfo{journal}{The Astrophysical Journal} \textbf{\bibinfo{volume}{844}},
  \bibinfo{pages}{139} (\bibinfo{year}{2017}),
  \urlprefix\url{https://doi.org/10.3847/1538-4357/aa7de9}.

\bibitem[{\citenamefont{Clark et~al.}(2007)\citenamefont{Clark, Sharma, Savard,
  Levand, Wang, Zhou, Blank, Buchinger, Crawford, Gulick et~al.}}]{Clark2007}
\bibinfo{author}{\bibfnamefont{J.~A.} \bibnamefont{Clark}},
  \bibinfo{author}{\bibfnamefont{K.~S.} \bibnamefont{Sharma}},
  \bibinfo{author}{\bibfnamefont{G.}~\bibnamefont{Savard}},
  \bibinfo{author}{\bibfnamefont{A.~F.} \bibnamefont{Levand}},
  \bibinfo{author}{\bibfnamefont{J.~C.} \bibnamefont{Wang}},
  \bibinfo{author}{\bibfnamefont{Z.}~\bibnamefont{Zhou}},
  \bibinfo{author}{\bibfnamefont{B.}~\bibnamefont{Blank}},
  \bibinfo{author}{\bibfnamefont{F.}~\bibnamefont{Buchinger}},
  \bibinfo{author}{\bibfnamefont{J.~E.} \bibnamefont{Crawford}},
  \bibinfo{author}{\bibfnamefont{S.}~\bibnamefont{Gulick}},
  \bibnamefont{et~al.}, \bibinfo{journal}{Phys. Rev. C}
  \textbf{\bibinfo{volume}{75}}, \bibinfo{pages}{032801}
  (\bibinfo{year}{2007}),
  \urlprefix\url{https://link.aps.org/doi/10.1103/PhysRevC.75.032801}.

\bibitem[{\citenamefont{Breitenfeldt et~al.}(2009)\citenamefont{Breitenfeldt,
  Audi, Beck, Blaum, George, Herfurth, Herlert, Kellerbauer, Kluge, Kowalska
  et~al.}}]{Breitenfeldt2009}
\bibinfo{author}{\bibfnamefont{M.}~\bibnamefont{Breitenfeldt}},
  \bibinfo{author}{\bibfnamefont{G.}~\bibnamefont{Audi}},
  \bibinfo{author}{\bibfnamefont{D.}~\bibnamefont{Beck}},
  \bibinfo{author}{\bibfnamefont{K.}~\bibnamefont{Blaum}},
  \bibinfo{author}{\bibfnamefont{S.}~\bibnamefont{George}},
  \bibinfo{author}{\bibfnamefont{F.}~\bibnamefont{Herfurth}},
  \bibinfo{author}{\bibfnamefont{A.}~\bibnamefont{Herlert}},
  \bibinfo{author}{\bibfnamefont{A.}~\bibnamefont{Kellerbauer}},
  \bibinfo{author}{\bibfnamefont{H.~J.} \bibnamefont{Kluge}},
  \bibinfo{author}{\bibfnamefont{M.}~\bibnamefont{Kowalska}},
  \bibnamefont{et~al.}, \bibinfo{journal}{Phys. Rev. C}
  \textbf{\bibinfo{volume}{80}}, \bibinfo{pages}{035805}
  (\bibinfo{year}{2009}),
  \urlprefix\url{https://link.aps.org/doi/10.1103/PhysRevC.80.035805}.

\bibitem[{\citenamefont{Kankainen et~al.}(2010)\citenamefont{Kankainen, Elomaa,
  Eronen, Gorelov, Hakala, Jokinen, Kessler, Kolhinen, Moore, Rahaman
  et~al.}}]{Kankainen2010}
\bibinfo{author}{\bibfnamefont{A.}~\bibnamefont{Kankainen}},
  \bibinfo{author}{\bibfnamefont{V.-V.} \bibnamefont{Elomaa}},
  \bibinfo{author}{\bibfnamefont{T.}~\bibnamefont{Eronen}},
  \bibinfo{author}{\bibfnamefont{D.}~\bibnamefont{Gorelov}},
  \bibinfo{author}{\bibfnamefont{J.}~\bibnamefont{Hakala}},
  \bibinfo{author}{\bibfnamefont{A.}~\bibnamefont{Jokinen}},
  \bibinfo{author}{\bibfnamefont{T.}~\bibnamefont{Kessler}},
  \bibinfo{author}{\bibfnamefont{V.~S.} \bibnamefont{Kolhinen}},
  \bibinfo{author}{\bibfnamefont{I.~D.} \bibnamefont{Moore}},
  \bibinfo{author}{\bibfnamefont{S.}~\bibnamefont{Rahaman}},
  \bibnamefont{et~al.}, \bibinfo{journal}{Phys. Rev. C}
  \textbf{\bibinfo{volume}{82}}, \bibinfo{pages}{034311}
  (\bibinfo{year}{2010}),
  \urlprefix\url{https://link.aps.org/doi/10.1103/PhysRevC.82.034311}.

\bibitem[{\citenamefont{Tu et~al.}(2011{\natexlab{a}})\citenamefont{Tu, Xu,
  Wang, Zhang, Litvinov, Sun, Schatz, Zhou, Yuan, Xia et~al.}}]{Tu2011b}
\bibinfo{author}{\bibfnamefont{X.~L.} \bibnamefont{Tu}},
  \bibinfo{author}{\bibfnamefont{H.~S.} \bibnamefont{Xu}},
  \bibinfo{author}{\bibfnamefont{M.}~\bibnamefont{Wang}},
  \bibinfo{author}{\bibfnamefont{Y.~H.} \bibnamefont{Zhang}},
  \bibinfo{author}{\bibfnamefont{Y.~A.} \bibnamefont{Litvinov}},
  \bibinfo{author}{\bibfnamefont{Y.}~\bibnamefont{Sun}},
  \bibinfo{author}{\bibfnamefont{H.}~\bibnamefont{Schatz}},
  \bibinfo{author}{\bibfnamefont{X.~H.} \bibnamefont{Zhou}},
  \bibinfo{author}{\bibfnamefont{Y.~J.} \bibnamefont{Yuan}},
  \bibinfo{author}{\bibfnamefont{J.~W.} \bibnamefont{Xia}},
  \bibnamefont{et~al.}, \bibinfo{journal}{Phys. Rev. Lett.}
  \textbf{\bibinfo{volume}{106}}, \bibinfo{pages}{112501}
  (\bibinfo{year}{2011}{\natexlab{a}}),
  \urlprefix\url{https://link.aps.org/doi/10.1103/PhysRevLett.106.112501}.

\bibitem[{\citenamefont{Valverde et~al.}(2018)\citenamefont{Valverde, Brodeur,
  Bollen, Eibach, Gulyuz, Hamaker, Izzo, Ong, Puentes, Redshaw
  et~al.}}]{Valverde2018}
\bibinfo{author}{\bibfnamefont{A.~A.} \bibnamefont{Valverde}},
  \bibinfo{author}{\bibfnamefont{M.}~\bibnamefont{Brodeur}},
  \bibinfo{author}{\bibfnamefont{G.}~\bibnamefont{Bollen}},
  \bibinfo{author}{\bibfnamefont{M.}~\bibnamefont{Eibach}},
  \bibinfo{author}{\bibfnamefont{K.}~\bibnamefont{Gulyuz}},
  \bibinfo{author}{\bibfnamefont{A.}~\bibnamefont{Hamaker}},
  \bibinfo{author}{\bibfnamefont{C.}~\bibnamefont{Izzo}},
  \bibinfo{author}{\bibfnamefont{W.-J.} \bibnamefont{Ong}},
  \bibinfo{author}{\bibfnamefont{D.}~\bibnamefont{Puentes}},
  \bibinfo{author}{\bibfnamefont{M.}~\bibnamefont{Redshaw}},
  \bibnamefont{et~al.}, \bibinfo{journal}{Phys. Rev. Lett.}
  \textbf{\bibinfo{volume}{120}}, \bibinfo{pages}{032701}
  (\bibinfo{year}{2018}),
  \urlprefix\url{https://link.aps.org/doi/10.1103/PhysRevLett.120.032701}.

\bibitem[{\citenamefont{Duy et~al.}(2020)\citenamefont{Duy, Ho, and
  Uyen}}]{Duy2020}
\bibinfo{author}{\bibfnamefont{N.~N.} \bibnamefont{Duy}},
  \bibinfo{author}{\bibfnamefont{P.-T.} \bibnamefont{Ho}}, \bibnamefont{and}
  \bibinfo{author}{\bibfnamefont{N.~K.} \bibnamefont{Uyen}},
  \bibinfo{journal}{Journal of the Korean Physical Society}
  \textbf{\bibinfo{volume}{76}}, \bibinfo{pages}{881 } (\bibinfo{year}{2020}),
  \urlprefix\url{https://doi.org/10.3938/jkps.76.881}.

\bibitem[{\citenamefont{Browne and Tuli}(2013)}]{Browne2013}
\bibinfo{author}{\bibfnamefont{E.}~\bibnamefont{Browne}} \bibnamefont{and}
  \bibinfo{author}{\bibfnamefont{J.}~\bibnamefont{Tuli}},
  \bibinfo{journal}{Nuclear Data Sheets} \textbf{\bibinfo{volume}{114}},
  \bibinfo{pages}{1849 } (\bibinfo{year}{2013}), ISSN
  \bibinfo{issn}{0090-3752},
  \urlprefix\url{https://www.sciencedirect.com/science/article/pii/S0090375213000823}.

\bibitem[{\citenamefont{{in 't Zand, J. J. M.} et~al.}(2017)\citenamefont{{in
  't Zand, J. J. M.}, {Visser, M. E. B.}, {Galloway, D. K.}, {Chenevez, J.},
  {Keek, L.}, {Kuulkers, E.}, {S\'anchez-Fern\'andez, C.}, and {W\"orpel,
  H.}}}]{in'tZand2017}
\bibinfo{author}{\bibnamefont{{in 't Zand, J. J. M.}}},
  \bibinfo{author}{\bibnamefont{{Visser, M. E. B.}}},
  \bibinfo{author}{\bibnamefont{{Galloway, D. K.}}},
  \bibinfo{author}{\bibnamefont{{Chenevez, J.}}},
  \bibinfo{author}{\bibnamefont{{Keek, L.}}},
  \bibinfo{author}{\bibnamefont{{Kuulkers, E.}}},
  \bibinfo{author}{\bibnamefont{{S\'anchez-Fern\'andez, C.}}},
  \bibnamefont{and} \bibinfo{author}{\bibnamefont{{W\"orpel, H.}}},
  \bibinfo{journal}{Astronomy \& Astrophysics} \textbf{\bibinfo{volume}{606}},
  \bibinfo{pages}{A130} (\bibinfo{year}{2017}),
  \urlprefix\url{https://doi.org/10.1051/0004-6361/201731281}.

\bibitem[{\citenamefont{Wang et~al.}(2021)\citenamefont{Wang, Huang, Kondev,
  Audi, and Naimi}}]{Wang2021}
\bibinfo{author}{\bibfnamefont{M.}~\bibnamefont{Wang}},
  \bibinfo{author}{\bibfnamefont{W.}~\bibnamefont{Huang}},
  \bibinfo{author}{\bibfnamefont{F.}~\bibnamefont{Kondev}},
  \bibinfo{author}{\bibfnamefont{G.}~\bibnamefont{Audi}}, \bibnamefont{and}
  \bibinfo{author}{\bibfnamefont{S.}~\bibnamefont{Naimi}},
  \bibinfo{journal}{Chinese Physics C} \textbf{\bibinfo{volume}{45}},
  \bibinfo{pages}{030003} (\bibinfo{year}{2021}),
  \urlprefix\url{https://doi.org/10.1088/1674-1137/abddaf}.

\bibitem[{\citenamefont{Parikh et~al.}(2008)\citenamefont{Parikh, Jos{\'{e}},
  Moreno, and Iliadis}}]{Parikh2008}
\bibinfo{author}{\bibfnamefont{A.}~\bibnamefont{Parikh}},
  \bibinfo{author}{\bibfnamefont{J.}~\bibnamefont{Jos{\'{e}}}},
  \bibinfo{author}{\bibfnamefont{F.}~\bibnamefont{Moreno}}, \bibnamefont{and}
  \bibinfo{author}{\bibfnamefont{C.}~\bibnamefont{Iliadis}},
  \bibinfo{journal}{The Astrophysical Journal Supplement Series}
  \textbf{\bibinfo{volume}{178}}, \bibinfo{pages}{110 } (\bibinfo{year}{2008}),
  \urlprefix\url{https://doi.org/10.1086/589879}.

\bibitem[{\citenamefont{Sobiczewski et~al.}(2018)\citenamefont{Sobiczewski,
  Litvinov, and Palczewski}}]{Sobiczewski2018}
\bibinfo{author}{\bibfnamefont{A.}~\bibnamefont{Sobiczewski}},
  \bibinfo{author}{\bibfnamefont{Y.}~\bibnamefont{Litvinov}}, \bibnamefont{and}
  \bibinfo{author}{\bibfnamefont{M.}~\bibnamefont{Palczewski}},
  \bibinfo{journal}{Atomic Data and Nuclear Data Tables}
  \textbf{\bibinfo{volume}{119}}, \bibinfo{pages}{1 } (\bibinfo{year}{2018}),
  ISSN \bibinfo{issn}{0092-640X},
  \urlprefix\url{https://www.sciencedirect.com/science/article/pii/S0092640X17300323}.

\bibitem[{\citenamefont{Brown et~al.}(2002)\citenamefont{Brown, Clement,
  Schatz, Volya, and Richter}}]{Brown2002}
\bibinfo{author}{\bibfnamefont{B.~A.} \bibnamefont{Brown}},
  \bibinfo{author}{\bibfnamefont{R.~R.~C.} \bibnamefont{Clement}},
  \bibinfo{author}{\bibfnamefont{H.}~\bibnamefont{Schatz}},
  \bibinfo{author}{\bibfnamefont{A.}~\bibnamefont{Volya}}, \bibnamefont{and}
  \bibinfo{author}{\bibfnamefont{W.~A.} \bibnamefont{Richter}},
  \bibinfo{journal}{Phys. Rev. C} \textbf{\bibinfo{volume}{65}},
  \bibinfo{pages}{045802} (\bibinfo{year}{2002}),
  \urlprefix\url{https://link.aps.org/doi/10.1103/PhysRevC.65.045802}.

\bibitem[{\citenamefont{Garvey and Kelson}(1966)}]{Garvey1966}
\bibinfo{author}{\bibfnamefont{G.~T.} \bibnamefont{Garvey}} \bibnamefont{and}
  \bibinfo{author}{\bibfnamefont{I.}~\bibnamefont{Kelson}},
  \bibinfo{journal}{Phys. Rev. Lett.} \textbf{\bibinfo{volume}{16}},
  \bibinfo{pages}{197 } (\bibinfo{year}{1966}),
  \urlprefix\url{https://link.aps.org/doi/10.1103/PhysRevLett.16.197}.

\bibitem[{\citenamefont{Zhang et~al.}(2018)\citenamefont{Zhang, Zhang, Zhou,
  Wang, Litvinov, Xu, Xu, Shuai, Lam, Chen et~al.}}]{Zhang2018}
\bibinfo{author}{\bibfnamefont{Y.~H.} \bibnamefont{Zhang}},
  \bibinfo{author}{\bibfnamefont{P.}~\bibnamefont{Zhang}},
  \bibinfo{author}{\bibfnamefont{X.~H.} \bibnamefont{Zhou}},
  \bibinfo{author}{\bibfnamefont{M.}~\bibnamefont{Wang}},
  \bibinfo{author}{\bibfnamefont{Y.~A.} \bibnamefont{Litvinov}},
  \bibinfo{author}{\bibfnamefont{H.~S.} \bibnamefont{Xu}},
  \bibinfo{author}{\bibfnamefont{X.}~\bibnamefont{Xu}},
  \bibinfo{author}{\bibfnamefont{P.}~\bibnamefont{Shuai}},
  \bibinfo{author}{\bibfnamefont{Y.~H.} \bibnamefont{Lam}},
  \bibinfo{author}{\bibfnamefont{R.~J.} \bibnamefont{Chen}},
  \bibnamefont{et~al.}, \bibinfo{journal}{Phys. Rev. C}
  \textbf{\bibinfo{volume}{98}}, \bibinfo{pages}{014319}
  (\bibinfo{year}{2018}),
  \urlprefix\url{https://link.aps.org/doi/10.1103/PhysRevC.98.014319}.

\bibitem[{\citenamefont{MacCormick and Audi}(2014)}]{MacCormick2014b}
\bibinfo{author}{\bibfnamefont{M.}~\bibnamefont{MacCormick}} \bibnamefont{and}
  \bibinfo{author}{\bibfnamefont{G.}~\bibnamefont{Audi}},
  \bibinfo{journal}{Nuclear Physics A} \textbf{\bibinfo{volume}{925}},
  \bibinfo{pages}{61 } (\bibinfo{year}{2014}), ISSN \bibinfo{issn}{0375-9474},
  \urlprefix\url{http://www.sciencedirect.com/science/article/pii/S0375947414000220}.

\bibitem[{\citenamefont{Ong et~al.}(2017)\citenamefont{Ong, Langer, Montes,
  Aprahamian, Bardayan, Bazin, Brown, Browne, Crawford, Cyburt
  et~al.}}]{Ong2017}
\bibinfo{author}{\bibfnamefont{W.-J.} \bibnamefont{Ong}},
  \bibinfo{author}{\bibfnamefont{C.}~\bibnamefont{Langer}},
  \bibinfo{author}{\bibfnamefont{F.}~\bibnamefont{Montes}},
  \bibinfo{author}{\bibfnamefont{A.}~\bibnamefont{Aprahamian}},
  \bibinfo{author}{\bibfnamefont{D.~W.} \bibnamefont{Bardayan}},
  \bibinfo{author}{\bibfnamefont{D.}~\bibnamefont{Bazin}},
  \bibinfo{author}{\bibfnamefont{B.~A.} \bibnamefont{Brown}},
  \bibinfo{author}{\bibfnamefont{J.}~\bibnamefont{Browne}},
  \bibinfo{author}{\bibfnamefont{H.}~\bibnamefont{Crawford}},
  \bibinfo{author}{\bibfnamefont{R.}~\bibnamefont{Cyburt}},
  \bibnamefont{et~al.}, \bibinfo{journal}{Phys. Rev. C}
  \textbf{\bibinfo{volume}{95}}, \bibinfo{pages}{055806}
  (\bibinfo{year}{2017}),
  \urlprefix\url{https://link.aps.org/doi/10.1103/PhysRevC.95.055806}.

\bibitem[{\citenamefont{Lam et~al.}(2013{\natexlab{a}})\citenamefont{Lam,
  Blank, Smirnova, Bueb, and Antony}}]{Lam2013}
\bibinfo{author}{\bibfnamefont{Y.~H.} \bibnamefont{Lam}},
  \bibinfo{author}{\bibfnamefont{B.}~\bibnamefont{Blank}},
  \bibinfo{author}{\bibfnamefont{N.~A.} \bibnamefont{Smirnova}},
  \bibinfo{author}{\bibfnamefont{J.~B.} \bibnamefont{Bueb}}, \bibnamefont{and}
  \bibinfo{author}{\bibfnamefont{M.~S.} \bibnamefont{Antony}},
  \bibinfo{journal}{Atomic Data and Nuclear Data Tables}
  \textbf{\bibinfo{volume}{99}}, \bibinfo{pages}{680 }
  (\bibinfo{year}{2013}{\natexlab{a}}), ISSN \bibinfo{issn}{0092-640X},
  \urlprefix\url{https://www.sciencedirect.com/science/article/pii/S0092640X13000569}.

\bibitem[{\citenamefont{Klochko and Smirnova}(2021)}]{Klochko2021}
\bibinfo{author}{\bibfnamefont{O.}~\bibnamefont{Klochko}} \bibnamefont{and}
  \bibinfo{author}{\bibfnamefont{N.~A.} \bibnamefont{Smirnova}},
  \bibinfo{journal}{Phys. Rev. C} \textbf{\bibinfo{volume}{103}},
  \bibinfo{pages}{024316} (\bibinfo{year}{2021}),
  \urlprefix\url{https://link.aps.org/doi/10.1103/PhysRevC.103.024316}.

\bibitem[{\citenamefont{Lam et~al.}(2013{\natexlab{b}})\citenamefont{Lam,
  Smirnova, and Caurier}}]{Lam2013b}
\bibinfo{author}{\bibfnamefont{Y.~H.} \bibnamefont{Lam}},
  \bibinfo{author}{\bibfnamefont{N.~A.} \bibnamefont{Smirnova}},
  \bibnamefont{and} \bibinfo{author}{\bibfnamefont{E.}~\bibnamefont{Caurier}},
  \bibinfo{journal}{Phys. Rev. C} \textbf{\bibinfo{volume}{87}},
  \bibinfo{pages}{054304} (\bibinfo{year}{2013}{\natexlab{b}}),
  \urlprefix\url{https://link.aps.org/doi/10.1103/PhysRevC.87.054304}.

\bibitem[{\citenamefont{Ormand and Brown}(1989)}]{Ormand1989}
\bibinfo{author}{\bibfnamefont{W.}~\bibnamefont{Ormand}} \bibnamefont{and}
  \bibinfo{author}{\bibfnamefont{B.}~\bibnamefont{Brown}},
  \bibinfo{journal}{Nuclear Physics A} \textbf{\bibinfo{volume}{491}},
  \bibinfo{pages}{1 } (\bibinfo{year}{1989}), ISSN \bibinfo{issn}{0375-9474},
  \urlprefix\url{https://www.sciencedirect.com/science/article/pii/0375947489902030}.

\bibitem[{\citenamefont{Kaneko et~al.}(2013)\citenamefont{Kaneko, Sun,
  Mizusaki, and Tazaki}}]{Kaneko2013}
\bibinfo{author}{\bibfnamefont{K.}~\bibnamefont{Kaneko}},
  \bibinfo{author}{\bibfnamefont{Y.}~\bibnamefont{Sun}},
  \bibinfo{author}{\bibfnamefont{T.}~\bibnamefont{Mizusaki}}, \bibnamefont{and}
  \bibinfo{author}{\bibfnamefont{S.}~\bibnamefont{Tazaki}},
  \bibinfo{journal}{Phys. Rev. Lett.} \textbf{\bibinfo{volume}{110}},
  \bibinfo{pages}{172505} (\bibinfo{year}{2013}),
  \urlprefix\url{https://link.aps.org/doi/10.1103/PhysRevLett.110.172505}.

\bibitem[{\citenamefont{Kaneko et~al.}(2014)\citenamefont{Kaneko, Sun,
  Mizusaki, and Tazaki}}]{Kaneko2014}
\bibinfo{author}{\bibfnamefont{K.}~\bibnamefont{Kaneko}},
  \bibinfo{author}{\bibfnamefont{Y.}~\bibnamefont{Sun}},
  \bibinfo{author}{\bibfnamefont{T.}~\bibnamefont{Mizusaki}}, \bibnamefont{and}
  \bibinfo{author}{\bibfnamefont{S.}~\bibnamefont{Tazaki}},
  \bibinfo{journal}{Phys. Rev. C} \textbf{\bibinfo{volume}{89}},
  \bibinfo{pages}{031302} (\bibinfo{year}{2014}),
  \urlprefix\url{https://link.aps.org/doi/10.1103/PhysRevC.89.031302}.

\bibitem[{\citenamefont{Kaneko et~al.}(2015)\citenamefont{Kaneko, Sun,
  Mizusaki, and Tazaki}}]{Kaneko2015}
\bibinfo{author}{\bibfnamefont{K.}~\bibnamefont{Kaneko}},
  \bibinfo{author}{\bibfnamefont{Y.}~\bibnamefont{Sun}},
  \bibinfo{author}{\bibfnamefont{T.}~\bibnamefont{Mizusaki}}, \bibnamefont{and}
  \bibinfo{author}{\bibfnamefont{S.}~\bibnamefont{Tazaki}},
  \bibinfo{journal}{Physica Scripta} \textbf{\bibinfo{volume}{T166}},
  \bibinfo{pages}{014011} (\bibinfo{year}{2015}),
  \urlprefix\url{https://doi.org/10.1088/0031-8949/2015/t166/014011}.

\bibitem[{\citenamefont{Puentes et~al.}(2020)\citenamefont{Puentes, Bollen,
  Brodeur, Eibach, Gulyuz, Hamaker, Izzo, Lenzi, MacCormick, Redshaw
  et~al.}}]{Puentes2020}
\bibinfo{author}{\bibfnamefont{D.}~\bibnamefont{Puentes}},
  \bibinfo{author}{\bibfnamefont{G.}~\bibnamefont{Bollen}},
  \bibinfo{author}{\bibfnamefont{M.}~\bibnamefont{Brodeur}},
  \bibinfo{author}{\bibfnamefont{M.}~\bibnamefont{Eibach}},
  \bibinfo{author}{\bibfnamefont{K.}~\bibnamefont{Gulyuz}},
  \bibinfo{author}{\bibfnamefont{A.}~\bibnamefont{Hamaker}},
  \bibinfo{author}{\bibfnamefont{C.}~\bibnamefont{Izzo}},
  \bibinfo{author}{\bibfnamefont{S.~M.} \bibnamefont{Lenzi}},
  \bibinfo{author}{\bibfnamefont{M.}~\bibnamefont{MacCormick}},
  \bibinfo{author}{\bibfnamefont{M.}~\bibnamefont{Redshaw}},
  \bibnamefont{et~al.}, \bibinfo{journal}{Phys. Rev. C}
  \textbf{\bibinfo{volume}{101}}, \bibinfo{pages}{064309}
  (\bibinfo{year}{2020}),
  \urlprefix\url{https://link.aps.org/doi/10.1103/PhysRevC.101.064309}.

\bibitem[{\citenamefont{Gadea et~al.}(2006)\citenamefont{Gadea, Lenzi, Lunardi,
  M\ifmmode~\u{a}\else \u{a}\fi{}rginean, Zuker, de~Angelis, Axiotis,
  Mart\'{\i}nez, Napoli, Farnea et~al.}}]{Gadea2006}
\bibinfo{author}{\bibfnamefont{A.}~\bibnamefont{Gadea}},
  \bibinfo{author}{\bibfnamefont{S.~M.} \bibnamefont{Lenzi}},
  \bibinfo{author}{\bibfnamefont{S.}~\bibnamefont{Lunardi}},
  \bibinfo{author}{\bibfnamefont{N.}~\bibnamefont{M\ifmmode~\u{a}\else
  \u{a}\fi{}rginean}}, \bibinfo{author}{\bibfnamefont{A.~P.}
  \bibnamefont{Zuker}},
  \bibinfo{author}{\bibfnamefont{G.}~\bibnamefont{de~Angelis}},
  \bibinfo{author}{\bibfnamefont{M.}~\bibnamefont{Axiotis}},
  \bibinfo{author}{\bibfnamefont{T.}~\bibnamefont{Mart\'{\i}nez}},
  \bibinfo{author}{\bibfnamefont{D.~R.} \bibnamefont{Napoli}},
  \bibinfo{author}{\bibfnamefont{E.}~\bibnamefont{Farnea}},
  \bibnamefont{et~al.}, \bibinfo{journal}{Phys. Rev. Lett.}
  \textbf{\bibinfo{volume}{97}}, \bibinfo{pages}{152501}
  (\bibinfo{year}{2006}),
  \urlprefix\url{https://link.aps.org/doi/10.1103/PhysRevLett.97.152501}.

\bibitem[{\citenamefont{Ormand et~al.}(2017)\citenamefont{Ormand, Brown, and
  Hjorth-Jensen}}]{Ormand2017}
\bibinfo{author}{\bibfnamefont{W.}~\bibnamefont{Ormand}},
  \bibinfo{author}{\bibfnamefont{B.}~\bibnamefont{Brown}}, \bibnamefont{and}
  \bibinfo{author}{\bibfnamefont{M.}~\bibnamefont{Hjorth-Jensen}},
  \bibinfo{journal}{Physical Review C} \textbf{\bibinfo{volume}{96}},
  \bibinfo{pages}{024323} (\bibinfo{year}{2017}).

\bibitem[{\citenamefont{Dilling et~al.}(2003)\citenamefont{Dilling, Bricault,
  Smith, and Kluge}}]{Dilling2003}
\bibinfo{author}{\bibfnamefont{J.}~\bibnamefont{Dilling}},
  \bibinfo{author}{\bibfnamefont{P.}~\bibnamefont{Bricault}},
  \bibinfo{author}{\bibfnamefont{M.}~\bibnamefont{Smith}}, \bibnamefont{and}
  \bibinfo{author}{\bibfnamefont{H.-J.} \bibnamefont{Kluge}},
  \bibinfo{journal}{Nuclear Instruments and Methods in Physics Research Section
  B: Beam Interactions with Materials and Atoms}
  \textbf{\bibinfo{volume}{204}}, \bibinfo{pages}{492 } (\bibinfo{year}{2003}),
  ISSN \bibinfo{issn}{0168-583X}, \bibinfo{note}{14th International Conference
  on Electromagnetic Isotope Separators and Techniques Related to their
  Applications},
  \urlprefix\url{http://www.sciencedirect.com/science/article/pii/S0168583X02021183}.

\bibitem[{\citenamefont{Jesch et~al.}(2017)\citenamefont{Jesch, Dickel,
  Pla{\ss}, Short, Andres, Dilling, Geissel, Greiner, Lang, Leach
  et~al.}}]{Jesch2017}
\bibinfo{author}{\bibfnamefont{C.}~\bibnamefont{Jesch}},
  \bibinfo{author}{\bibfnamefont{T.}~\bibnamefont{Dickel}},
  \bibinfo{author}{\bibfnamefont{W.~R.} \bibnamefont{Pla{\ss}}},
  \bibinfo{author}{\bibfnamefont{D.}~\bibnamefont{Short}},
  \bibinfo{author}{\bibfnamefont{S.~A.~S.} \bibnamefont{Andres}},
  \bibinfo{author}{\bibfnamefont{J.}~\bibnamefont{Dilling}},
  \bibinfo{author}{\bibfnamefont{H.}~\bibnamefont{Geissel}},
  \bibinfo{author}{\bibfnamefont{F.}~\bibnamefont{Greiner}},
  \bibinfo{author}{\bibfnamefont{J.}~\bibnamefont{Lang}},
  \bibinfo{author}{\bibfnamefont{K.~G.} \bibnamefont{Leach}},
  \bibnamefont{et~al.}, in \emph{\bibinfo{booktitle}{TCP 2014}}, edited by
  \bibinfo{editor}{\bibfnamefont{M.}~\bibnamefont{Wada}},
  \bibinfo{editor}{\bibfnamefont{P.}~\bibnamefont{Schury}}, \bibnamefont{and}
  \bibinfo{editor}{\bibfnamefont{Y.}~\bibnamefont{Ichikawa}}
  (\bibinfo{publisher}{Springer International Publishing},
  \bibinfo{address}{Cham}, \bibinfo{year}{2017}), pp. \bibinfo{pages}{175 --
  184}, ISBN \bibinfo{isbn}{978-3-319-61588-2}.

\bibitem[{\citenamefont{Dilling et~al.}(2014)\citenamefont{Dilling,
  Kr{\"u}cken, and Merminga}}]{Dilling2014ISAC}
\bibinfo{author}{\bibfnamefont{J.}~\bibnamefont{Dilling}},
  \bibinfo{author}{\bibfnamefont{R.}~\bibnamefont{Kr{\"u}cken}},
  \bibnamefont{and} \bibinfo{author}{\bibfnamefont{L.}~\bibnamefont{Merminga}},
  \emph{\bibinfo{title}{ISAC and ARIEL: The TRIUMF Radioactive Beam Facilities
  and the Scientific Program}} (\bibinfo{publisher}{Springer},
  \bibinfo{year}{2014}).

\bibitem[{\citenamefont{Lassen et~al.}(2009)\citenamefont{Lassen, Bricault,
  Dombsky, Lavoie, Gillner, Gottwald, Hellbusch, Teigelhöfer, Voss, and
  Wendt}}]{Lassen2009}
\bibinfo{author}{\bibfnamefont{J.}~\bibnamefont{Lassen}},
  \bibinfo{author}{\bibfnamefont{P.}~\bibnamefont{Bricault}},
  \bibinfo{author}{\bibfnamefont{M.}~\bibnamefont{Dombsky}},
  \bibinfo{author}{\bibfnamefont{J.~P.} \bibnamefont{Lavoie}},
  \bibinfo{author}{\bibfnamefont{M.}~\bibnamefont{Gillner}},
  \bibinfo{author}{\bibfnamefont{T.}~\bibnamefont{Gottwald}},
  \bibinfo{author}{\bibfnamefont{F.}~\bibnamefont{Hellbusch}},
  \bibinfo{author}{\bibfnamefont{A.}~\bibnamefont{Teigelhöfer}},
  \bibinfo{author}{\bibfnamefont{A.}~\bibnamefont{Voss}}, \bibnamefont{and}
  \bibinfo{author}{\bibfnamefont{K.~D.~A.} \bibnamefont{Wendt}},
  \bibinfo{journal}{AIP Conference Proceedings}
  \textbf{\bibinfo{volume}{1104}}, \bibinfo{pages}{9 } (\bibinfo{year}{2009}),
  \eprint{https://aip.scitation.org/doi/pdf/10.1063/1.3115616},
  \urlprefix\url{https://aip.scitation.org/doi/abs/10.1063/1.3115616}.

\bibitem[{\citenamefont{Bricault et~al.}(2002)\citenamefont{Bricault, Baartman,
  Dombsky, Hurst, Mark, Stanford, and Schmor}}]{Bricault2002}
\bibinfo{author}{\bibfnamefont{P.}~\bibnamefont{Bricault}},
  \bibinfo{author}{\bibfnamefont{R.}~\bibnamefont{Baartman}},
  \bibinfo{author}{\bibfnamefont{M.}~\bibnamefont{Dombsky}},
  \bibinfo{author}{\bibfnamefont{A.}~\bibnamefont{Hurst}},
  \bibinfo{author}{\bibfnamefont{C.}~\bibnamefont{Mark}},
  \bibinfo{author}{\bibfnamefont{G.}~\bibnamefont{Stanford}}, \bibnamefont{and}
  \bibinfo{author}{\bibfnamefont{P.}~\bibnamefont{Schmor}},
  \bibinfo{journal}{Nuclear Physics A} \textbf{\bibinfo{volume}{701}},
  \bibinfo{pages}{49 } (\bibinfo{year}{2002}), ISSN \bibinfo{issn}{0375-9474},
  \bibinfo{note}{5th International Conference on Radioactive Nuclear Beams},
  \urlprefix\url{https://www.sciencedirect.com/science/article/pii/S0375947401015469}.

\bibitem[{\citenamefont{Brunner et~al.}(2012)\citenamefont{Brunner, Smith,
  Brodeur, Ettenauer, Gallant, Simon, Chaudhuri, Lapierre, Mané, Ringle
  et~al.}}]{Brunner2012}
\bibinfo{author}{\bibfnamefont{T.}~\bibnamefont{Brunner}},
  \bibinfo{author}{\bibfnamefont{M.}~\bibnamefont{Smith}},
  \bibinfo{author}{\bibfnamefont{M.}~\bibnamefont{Brodeur}},
  \bibinfo{author}{\bibfnamefont{S.}~\bibnamefont{Ettenauer}},
  \bibinfo{author}{\bibfnamefont{A.}~\bibnamefont{Gallant}},
  \bibinfo{author}{\bibfnamefont{V.}~\bibnamefont{Simon}},
  \bibinfo{author}{\bibfnamefont{A.}~\bibnamefont{Chaudhuri}},
  \bibinfo{author}{\bibfnamefont{A.}~\bibnamefont{Lapierre}},
  \bibinfo{author}{\bibfnamefont{E.}~\bibnamefont{Mané}},
  \bibinfo{author}{\bibfnamefont{R.}~\bibnamefont{Ringle}},
  \bibnamefont{et~al.}, \bibinfo{journal}{Nuclear Instruments and Methods in
  Physics Research Section A: Accelerators, Spectrometers, Detectors and
  Associated Equipment} \textbf{\bibinfo{volume}{676}} (\bibinfo{year}{2012}).

\bibitem[{\citenamefont{Yavor et~al.}(2015)\citenamefont{Yavor, Plaß, Dickel,
  Geissel, and Scheidenberger}}]{Yavor2015}
\bibinfo{author}{\bibfnamefont{M.~I.} \bibnamefont{Yavor}},
  \bibinfo{author}{\bibfnamefont{W.~R.} \bibnamefont{Plaß}},
  \bibinfo{author}{\bibfnamefont{T.}~\bibnamefont{Dickel}},
  \bibinfo{author}{\bibfnamefont{H.}~\bibnamefont{Geissel}}, \bibnamefont{and}
  \bibinfo{author}{\bibfnamefont{C.}~\bibnamefont{Scheidenberger}},
  \bibinfo{journal}{International Journal of Mass Spectrometry}
  \textbf{\bibinfo{volume}{381-382}}, \bibinfo{pages}{1 }
  (\bibinfo{year}{2015}), ISSN \bibinfo{issn}{1387-3806},
  \urlprefix\url{https://www.sciencedirect.com/science/article/pii/S1387380615000202}.

\bibitem[{\citenamefont{Dickel et~al.}(2017{\natexlab{a}})\citenamefont{Dickel,
  Yavor, Lang, Plaß, Lippert, Geissel, and Scheidenberger}}]{Dickel2017b}
\bibinfo{author}{\bibfnamefont{T.}~\bibnamefont{Dickel}},
  \bibinfo{author}{\bibfnamefont{M.~I.} \bibnamefont{Yavor}},
  \bibinfo{author}{\bibfnamefont{J.}~\bibnamefont{Lang}},
  \bibinfo{author}{\bibfnamefont{W.~R.} \bibnamefont{Plaß}},
  \bibinfo{author}{\bibfnamefont{W.}~\bibnamefont{Lippert}},
  \bibinfo{author}{\bibfnamefont{H.}~\bibnamefont{Geissel}}, \bibnamefont{and}
  \bibinfo{author}{\bibfnamefont{C.}~\bibnamefont{Scheidenberger}},
  \bibinfo{journal}{International Journal of Mass Spectrometry}
  \textbf{\bibinfo{volume}{412}}, \bibinfo{pages}{1 }
  (\bibinfo{year}{2017}{\natexlab{a}}), ISSN \bibinfo{issn}{1387-3806},
  \urlprefix\url{https://www.sciencedirect.com/science/article/pii/S1387380616302664}.

\bibitem[{\citenamefont{Bergmann}(2015)}]{BergmannMSc2015}
\bibinfo{author}{\bibfnamefont{J.}~\bibnamefont{Bergmann}}, Master's thesis,
  \bibinfo{school}{Justus-Liebig-Universit{\"a}t Gie{\ss}en}
  (\bibinfo{year}{2015}).

\bibitem[{\citenamefont{Dickel et~al.}(2019)\citenamefont{Dickel,
  San~Andr{\'e}s, Beck, Bergmann, Dilling, Greiner, Hornung, Jacobs,
  Kripko-Koncz, Kwiatkowski et~al.}}]{Dickel2019}
\bibinfo{author}{\bibfnamefont{T.}~\bibnamefont{Dickel}},
  \bibinfo{author}{\bibfnamefont{S.~A.} \bibnamefont{San~Andr{\'e}s}},
  \bibinfo{author}{\bibfnamefont{S.}~\bibnamefont{Beck}},
  \bibinfo{author}{\bibfnamefont{J.}~\bibnamefont{Bergmann}},
  \bibinfo{author}{\bibfnamefont{J.}~\bibnamefont{Dilling}},
  \bibinfo{author}{\bibfnamefont{F.}~\bibnamefont{Greiner}},
  \bibinfo{author}{\bibfnamefont{C.}~\bibnamefont{Hornung}},
  \bibinfo{author}{\bibfnamefont{A.}~\bibnamefont{Jacobs}},
  \bibinfo{author}{\bibfnamefont{G.}~\bibnamefont{Kripko-Koncz}},
  \bibinfo{author}{\bibfnamefont{A.}~\bibnamefont{Kwiatkowski}},
  \bibnamefont{et~al.}, \bibinfo{journal}{Hyperfine Interactions}
  \textbf{\bibinfo{volume}{240}}, \bibinfo{pages}{1 } (\bibinfo{year}{2019}),
  \urlprefix\url{https://doi.org/10.1007/s10751-019-1610-y}.

\bibitem[{\citenamefont{Reiter et~al.}(2021)\citenamefont{Reiter, Andrés,
  Bergmann, Dickel, Dilling, Jacobs, Kwiatkowski, Plaß, Scheidenberger, Short
  et~al.}}]{Reiter2021}
\bibinfo{author}{\bibfnamefont{M.}~\bibnamefont{Reiter}},
  \bibinfo{author}{\bibfnamefont{S.~A.~S.} \bibnamefont{Andrés}},
  \bibinfo{author}{\bibfnamefont{J.}~\bibnamefont{Bergmann}},
  \bibinfo{author}{\bibfnamefont{T.}~\bibnamefont{Dickel}},
  \bibinfo{author}{\bibfnamefont{J.}~\bibnamefont{Dilling}},
  \bibinfo{author}{\bibfnamefont{A.}~\bibnamefont{Jacobs}},
  \bibinfo{author}{\bibfnamefont{A.}~\bibnamefont{Kwiatkowski}},
  \bibinfo{author}{\bibfnamefont{W.}~\bibnamefont{Plaß}},
  \bibinfo{author}{\bibfnamefont{C.}~\bibnamefont{Scheidenberger}},
  \bibinfo{author}{\bibfnamefont{D.}~\bibnamefont{Short}},
  \bibnamefont{et~al.}, \bibinfo{journal}{Nuclear Instruments and Methods in
  Physics Research Section A: Accelerators, Spectrometers, Detectors and
  Associated Equipment} \textbf{\bibinfo{volume}{1018}},
  \bibinfo{pages}{165823} (\bibinfo{year}{2021}), ISSN
  \bibinfo{issn}{0168-9002},
  \urlprefix\url{https://www.sciencedirect.com/science/article/pii/S0168900221008081}.

\bibitem[{\citenamefont{Dickel et~al.}(2017{\natexlab{b}})\citenamefont{Dickel,
  Pla{\ss}, Lippert, Lang, Yavor, Geissel, and Scheidenberger}}]{Dickel2017}
\bibinfo{author}{\bibfnamefont{T.}~\bibnamefont{Dickel}},
  \bibinfo{author}{\bibfnamefont{W.~R.} \bibnamefont{Pla{\ss}}},
  \bibinfo{author}{\bibfnamefont{W.}~\bibnamefont{Lippert}},
  \bibinfo{author}{\bibfnamefont{J.}~\bibnamefont{Lang}},
  \bibinfo{author}{\bibfnamefont{M.~I.} \bibnamefont{Yavor}},
  \bibinfo{author}{\bibfnamefont{H.}~\bibnamefont{Geissel}}, \bibnamefont{and}
  \bibinfo{author}{\bibfnamefont{C.}~\bibnamefont{Scheidenberger}},
  \bibinfo{journal}{Journal of The American Society for Mass Spectrometry}
  \textbf{\bibinfo{volume}{28}}, \bibinfo{pages}{1079 }
  (\bibinfo{year}{2017}{\natexlab{b}}),
  \urlprefix\url{https://doi.org/10.1007/s13361-017-1617-z}.

\bibitem[{\citenamefont{Beck et~al.}(2021)\citenamefont{Beck, Kootte, Dedes,
  Dickel, Kwiatkowski, Lykiardopoulou, Pla\ss{}, Reiter, Andreoiu, Bergmann
  et~al.}}]{Beck2021}
\bibinfo{author}{\bibfnamefont{S.}~\bibnamefont{Beck}},
  \bibinfo{author}{\bibfnamefont{B.}~\bibnamefont{Kootte}},
  \bibinfo{author}{\bibfnamefont{I.}~\bibnamefont{Dedes}},
  \bibinfo{author}{\bibfnamefont{T.}~\bibnamefont{Dickel}},
  \bibinfo{author}{\bibfnamefont{A.~A.} \bibnamefont{Kwiatkowski}},
  \bibinfo{author}{\bibfnamefont{E.~M.} \bibnamefont{Lykiardopoulou}},
  \bibinfo{author}{\bibfnamefont{W.~R.} \bibnamefont{Pla\ss{}}},
  \bibinfo{author}{\bibfnamefont{M.~P.} \bibnamefont{Reiter}},
  \bibinfo{author}{\bibfnamefont{C.}~\bibnamefont{Andreoiu}},
  \bibinfo{author}{\bibfnamefont{J.}~\bibnamefont{Bergmann}},
  \bibnamefont{et~al.}, \bibinfo{journal}{Phys. Rev. Lett.}
  \textbf{\bibinfo{volume}{127}}, \bibinfo{pages}{112501}
  (\bibinfo{year}{2021}),
  \urlprefix\url{https://link.aps.org/doi/10.1103/PhysRevLett.127.112501}.

\bibitem[{\citenamefont{Mukul et~al.}(2021)\citenamefont{Mukul, Andreoiu,
  Bergmann, Brodeur, Brunner, Dietrich, Dickel, Dillmann, Dunling, Fusco
  et~al.}}]{Mukul2021}
\bibinfo{author}{\bibfnamefont{I.}~\bibnamefont{Mukul}},
  \bibinfo{author}{\bibfnamefont{C.}~\bibnamefont{Andreoiu}},
  \bibinfo{author}{\bibfnamefont{J.}~\bibnamefont{Bergmann}},
  \bibinfo{author}{\bibfnamefont{M.}~\bibnamefont{Brodeur}},
  \bibinfo{author}{\bibfnamefont{T.}~\bibnamefont{Brunner}},
  \bibinfo{author}{\bibfnamefont{K.~A.} \bibnamefont{Dietrich}},
  \bibinfo{author}{\bibfnamefont{T.}~\bibnamefont{Dickel}},
  \bibinfo{author}{\bibfnamefont{I.}~\bibnamefont{Dillmann}},
  \bibinfo{author}{\bibfnamefont{E.}~\bibnamefont{Dunling}},
  \bibinfo{author}{\bibfnamefont{D.}~\bibnamefont{Fusco}},
  \bibnamefont{et~al.}, \bibinfo{journal}{Phys. Rev. C}
  \textbf{\bibinfo{volume}{103}}, \bibinfo{pages}{044320}
  (\bibinfo{year}{2021}),
  \urlprefix\url{https://link.aps.org/doi/10.1103/PhysRevC.103.044320}.

\bibitem[{\citenamefont{Izzo et~al.}(2021)\citenamefont{Izzo, Bergmann,
  Dietrich, Dunling, Fusco, Jacobs, Kootte, Kripk\'o-Koncz, Lan,
  Leistenschneider et~al.}}]{Izzo2021}
\bibinfo{author}{\bibfnamefont{C.}~\bibnamefont{Izzo}},
  \bibinfo{author}{\bibfnamefont{J.}~\bibnamefont{Bergmann}},
  \bibinfo{author}{\bibfnamefont{K.~A.} \bibnamefont{Dietrich}},
  \bibinfo{author}{\bibfnamefont{E.}~\bibnamefont{Dunling}},
  \bibinfo{author}{\bibfnamefont{D.}~\bibnamefont{Fusco}},
  \bibinfo{author}{\bibfnamefont{A.}~\bibnamefont{Jacobs}},
  \bibinfo{author}{\bibfnamefont{B.}~\bibnamefont{Kootte}},
  \bibinfo{author}{\bibfnamefont{G.}~\bibnamefont{Kripk\'o-Koncz}},
  \bibinfo{author}{\bibfnamefont{Y.}~\bibnamefont{Lan}},
  \bibinfo{author}{\bibfnamefont{E.}~\bibnamefont{Leistenschneider}},
  \bibnamefont{et~al.}, \bibinfo{journal}{Phys. Rev. C}
  \textbf{\bibinfo{volume}{103}}, \bibinfo{pages}{025811}
  (\bibinfo{year}{2021}),
  \urlprefix\url{https://link.aps.org/doi/10.1103/PhysRevC.103.025811}.

\bibitem[{\citenamefont{Purushothaman et~al.}(2017)\citenamefont{Purushothaman,
  Andrés, Bergmann, Dickel, Ebert, Geissel, Hornung, Plaß, Rappold,
  Scheidenberger et~al.}}]{Purushothaman2017}
\bibinfo{author}{\bibfnamefont{S.}~\bibnamefont{Purushothaman}},
  \bibinfo{author}{\bibfnamefont{S.~A.~S.} \bibnamefont{Andrés}},
  \bibinfo{author}{\bibfnamefont{J.}~\bibnamefont{Bergmann}},
  \bibinfo{author}{\bibfnamefont{T.}~\bibnamefont{Dickel}},
  \bibinfo{author}{\bibfnamefont{J.}~\bibnamefont{Ebert}},
  \bibinfo{author}{\bibfnamefont{H.}~\bibnamefont{Geissel}},
  \bibinfo{author}{\bibfnamefont{C.}~\bibnamefont{Hornung}},
  \bibinfo{author}{\bibfnamefont{W.}~\bibnamefont{Plaß}},
  \bibinfo{author}{\bibfnamefont{C.}~\bibnamefont{Rappold}},
  \bibinfo{author}{\bibfnamefont{C.}~\bibnamefont{Scheidenberger}},
  \bibnamefont{et~al.}, \bibinfo{journal}{International Journal of Mass
  Spectrometry} \textbf{\bibinfo{volume}{421}}, \bibinfo{pages}{245 }
  (\bibinfo{year}{2017}), ISSN \bibinfo{issn}{1387-3806},
  \urlprefix\url{http://www.sciencedirect.com/science/article/pii/S1387380616302913}.

\bibitem[{\citenamefont{San~Andr\'es et~al.}(2019)\citenamefont{San~Andr\'es,
  Hornung, Ebert, Pla\ss{}, Dickel, Geissel, Scheidenberger, Bergmann, Greiner,
  Haettner et~al.}}]{Ayet2019}
\bibinfo{author}{\bibfnamefont{A.~S.} \bibnamefont{San~Andr\'es}},
  \bibinfo{author}{\bibfnamefont{C.}~\bibnamefont{Hornung}},
  \bibinfo{author}{\bibfnamefont{J.}~\bibnamefont{Ebert}},
  \bibinfo{author}{\bibfnamefont{W.~R.} \bibnamefont{Pla\ss{}}},
  \bibinfo{author}{\bibfnamefont{T.}~\bibnamefont{Dickel}},
  \bibinfo{author}{\bibfnamefont{H.}~\bibnamefont{Geissel}},
  \bibinfo{author}{\bibfnamefont{C.}~\bibnamefont{Scheidenberger}},
  \bibinfo{author}{\bibfnamefont{J.}~\bibnamefont{Bergmann}},
  \bibinfo{author}{\bibfnamefont{F.}~\bibnamefont{Greiner}},
  \bibinfo{author}{\bibfnamefont{E.}~\bibnamefont{Haettner}},
  \bibnamefont{et~al.}, \bibinfo{journal}{Phys. Rev. C}
  \textbf{\bibinfo{volume}{99}}, \bibinfo{pages}{064313}
  (\bibinfo{year}{2019}),
  \urlprefix\url{https://link.aps.org/doi/10.1103/PhysRevC.99.064313}.

\bibitem[{\citenamefont{Paul}(2020)}]{emgfit_v0_3_5}
\bibinfo{author}{\bibfnamefont{S.~F.} \bibnamefont{Paul}},
  \emph{\bibinfo{title}{{emgfit v0.3.5 - Fitting of time-of-flight mass spectra
  with hyper-EMG models}}} (\bibinfo{year}{2020}),
  \urlprefix\url{https://doi.org/10.5281/zenodo.4731019}.

\bibitem[{\citenamefont{Paul}(2021)}]{emgfit}
\bibinfo{author}{\bibfnamefont{S.~F.} \bibnamefont{Paul}},
  \emph{\bibinfo{title}{{emgfit - Fitting of time-of-flight mass spectra with
  hyper-EMG models}}} (\bibinfo{year}{2021}),
  \urlprefix\url{https://doi.org/10.5281/zenodo.4731018}.

\bibitem[{\citenamefont{Mighell}(1999)}]{Mighell1999}
\bibinfo{author}{\bibfnamefont{K.~J.} \bibnamefont{Mighell}},
  \bibinfo{journal}{The Astrophysical Journal} \textbf{\bibinfo{volume}{518}},
  \bibinfo{pages}{380} (\bibinfo{year}{1999}),
  \urlprefix\url{https://doi.org/10.1086/307253}.

\bibitem[{\citenamefont{Dembinski et~al.}(2019)\citenamefont{Dembinski,
  Schmelling, and Waldi}}]{Dembinski2019}
\bibinfo{author}{\bibfnamefont{H.}~\bibnamefont{Dembinski}},
  \bibinfo{author}{\bibfnamefont{M.}~\bibnamefont{Schmelling}},
  \bibnamefont{and} \bibinfo{author}{\bibfnamefont{R.}~\bibnamefont{Waldi}},
  \bibinfo{journal}{Nuclear Instruments and Methods in Physics Research Section
  A: Accelerators, Spectrometers, Detectors and Associated Equipment}
  \textbf{\bibinfo{volume}{940}}, \bibinfo{pages}{135 } (\bibinfo{year}{2019}),
  ISSN \bibinfo{issn}{0168-9002},
  \urlprefix\url{https://www.sciencedirect.com/science/article/pii/S0168900219307508}.

\bibitem[{\citenamefont{Cash}(1979)}]{Cash1979}
\bibinfo{author}{\bibfnamefont{W.}~\bibnamefont{Cash}}, \bibinfo{journal}{The
  Astrophysical Journal} \textbf{\bibinfo{volume}{228}}, \bibinfo{pages}{939 }
  (\bibinfo{year}{1979}),
  \urlprefix\url{https://ui.adsabs.harvard.edu/link_gateway/1979ApJ...228..939C/doi:10.1086/156922}.

\bibitem[{\citenamefont{Kaastra}(2017)}]{Kaastra2017}
\bibinfo{author}{\bibfnamefont{J.}~\bibnamefont{Kaastra}},
  \bibinfo{journal}{Astronomy \& Astrophysics} \textbf{\bibinfo{volume}{605}},
  \bibinfo{pages}{A51} (\bibinfo{year}{2017}),
  \urlprefix\url{https://doi.org/10.1051/0004-6361/201629319}.

\bibitem[{\citenamefont{Smith et~al.}(2020)\citenamefont{Smith, Murb{\"o}ck,
  Dunling, Jacobs, Kootte, Lan, Leistenschneider, Lunney, Lykiardopoulou, Mukul
  et~al.}}]{Smith2020}
\bibinfo{author}{\bibfnamefont{M.~B.} \bibnamefont{Smith}},
  \bibinfo{author}{\bibfnamefont{T.}~\bibnamefont{Murb{\"o}ck}},
  \bibinfo{author}{\bibfnamefont{E.}~\bibnamefont{Dunling}},
  \bibinfo{author}{\bibfnamefont{A.}~\bibnamefont{Jacobs}},
  \bibinfo{author}{\bibfnamefont{B.}~\bibnamefont{Kootte}},
  \bibinfo{author}{\bibfnamefont{Y.}~\bibnamefont{Lan}},
  \bibinfo{author}{\bibfnamefont{E.}~\bibnamefont{Leistenschneider}},
  \bibinfo{author}{\bibfnamefont{D.}~\bibnamefont{Lunney}},
  \bibinfo{author}{\bibfnamefont{E.~M.} \bibnamefont{Lykiardopoulou}},
  \bibinfo{author}{\bibfnamefont{I.}~\bibnamefont{Mukul}},
  \bibnamefont{et~al.}, \bibinfo{journal}{Hyperfine Interactions}
  \textbf{\bibinfo{volume}{241}}, \bibinfo{pages}{1 } (\bibinfo{year}{2020}),
  \urlprefix\url{https://doi.org/10.1007/s10751-020-01722-2}.

\bibitem[{\citenamefont{Jayamanna et~al.}(2008)\citenamefont{Jayamanna, Ames,
  Cojocaru, Baartman, Bricault, Dube, Laxdal, Marchetto, MacDonald, Schmor
  et~al.}}]{Jayamanna2008}
\bibinfo{author}{\bibfnamefont{K.}~\bibnamefont{Jayamanna}},
  \bibinfo{author}{\bibfnamefont{F.}~\bibnamefont{Ames}},
  \bibinfo{author}{\bibfnamefont{G.}~\bibnamefont{Cojocaru}},
  \bibinfo{author}{\bibfnamefont{R.}~\bibnamefont{Baartman}},
  \bibinfo{author}{\bibfnamefont{P.}~\bibnamefont{Bricault}},
  \bibinfo{author}{\bibfnamefont{R.}~\bibnamefont{Dube}},
  \bibinfo{author}{\bibfnamefont{R.}~\bibnamefont{Laxdal}},
  \bibinfo{author}{\bibfnamefont{M.}~\bibnamefont{Marchetto}},
  \bibinfo{author}{\bibfnamefont{M.}~\bibnamefont{MacDonald}},
  \bibinfo{author}{\bibfnamefont{P.}~\bibnamefont{Schmor}},
  \bibnamefont{et~al.}, \bibinfo{journal}{Review of Scientific Instruments}
  \textbf{\bibinfo{volume}{79}}, \bibinfo{pages}{02C711}
  (\bibinfo{year}{2008}),
  \eprint{https://aip.scitation.org/doi/pdf/10.1063/1.2816928},
  \urlprefix\url{https://aip.scitation.org/doi/abs/10.1063/1.2816928}.

\bibitem[{\citenamefont{Mazzocchi et~al.}(2001)\citenamefont{Mazzocchi, Janas,
  D{\"o}ring, Axiotis, Batist, Borcea, Cano-Ott, Caurier, De~Angelis, Farnea
  et~al.}}]{Mazzocchi2001}
\bibinfo{author}{\bibfnamefont{C.}~\bibnamefont{Mazzocchi}},
  \bibinfo{author}{\bibfnamefont{Z.}~\bibnamefont{Janas}},
  \bibinfo{author}{\bibfnamefont{J.}~\bibnamefont{D{\"o}ring}},
  \bibinfo{author}{\bibfnamefont{M.}~\bibnamefont{Axiotis}},
  \bibinfo{author}{\bibfnamefont{L.}~\bibnamefont{Batist}},
  \bibinfo{author}{\bibfnamefont{R.}~\bibnamefont{Borcea}},
  \bibinfo{author}{\bibfnamefont{D.}~\bibnamefont{Cano-Ott}},
  \bibinfo{author}{\bibfnamefont{E.}~\bibnamefont{Caurier}},
  \bibinfo{author}{\bibfnamefont{G.}~\bibnamefont{De~Angelis}},
  \bibinfo{author}{\bibfnamefont{E.}~\bibnamefont{Farnea}},
  \bibnamefont{et~al.}, \bibinfo{journal}{The European Physical Journal
  A-Hadrons and Nuclei} \textbf{\bibinfo{volume}{12}}, \bibinfo{pages}{269 }
  (\bibinfo{year}{2001}),
  \urlprefix\url{https://doi.org/10.1007/s100500170004}.

\bibitem[{\citenamefont{Orrigo et~al.}(2021)\citenamefont{Orrigo, Rubio,
  Gelletly, Aguilera, Algora, Morales, Agramunt, Ahn, Ascher, Blank
  et~al.}}]{Orrigo2021}
\bibinfo{author}{\bibfnamefont{S.~E.~A.} \bibnamefont{Orrigo}},
  \bibinfo{author}{\bibfnamefont{B.}~\bibnamefont{Rubio}},
  \bibinfo{author}{\bibfnamefont{W.}~\bibnamefont{Gelletly}},
  \bibinfo{author}{\bibfnamefont{P.}~\bibnamefont{Aguilera}},
  \bibinfo{author}{\bibfnamefont{A.}~\bibnamefont{Algora}},
  \bibinfo{author}{\bibfnamefont{A.~I.} \bibnamefont{Morales}},
  \bibinfo{author}{\bibfnamefont{J.}~\bibnamefont{Agramunt}},
  \bibinfo{author}{\bibfnamefont{D.~S.} \bibnamefont{Ahn}},
  \bibinfo{author}{\bibfnamefont{P.}~\bibnamefont{Ascher}},
  \bibinfo{author}{\bibfnamefont{B.}~\bibnamefont{Blank}},
  \bibnamefont{et~al.}, \bibinfo{journal}{Phys. Rev. C}
  \textbf{\bibinfo{volume}{103}}, \bibinfo{pages}{014324}
  (\bibinfo{year}{2021}),
  \urlprefix\url{https://link.aps.org/doi/10.1103/PhysRevC.103.014324}.

\bibitem[{\citenamefont{Zuber and Singh}(2015)}]{Zuber2015}
\bibinfo{author}{\bibfnamefont{K.}~\bibnamefont{Zuber}} \bibnamefont{and}
  \bibinfo{author}{\bibfnamefont{B.}~\bibnamefont{Singh}},
  \bibinfo{journal}{Nuclear Data Sheets} \textbf{\bibinfo{volume}{125}},
  \bibinfo{pages}{1 } (\bibinfo{year}{2015}), ISSN \bibinfo{issn}{0090-3752},
  \urlprefix\url{https://www.sciencedirect.com/science/article/pii/S0090375215000022}.

\bibitem[{\citenamefont{Nichols et~al.}(2012)\citenamefont{Nichols, Singh, and
  Tuli}}]{Nichols2012}
\bibinfo{author}{\bibfnamefont{A.~L.} \bibnamefont{Nichols}},
  \bibinfo{author}{\bibfnamefont{B.}~\bibnamefont{Singh}}, \bibnamefont{and}
  \bibinfo{author}{\bibfnamefont{J.~K.} \bibnamefont{Tuli}},
  \bibinfo{journal}{Nuclear Data Sheets} \textbf{\bibinfo{volume}{113}},
  \bibinfo{pages}{973 } (\bibinfo{year}{2012}), ISSN \bibinfo{issn}{0090-3752},
  \urlprefix\url{https://www.sciencedirect.com/science/article/pii/S0090375212000312}.

\bibitem[{\citenamefont{Erjun and Junde}(2001)}]{Erjun2001}
\bibinfo{author}{\bibfnamefont{B.}~\bibnamefont{Erjun}} \bibnamefont{and}
  \bibinfo{author}{\bibfnamefont{H.}~\bibnamefont{Junde}},
  \bibinfo{journal}{Nuclear Data Sheets} \textbf{\bibinfo{volume}{92}},
  \bibinfo{pages}{147 } (\bibinfo{year}{2001}), ISSN \bibinfo{issn}{0090-3752},
  \urlprefix\url{https://www.sciencedirect.com/science/article/pii/S009037520190002X}.

\bibitem[{\citenamefont{Huang et~al.}(2021)\citenamefont{Huang, Wang, Kondev,
  Audi, and Naimi}}]{Huang2021}
\bibinfo{author}{\bibfnamefont{W.}~\bibnamefont{Huang}},
  \bibinfo{author}{\bibfnamefont{M.}~\bibnamefont{Wang}},
  \bibinfo{author}{\bibfnamefont{F.}~\bibnamefont{Kondev}},
  \bibinfo{author}{\bibfnamefont{G.}~\bibnamefont{Audi}}, \bibnamefont{and}
  \bibinfo{author}{\bibfnamefont{S.}~\bibnamefont{Naimi}},
  \bibinfo{journal}{Chinese Physics C} \textbf{\bibinfo{volume}{45}},
  \bibinfo{pages}{030002} (\bibinfo{year}{2021}),
  \urlprefix\url{https://doi.org/10.1088/1674-1137/abddb0}.

\bibitem[{\citenamefont{Gu\'enaut et~al.}(2007)\citenamefont{Gu\'enaut, Audi,
  Beck, Blaum, Bollen, Delahaye, Herfurth, Kellerbauer, Kluge, Libert
  et~al.}}]{Guenaut2007}
\bibinfo{author}{\bibfnamefont{C.}~\bibnamefont{Gu\'enaut}},
  \bibinfo{author}{\bibfnamefont{G.}~\bibnamefont{Audi}},
  \bibinfo{author}{\bibfnamefont{D.}~\bibnamefont{Beck}},
  \bibinfo{author}{\bibfnamefont{K.}~\bibnamefont{Blaum}},
  \bibinfo{author}{\bibfnamefont{G.}~\bibnamefont{Bollen}},
  \bibinfo{author}{\bibfnamefont{P.}~\bibnamefont{Delahaye}},
  \bibinfo{author}{\bibfnamefont{F.}~\bibnamefont{Herfurth}},
  \bibinfo{author}{\bibfnamefont{A.}~\bibnamefont{Kellerbauer}},
  \bibinfo{author}{\bibfnamefont{H.-J.} \bibnamefont{Kluge}},
  \bibinfo{author}{\bibfnamefont{J.}~\bibnamefont{Libert}},
  \bibnamefont{et~al.}, \bibinfo{journal}{Phys. Rev. C}
  \textbf{\bibinfo{volume}{75}}, \bibinfo{pages}{044303}
  (\bibinfo{year}{2007}),
  \urlprefix\url{https://link.aps.org/doi/10.1103/PhysRevC.75.044303}.

\bibitem[{\citenamefont{Davids et~al.}(1979)\citenamefont{Davids, Gagliardi,
  Murphy, and Norman}}]{Davids1979}
\bibinfo{author}{\bibfnamefont{C.~N.} \bibnamefont{Davids}},
  \bibinfo{author}{\bibfnamefont{C.~A.} \bibnamefont{Gagliardi}},
  \bibinfo{author}{\bibfnamefont{M.~J.} \bibnamefont{Murphy}},
  \bibnamefont{and} \bibinfo{author}{\bibfnamefont{E.~B.}
  \bibnamefont{Norman}}, \bibinfo{journal}{Phys. Rev. C}
  \textbf{\bibinfo{volume}{19}}, \bibinfo{pages}{1463} (\bibinfo{year}{1979}),
  \urlprefix\url{https://link.aps.org/doi/10.1103/PhysRevC.19.1463}.

\bibitem[{\citenamefont{Hyland et~al.}(2006)\citenamefont{Hyland, Svensson,
  Ball, Leslie, Achtzehn, Albers, Andreoiu, Bricault, Churchman, Cross
  et~al.}}]{Hyland2006}
\bibinfo{author}{\bibfnamefont{B.}~\bibnamefont{Hyland}},
  \bibinfo{author}{\bibfnamefont{C.~E.} \bibnamefont{Svensson}},
  \bibinfo{author}{\bibfnamefont{G.~C.} \bibnamefont{Ball}},
  \bibinfo{author}{\bibfnamefont{J.~R.} \bibnamefont{Leslie}},
  \bibinfo{author}{\bibfnamefont{T.}~\bibnamefont{Achtzehn}},
  \bibinfo{author}{\bibfnamefont{D.}~\bibnamefont{Albers}},
  \bibinfo{author}{\bibfnamefont{C.}~\bibnamefont{Andreoiu}},
  \bibinfo{author}{\bibfnamefont{P.}~\bibnamefont{Bricault}},
  \bibinfo{author}{\bibfnamefont{R.}~\bibnamefont{Churchman}},
  \bibinfo{author}{\bibfnamefont{D.}~\bibnamefont{Cross}},
  \bibnamefont{et~al.}, \bibinfo{journal}{Phys. Rev. Lett.}
  \textbf{\bibinfo{volume}{97}}, \bibinfo{pages}{102501}
  (\bibinfo{year}{2006}),
  \urlprefix\url{https://link.aps.org/doi/10.1103/PhysRevLett.97.102501}.

\bibitem[{\citenamefont{Finlay et~al.}(2008)\citenamefont{Finlay, Ball, Leslie,
  Svensson, Towner, Austin, Bandyopadhyay, Chaffey, Chakrawarthy, Garrett
  et~al.}}]{Finlay2008}
\bibinfo{author}{\bibfnamefont{P.}~\bibnamefont{Finlay}},
  \bibinfo{author}{\bibfnamefont{G.~C.} \bibnamefont{Ball}},
  \bibinfo{author}{\bibfnamefont{J.~R.} \bibnamefont{Leslie}},
  \bibinfo{author}{\bibfnamefont{C.~E.} \bibnamefont{Svensson}},
  \bibinfo{author}{\bibfnamefont{I.~S.} \bibnamefont{Towner}},
  \bibinfo{author}{\bibfnamefont{R.~A.~E.} \bibnamefont{Austin}},
  \bibinfo{author}{\bibfnamefont{D.}~\bibnamefont{Bandyopadhyay}},
  \bibinfo{author}{\bibfnamefont{A.}~\bibnamefont{Chaffey}},
  \bibinfo{author}{\bibfnamefont{R.~S.} \bibnamefont{Chakrawarthy}},
  \bibinfo{author}{\bibfnamefont{P.~E.} \bibnamefont{Garrett}},
  \bibnamefont{et~al.}, \bibinfo{journal}{Phys. Rev. C}
  \textbf{\bibinfo{volume}{78}}, \bibinfo{pages}{025502}
  (\bibinfo{year}{2008}),
  \urlprefix\url{https://link.aps.org/doi/10.1103/PhysRevC.78.025502}.

\bibitem[{\citenamefont{MacLean et~al.}(2020)\citenamefont{MacLean, Laffoley,
  Svensson, Ball, Leslie, Andreoiu, Babu, Bhattacharjee, Bidaman, Bildstein
  et~al.}}]{MacLean2020}
\bibinfo{author}{\bibfnamefont{A.~D.} \bibnamefont{MacLean}},
  \bibinfo{author}{\bibfnamefont{A.~T.} \bibnamefont{Laffoley}},
  \bibinfo{author}{\bibfnamefont{C.~E.} \bibnamefont{Svensson}},
  \bibinfo{author}{\bibfnamefont{G.~C.} \bibnamefont{Ball}},
  \bibinfo{author}{\bibfnamefont{J.~R.} \bibnamefont{Leslie}},
  \bibinfo{author}{\bibfnamefont{C.}~\bibnamefont{Andreoiu}},
  \bibinfo{author}{\bibfnamefont{A.}~\bibnamefont{Babu}},
  \bibinfo{author}{\bibfnamefont{S.~S.} \bibnamefont{Bhattacharjee}},
  \bibinfo{author}{\bibfnamefont{H.}~\bibnamefont{Bidaman}},
  \bibinfo{author}{\bibfnamefont{V.}~\bibnamefont{Bildstein}},
  \bibnamefont{et~al.}, \bibinfo{journal}{Phys. Rev. C}
  \textbf{\bibinfo{volume}{102}}, \bibinfo{pages}{054325}
  (\bibinfo{year}{2020}),
  \urlprefix\url{https://link.aps.org/doi/10.1103/PhysRevC.102.054325}.

\bibitem[{\citenamefont{Eronen et~al.}(2006)\citenamefont{Eronen, Elomaa,
  Hager, Hakala, Jokinen, Kankainen, Moore, Penttilä, Rahaman, Rinta-Antila
  et~al.}}]{Eronen2006}
\bibinfo{author}{\bibfnamefont{T.}~\bibnamefont{Eronen}},
  \bibinfo{author}{\bibfnamefont{V.}~\bibnamefont{Elomaa}},
  \bibinfo{author}{\bibfnamefont{U.}~\bibnamefont{Hager}},
  \bibinfo{author}{\bibfnamefont{J.}~\bibnamefont{Hakala}},
  \bibinfo{author}{\bibfnamefont{A.}~\bibnamefont{Jokinen}},
  \bibinfo{author}{\bibfnamefont{A.}~\bibnamefont{Kankainen}},
  \bibinfo{author}{\bibfnamefont{I.}~\bibnamefont{Moore}},
  \bibinfo{author}{\bibfnamefont{H.}~\bibnamefont{Penttilä}},
  \bibinfo{author}{\bibfnamefont{S.}~\bibnamefont{Rahaman}},
  \bibinfo{author}{\bibfnamefont{S.}~\bibnamefont{Rinta-Antila}},
  \bibnamefont{et~al.}, \bibinfo{journal}{Physics Letters B}
  \textbf{\bibinfo{volume}{636}}, \bibinfo{pages}{191 } (\bibinfo{year}{2006}),
  ISSN \bibinfo{issn}{0370-2693},
  \urlprefix\url{https://www.sciencedirect.com/science/article/pii/S0370269306004023}.

\bibitem[{\citenamefont{Weissman et~al.}(2002)\citenamefont{Weissman,
  Cederkall, \"Ayst\"o, Fynbo, Fraile, Fedosseyev, Franchoo, Jokinen, K\"oster,
  Mart\'{\i}nez-Pinedo et~al.}}]{Weissman2002}
\bibinfo{author}{\bibfnamefont{L.}~\bibnamefont{Weissman}},
  \bibinfo{author}{\bibfnamefont{J.}~\bibnamefont{Cederkall}},
  \bibinfo{author}{\bibfnamefont{J.}~\bibnamefont{\"Ayst\"o}},
  \bibinfo{author}{\bibfnamefont{H.}~\bibnamefont{Fynbo}},
  \bibinfo{author}{\bibfnamefont{L.}~\bibnamefont{Fraile}},
  \bibinfo{author}{\bibfnamefont{V.}~\bibnamefont{Fedosseyev}},
  \bibinfo{author}{\bibfnamefont{S.}~\bibnamefont{Franchoo}},
  \bibinfo{author}{\bibfnamefont{A.}~\bibnamefont{Jokinen}},
  \bibinfo{author}{\bibfnamefont{U.}~\bibnamefont{K\"oster}},
  \bibinfo{author}{\bibfnamefont{G.}~\bibnamefont{Mart\'{\i}nez-Pinedo}},
  \bibnamefont{et~al.}, \bibinfo{journal}{Phys. Rev. C}
  \textbf{\bibinfo{volume}{65}}, \bibinfo{pages}{044321}
  (\bibinfo{year}{2002}),
  \urlprefix\url{https://link.aps.org/doi/10.1103/PhysRevC.65.044321}.

\bibitem[{\citenamefont{Tu et~al.}(2011{\natexlab{b}})\citenamefont{Tu, Wang,
  Litvinov, Zhang, Xu, Sun, Audi, Blaum, Du, Huang et~al.}}]{Tu2011}
\bibinfo{author}{\bibfnamefont{X.}~\bibnamefont{Tu}},
  \bibinfo{author}{\bibfnamefont{M.}~\bibnamefont{Wang}},
  \bibinfo{author}{\bibfnamefont{Y.}~\bibnamefont{Litvinov}},
  \bibinfo{author}{\bibfnamefont{Y.}~\bibnamefont{Zhang}},
  \bibinfo{author}{\bibfnamefont{H.}~\bibnamefont{Xu}},
  \bibinfo{author}{\bibfnamefont{Z.}~\bibnamefont{Sun}},
  \bibinfo{author}{\bibfnamefont{G.}~\bibnamefont{Audi}},
  \bibinfo{author}{\bibfnamefont{K.}~\bibnamefont{Blaum}},
  \bibinfo{author}{\bibfnamefont{C.}~\bibnamefont{Du}},
  \bibinfo{author}{\bibfnamefont{W.}~\bibnamefont{Huang}},
  \bibnamefont{et~al.}, \bibinfo{journal}{Nuclear Instruments and Methods in
  Physics Research Section A: Accelerators, Spectrometers, Detectors and
  Associated Equipment} \textbf{\bibinfo{volume}{654}}, \bibinfo{pages}{213 }
  (\bibinfo{year}{2011}{\natexlab{b}}), ISSN \bibinfo{issn}{0168-9002},
  \urlprefix\url{https://www.sciencedirect.com/science/article/pii/S0168900211014471}.

\bibitem[{\citenamefont{Ormand}(1997)}]{Ormand1997}
\bibinfo{author}{\bibfnamefont{W.~E.} \bibnamefont{Ormand}},
  \bibinfo{journal}{Phys. Rev. C} \textbf{\bibinfo{volume}{55}},
  \bibinfo{pages}{2407 } (\bibinfo{year}{1997}),
  \urlprefix\url{https://link.aps.org/doi/10.1103/PhysRevC.55.2407}.

\bibitem[{\citenamefont{Cole}(1999)}]{Cole1999}
\bibinfo{author}{\bibfnamefont{B.~J.} \bibnamefont{Cole}},
  \bibinfo{journal}{Phys. Rev. C} \textbf{\bibinfo{volume}{59}},
  \bibinfo{pages}{726 } (\bibinfo{year}{1999}),
  \urlprefix\url{https://link.aps.org/doi/10.1103/PhysRevC.59.726}.

\bibitem[{\citenamefont{Antony et~al.}(1997)\citenamefont{Antony, Pape, and
  Britz}}]{Antony1997}
\bibinfo{author}{\bibfnamefont{M.}~\bibnamefont{Antony}},
  \bibinfo{author}{\bibfnamefont{A.}~\bibnamefont{Pape}}, \bibnamefont{and}
  \bibinfo{author}{\bibfnamefont{J.}~\bibnamefont{Britz}},
  \bibinfo{journal}{Atomic Data and Nuclear Data Tables}
  \textbf{\bibinfo{volume}{66}}, \bibinfo{pages}{1} (\bibinfo{year}{1997}),
  ISSN \bibinfo{issn}{0092-640X},
  \urlprefix\url{https://www.sciencedirect.com/science/article/pii/S0092640X97907403}.

\bibitem[{\citenamefont{Audi et~al.}(2003)\citenamefont{Audi, Wapstra, and
  Thibault}}]{Audi2003}
\bibinfo{author}{\bibfnamefont{G.}~\bibnamefont{Audi}},
  \bibinfo{author}{\bibfnamefont{A.~H.} \bibnamefont{Wapstra}},
  \bibnamefont{and} \bibinfo{author}{\bibfnamefont{C.}~\bibnamefont{Thibault}},
  \bibinfo{journal}{Nuclear Physics A} \textbf{\bibinfo{volume}{729}},
  \bibinfo{pages}{337} (\bibinfo{year}{2003}), ISSN \bibinfo{issn}{0375-9474},
  \bibinfo{note}{the 2003 NUBASE and Atomic Mass Evaluations},
  \urlprefix\url{https://www.sciencedirect.com/science/article/pii/S0375947403018098}.

\bibitem[{\citenamefont{{Del Santo} et~al.}(2014)\citenamefont{{Del Santo},
  Meisel, Bazin, Becerril, Brown, Crawford, Cyburt, George, Grinyer, Lorusso
  et~al.}}]{DelSanto2014}
\bibinfo{author}{\bibfnamefont{M.}~\bibnamefont{{Del Santo}}},
  \bibinfo{author}{\bibfnamefont{Z.}~\bibnamefont{Meisel}},
  \bibinfo{author}{\bibfnamefont{D.}~\bibnamefont{Bazin}},
  \bibinfo{author}{\bibfnamefont{A.}~\bibnamefont{Becerril}},
  \bibinfo{author}{\bibfnamefont{B.}~\bibnamefont{Brown}},
  \bibinfo{author}{\bibfnamefont{H.}~\bibnamefont{Crawford}},
  \bibinfo{author}{\bibfnamefont{R.}~\bibnamefont{Cyburt}},
  \bibinfo{author}{\bibfnamefont{S.}~\bibnamefont{George}},
  \bibinfo{author}{\bibfnamefont{G.}~\bibnamefont{Grinyer}},
  \bibinfo{author}{\bibfnamefont{G.}~\bibnamefont{Lorusso}},
  \bibnamefont{et~al.}, \bibinfo{journal}{Physics Letters B}
  \textbf{\bibinfo{volume}{738}}, \bibinfo{pages}{453 } (\bibinfo{year}{2014}),
  ISSN \bibinfo{issn}{0370-2693},
  \urlprefix\url{https://www.sciencedirect.com/science/article/pii/S0370269314007461}.

\bibitem[{\citenamefont{Orrigo et~al.}(2014)\citenamefont{Orrigo, Rubio,
  Fujita, Blank, Gelletly, Agramunt, Algora, Ascher, Bilgier, C\'aceres
  et~al.}}]{Orrigo2014}
\bibinfo{author}{\bibfnamefont{S.~E.~A.} \bibnamefont{Orrigo}},
  \bibinfo{author}{\bibfnamefont{B.}~\bibnamefont{Rubio}},
  \bibinfo{author}{\bibfnamefont{Y.}~\bibnamefont{Fujita}},
  \bibinfo{author}{\bibfnamefont{B.}~\bibnamefont{Blank}},
  \bibinfo{author}{\bibfnamefont{W.}~\bibnamefont{Gelletly}},
  \bibinfo{author}{\bibfnamefont{J.}~\bibnamefont{Agramunt}},
  \bibinfo{author}{\bibfnamefont{A.}~\bibnamefont{Algora}},
  \bibinfo{author}{\bibfnamefont{P.}~\bibnamefont{Ascher}},
  \bibinfo{author}{\bibfnamefont{B.}~\bibnamefont{Bilgier}},
  \bibinfo{author}{\bibfnamefont{L.}~\bibnamefont{C\'aceres}},
  \bibnamefont{et~al.}, \bibinfo{journal}{Phys. Rev. Lett.}
  \textbf{\bibinfo{volume}{112}}, \bibinfo{pages}{222501}
  (\bibinfo{year}{2014}),
  \urlprefix\url{https://link.aps.org/doi/10.1103/PhysRevLett.112.222501}.

\bibitem[{\citenamefont{Orrigo et~al.}(2016)\citenamefont{Orrigo, Rubio,
  Fujita, Gelletly, Agramunt, Algora, Ascher, Bilgier, Blank, C\'aceres
  et~al.}}]{Orrigo2016}
\bibinfo{author}{\bibfnamefont{S.~E.~A.} \bibnamefont{Orrigo}},
  \bibinfo{author}{\bibfnamefont{B.}~\bibnamefont{Rubio}},
  \bibinfo{author}{\bibfnamefont{Y.}~\bibnamefont{Fujita}},
  \bibinfo{author}{\bibfnamefont{W.}~\bibnamefont{Gelletly}},
  \bibinfo{author}{\bibfnamefont{J.}~\bibnamefont{Agramunt}},
  \bibinfo{author}{\bibfnamefont{A.}~\bibnamefont{Algora}},
  \bibinfo{author}{\bibfnamefont{P.}~\bibnamefont{Ascher}},
  \bibinfo{author}{\bibfnamefont{B.}~\bibnamefont{Bilgier}},
  \bibinfo{author}{\bibfnamefont{B.}~\bibnamefont{Blank}},
  \bibinfo{author}{\bibfnamefont{L.}~\bibnamefont{C\'aceres}},
  \bibnamefont{et~al.}, \bibinfo{journal}{Phys. Rev. C}
  \textbf{\bibinfo{volume}{93}}, \bibinfo{pages}{044336}
  (\bibinfo{year}{2016}),
  \urlprefix\url{https://link.aps.org/doi/10.1103/PhysRevC.93.044336}.

\bibitem[{\citenamefont{Wang et~al.}(2017)\citenamefont{Wang, Audi, Kondev,
  Huang, Naimi, and Xu}}]{Wang2017}
\bibinfo{author}{\bibfnamefont{M.}~\bibnamefont{Wang}},
  \bibinfo{author}{\bibfnamefont{G.}~\bibnamefont{Audi}},
  \bibinfo{author}{\bibfnamefont{F.}~\bibnamefont{Kondev}},
  \bibinfo{author}{\bibfnamefont{W.}~\bibnamefont{Huang}},
  \bibinfo{author}{\bibfnamefont{S.}~\bibnamefont{Naimi}}, \bibnamefont{and}
  \bibinfo{author}{\bibfnamefont{X.}~\bibnamefont{Xu}},
  \bibinfo{journal}{Chinese Physics C} \textbf{\bibinfo{volume}{41}},
  \bibinfo{pages}{030003} (\bibinfo{year}{2017}),
  \urlprefix\url{http://stacks.iop.org/1674-1137/41/i=3/a=030003}.

\bibitem[{\citenamefont{Rauth et~al.}(2008)\citenamefont{Rauth, Ackermann,
  Blaum, Block, Chaudhuri, Di, Eliseev, Ferrer, Habs, Herfurth
  et~al.}}]{Rauth2008}
\bibinfo{author}{\bibfnamefont{C.}~\bibnamefont{Rauth}},
  \bibinfo{author}{\bibfnamefont{D.}~\bibnamefont{Ackermann}},
  \bibinfo{author}{\bibfnamefont{K.}~\bibnamefont{Blaum}},
  \bibinfo{author}{\bibfnamefont{M.}~\bibnamefont{Block}},
  \bibinfo{author}{\bibfnamefont{A.}~\bibnamefont{Chaudhuri}},
  \bibinfo{author}{\bibfnamefont{Z.}~\bibnamefont{Di}},
  \bibinfo{author}{\bibfnamefont{S.}~\bibnamefont{Eliseev}},
  \bibinfo{author}{\bibfnamefont{R.}~\bibnamefont{Ferrer}},
  \bibinfo{author}{\bibfnamefont{D.}~\bibnamefont{Habs}},
  \bibinfo{author}{\bibfnamefont{F.}~\bibnamefont{Herfurth}},
  \bibnamefont{et~al.}, \bibinfo{journal}{Phys. Rev. Lett.}
  \textbf{\bibinfo{volume}{100}}, \bibinfo{pages}{012501}
  (\bibinfo{year}{2008}),
  \urlprefix\url{https://link.aps.org/doi/10.1103/PhysRevLett.100.012501}.

\bibitem[{\citenamefont{Thoennessen}(2004)}]{Thoennessen2004}
\bibinfo{author}{\bibfnamefont{M.}~\bibnamefont{Thoennessen}},
  \bibinfo{journal}{Reports on Progress in Physics}
  \textbf{\bibinfo{volume}{67}}, \bibinfo{pages}{1187 } (\bibinfo{year}{2004}),
  \urlprefix\url{https://doi.org/10.1088/0034-4885/67/7/r04}.

\bibitem[{\citenamefont{M\"oller et~al.}(2012)\citenamefont{M\"oller, Myers,
  Sagawa, and Yoshida}}]{Moller2012}
\bibinfo{author}{\bibfnamefont{P.}~\bibnamefont{M\"oller}},
  \bibinfo{author}{\bibfnamefont{W.~D.} \bibnamefont{Myers}},
  \bibinfo{author}{\bibfnamefont{H.}~\bibnamefont{Sagawa}}, \bibnamefont{and}
  \bibinfo{author}{\bibfnamefont{S.}~\bibnamefont{Yoshida}},
  \bibinfo{journal}{Phys. Rev. Lett.} \textbf{\bibinfo{volume}{108}},
  \bibinfo{pages}{052501} (\bibinfo{year}{2012}),
  \urlprefix\url{https://link.aps.org/doi/10.1103/PhysRevLett.108.052501}.

\bibitem[{\citenamefont{Goriely et~al.}(2016)\citenamefont{Goriely, Chamel, and
  Pearson}}]{Goriely2016}
\bibinfo{author}{\bibfnamefont{S.}~\bibnamefont{Goriely}},
  \bibinfo{author}{\bibfnamefont{N.}~\bibnamefont{Chamel}}, \bibnamefont{and}
  \bibinfo{author}{\bibfnamefont{J.~M.} \bibnamefont{Pearson}},
  \bibinfo{journal}{Phys. Rev. C} \textbf{\bibinfo{volume}{93}},
  \bibinfo{pages}{034337} (\bibinfo{year}{2016}),
  \urlprefix\url{https://link.aps.org/doi/10.1103/PhysRevC.93.034337}.

\bibitem[{\citenamefont{Bartel et~al.}(1982)\citenamefont{Bartel, Quentin,
  Brack, Guet, and Håkansson}}]{Bartel1982}
\bibinfo{author}{\bibfnamefont{J.}~\bibnamefont{Bartel}},
  \bibinfo{author}{\bibfnamefont{P.}~\bibnamefont{Quentin}},
  \bibinfo{author}{\bibfnamefont{M.}~\bibnamefont{Brack}},
  \bibinfo{author}{\bibfnamefont{C.}~\bibnamefont{Guet}}, \bibnamefont{and}
  \bibinfo{author}{\bibfnamefont{H.-B.} \bibnamefont{Håkansson}},
  \bibinfo{journal}{Nuclear Physics A} \textbf{\bibinfo{volume}{386}},
  \bibinfo{pages}{79 } (\bibinfo{year}{1982}), ISSN \bibinfo{issn}{0375-9474},
  \urlprefix\url{https://www.sciencedirect.com/science/article/pii/0375947482904031}.

\bibitem[{\citenamefont{Wang et~al.}(2014)\citenamefont{Wang, Liu, Wu, and
  Meng}}]{Wang2014}
\bibinfo{author}{\bibfnamefont{N.}~\bibnamefont{Wang}},
  \bibinfo{author}{\bibfnamefont{M.}~\bibnamefont{Liu}},
  \bibinfo{author}{\bibfnamefont{X.}~\bibnamefont{Wu}}, \bibnamefont{and}
  \bibinfo{author}{\bibfnamefont{J.}~\bibnamefont{Meng}},
  \bibinfo{journal}{Physics Letters B} \textbf{\bibinfo{volume}{734}},
  \bibinfo{pages}{215} (\bibinfo{year}{2014}), ISSN \bibinfo{issn}{0370-2693},
  \urlprefix\url{https://www.sciencedirect.com/science/article/pii/S037026931400358X}.

\bibitem[{\citenamefont{Niu et~al.}(2018)\citenamefont{Niu, Liang, Sun, Niu,
  Guo, and Meng}}]{Niu2018}
\bibinfo{author}{\bibfnamefont{Z.}~\bibnamefont{Niu}},
  \bibinfo{author}{\bibfnamefont{H.}~\bibnamefont{Liang}},
  \bibinfo{author}{\bibfnamefont{B.}~\bibnamefont{Sun}},
  \bibinfo{author}{\bibfnamefont{Y.}~\bibnamefont{Niu}},
  \bibinfo{author}{\bibfnamefont{J.}~\bibnamefont{Guo}}, \bibnamefont{and}
  \bibinfo{author}{\bibfnamefont{J.}~\bibnamefont{Meng}},
  \bibinfo{journal}{Science Bulletin} \textbf{\bibinfo{volume}{63}},
  \bibinfo{pages}{759 } (\bibinfo{year}{2018}), ISSN \bibinfo{issn}{2095-9273},
  \urlprefix\url{https://www.sciencedirect.com/science/article/pii/S2095927318302135}.

\bibitem[{\citenamefont{Neufcourt
  et~al.}(2020{\natexlab{a}})\citenamefont{Neufcourt, Cao, Giuliani,
  Nazarewicz, Olsen, and Tarasov}}]{Neufcourt2020}
\bibinfo{author}{\bibfnamefont{L.}~\bibnamefont{Neufcourt}},
  \bibinfo{author}{\bibfnamefont{Y.}~\bibnamefont{Cao}},
  \bibinfo{author}{\bibfnamefont{S.~A.} \bibnamefont{Giuliani}},
  \bibinfo{author}{\bibfnamefont{W.}~\bibnamefont{Nazarewicz}},
  \bibinfo{author}{\bibfnamefont{E.}~\bibnamefont{Olsen}}, \bibnamefont{and}
  \bibinfo{author}{\bibfnamefont{O.~B.} \bibnamefont{Tarasov}},
  \bibinfo{journal}{Phys. Rev. C} \textbf{\bibinfo{volume}{101}},
  \bibinfo{pages}{044307} (\bibinfo{year}{2020}{\natexlab{a}}),
  \urlprefix\url{https://link.aps.org/doi/10.1103/PhysRevC.101.044307}.

\bibitem[{\citenamefont{Shelley and Pastore}(2021)}]{Shelley2021}
\bibinfo{author}{\bibfnamefont{M.}~\bibnamefont{Shelley}} \bibnamefont{and}
  \bibinfo{author}{\bibfnamefont{A.}~\bibnamefont{Pastore}},
  \bibinfo{journal}{Universe} \textbf{\bibinfo{volume}{7}}
  (\bibinfo{year}{2021}), ISSN \bibinfo{issn}{2218-1997},
  \urlprefix\url{https://www.mdpi.com/2218-1997/7/5/131}.

\bibitem[{\citenamefont{Neufcourt
  et~al.}(2020{\natexlab{b}})\citenamefont{Neufcourt, Cao, Giuliani,
  Nazarewicz, Olsen, and Tarasov}}]{Neufcourt2020b}
\bibinfo{author}{\bibfnamefont{L.}~\bibnamefont{Neufcourt}},
  \bibinfo{author}{\bibfnamefont{Y.}~\bibnamefont{Cao}},
  \bibinfo{author}{\bibfnamefont{S.}~\bibnamefont{Giuliani}},
  \bibinfo{author}{\bibfnamefont{W.}~\bibnamefont{Nazarewicz}},
  \bibinfo{author}{\bibfnamefont{E.}~\bibnamefont{Olsen}}, \bibnamefont{and}
  \bibinfo{author}{\bibfnamefont{O.~B.} \bibnamefont{Tarasov}},
  \bibinfo{journal}{Phys. Rev. C} \textbf{\bibinfo{volume}{101}},
  \bibinfo{pages}{014319} (\bibinfo{year}{2020}{\natexlab{b}}),
  \urlprefix\url{https://link.aps.org/doi/10.1103/PhysRevC.101.014319}.

\bibitem[{\citenamefont{Moller et~al.}(1995)\citenamefont{Moller, Nix, Myers,
  and Swiatecki}}]{Moller1995}
\bibinfo{author}{\bibfnamefont{P.}~\bibnamefont{Moller}},
  \bibinfo{author}{\bibfnamefont{J.}~\bibnamefont{Nix}},
  \bibinfo{author}{\bibfnamefont{W.}~\bibnamefont{Myers}}, \bibnamefont{and}
  \bibinfo{author}{\bibfnamefont{W.}~\bibnamefont{Swiatecki}},
  \bibinfo{journal}{Atomic Data and Nuclear Data Tables}
  \textbf{\bibinfo{volume}{59}}, \bibinfo{pages}{185 } (\bibinfo{year}{1995}),
  ISSN \bibinfo{issn}{0092-640X},
  \urlprefix\url{https://www.sciencedirect.com/science/article/pii/S0092640X85710029}.

\bibitem[{\citenamefont{Rodr{\i}guez et~al.}(2005)\citenamefont{Rodr{\i}guez,
  Kolhinen, Audi, {\"A}yst{\"o}, Beck, Blaum, Bollen, Herfurth, Jokinen,
  Kellerbauer et~al.}}]{Rodriguez2005}
\bibinfo{author}{\bibfnamefont{D.}~\bibnamefont{Rodr{\i}guez}},
  \bibinfo{author}{\bibfnamefont{V.}~\bibnamefont{Kolhinen}},
  \bibinfo{author}{\bibfnamefont{G.}~\bibnamefont{Audi}},
  \bibinfo{author}{\bibfnamefont{J.}~\bibnamefont{{\"A}yst{\"o}}},
  \bibinfo{author}{\bibfnamefont{D.}~\bibnamefont{Beck}},
  \bibinfo{author}{\bibfnamefont{K.}~\bibnamefont{Blaum}},
  \bibinfo{author}{\bibfnamefont{G.}~\bibnamefont{Bollen}},
  \bibinfo{author}{\bibfnamefont{F.}~\bibnamefont{Herfurth}},
  \bibinfo{author}{\bibfnamefont{A.}~\bibnamefont{Jokinen}},
  \bibinfo{author}{\bibfnamefont{A.}~\bibnamefont{Kellerbauer}},
  \bibnamefont{et~al.}, \bibinfo{journal}{The European Physical Journal
  A-Hadrons and Nuclei} \textbf{\bibinfo{volume}{25}}, \bibinfo{pages}{41 }
  (\bibinfo{year}{2005}),
  \urlprefix\url{https://doi.org/10.1140/epjad/i2005-06-164-3}.

\bibitem[{\citenamefont{Stolz et~al.}(2005{\natexlab{a}})\citenamefont{Stolz,
  Baumann, Frank, Ginter, Hitt, Kwan, Mocko, Peters, Schiller, Sumithrarachchi
  et~al.}}]{Stolz2005}
\bibinfo{author}{\bibfnamefont{A.}~\bibnamefont{Stolz}},
  \bibinfo{author}{\bibfnamefont{T.}~\bibnamefont{Baumann}},
  \bibinfo{author}{\bibfnamefont{N.}~\bibnamefont{Frank}},
  \bibinfo{author}{\bibfnamefont{T.}~\bibnamefont{Ginter}},
  \bibinfo{author}{\bibfnamefont{G.}~\bibnamefont{Hitt}},
  \bibinfo{author}{\bibfnamefont{E.}~\bibnamefont{Kwan}},
  \bibinfo{author}{\bibfnamefont{M.}~\bibnamefont{Mocko}},
  \bibinfo{author}{\bibfnamefont{W.}~\bibnamefont{Peters}},
  \bibinfo{author}{\bibfnamefont{A.}~\bibnamefont{Schiller}},
  \bibinfo{author}{\bibfnamefont{C.}~\bibnamefont{Sumithrarachchi}},
  \bibnamefont{et~al.}, \bibinfo{journal}{Physics Letters B}
  \textbf{\bibinfo{volume}{627}}, \bibinfo{pages}{32 }
  (\bibinfo{year}{2005}{\natexlab{a}}), ISSN \bibinfo{issn}{0370-2693},
  \urlprefix\url{https://www.sciencedirect.com/science/article/pii/S0370269305012748}.

\bibitem[{\citenamefont{Stolz et~al.}(2005{\natexlab{b}})\citenamefont{Stolz,
  Baumann, Frank, Ginter, Hitt, Kwan, Mocko, Peters, Schiller, Sumithrarachchi
  et~al.}}]{Stolz2005b}
\bibinfo{author}{\bibfnamefont{A.}~\bibnamefont{Stolz}},
  \bibinfo{author}{\bibfnamefont{T.}~\bibnamefont{Baumann}},
  \bibinfo{author}{\bibfnamefont{N.}~\bibnamefont{Frank}},
  \bibinfo{author}{\bibfnamefont{T.}~\bibnamefont{Ginter}},
  \bibinfo{author}{\bibfnamefont{G.}~\bibnamefont{Hitt}},
  \bibinfo{author}{\bibfnamefont{E.}~\bibnamefont{Kwan}},
  \bibinfo{author}{\bibfnamefont{M.}~\bibnamefont{Mocko}},
  \bibinfo{author}{\bibfnamefont{W.}~\bibnamefont{Peters}},
  \bibinfo{author}{\bibfnamefont{A.}~\bibnamefont{Schiller}},
  \bibinfo{author}{\bibfnamefont{C.}~\bibnamefont{Sumithrarachchi}},
  \bibnamefont{et~al.}, in \emph{\bibinfo{booktitle}{The 4th International
  Conference on Exotic Nuclei and Atomic Masses}}
  (\bibinfo{organization}{Springer}, \bibinfo{year}{2005}{\natexlab{b}}), pp.
  \bibinfo{pages}{335 -- 338},
  \urlprefix\url{https://doi.org/10.1007/3-540-37642-9_93}.

\bibitem[{\citenamefont{Wang et~al.}(2012)\citenamefont{Wang, Audi, Wapstra,
  Kondev, MacCormick, Xu, and Pfeiffer}}]{Wang2012}
\bibinfo{author}{\bibfnamefont{M.}~\bibnamefont{Wang}},
  \bibinfo{author}{\bibfnamefont{G.}~\bibnamefont{Audi}},
  \bibinfo{author}{\bibfnamefont{A.}~\bibnamefont{Wapstra}},
  \bibinfo{author}{\bibfnamefont{F.}~\bibnamefont{Kondev}},
  \bibinfo{author}{\bibfnamefont{M.}~\bibnamefont{MacCormick}},
  \bibinfo{author}{\bibfnamefont{X.}~\bibnamefont{Xu}}, \bibnamefont{and}
  \bibinfo{author}{\bibfnamefont{B.}~\bibnamefont{Pfeiffer}},
  \bibinfo{journal}{Chinese Physics C} \textbf{\bibinfo{volume}{36}},
  \bibinfo{pages}{1603 } (\bibinfo{year}{2012}),
  \urlprefix\url{https://doi.org/10.1088/1674-1137/36/12/003}.

\bibitem[{\citenamefont{Greenfield et~al.}(1972)\citenamefont{Greenfield,
  Bingham, Newman, and Saltmarsh}}]{Greenfield1972}
\bibinfo{author}{\bibfnamefont{M.~B.} \bibnamefont{Greenfield}},
  \bibinfo{author}{\bibfnamefont{C.~R.} \bibnamefont{Bingham}},
  \bibinfo{author}{\bibfnamefont{E.}~\bibnamefont{Newman}}, \bibnamefont{and}
  \bibinfo{author}{\bibfnamefont{M.~J.} \bibnamefont{Saltmarsh}},
  \bibinfo{journal}{Phys. Rev. C} \textbf{\bibinfo{volume}{6}},
  \bibinfo{pages}{1756 } (\bibinfo{year}{1972}),
  \urlprefix\url{https://link.aps.org/doi/10.1103/PhysRevC.6.1756}.

\bibitem[{\citenamefont{Schubank et~al.}(1989)\citenamefont{Schubank, Cameron,
  and Janzen}}]{Schubank1989}
\bibinfo{author}{\bibfnamefont{R.~B.} \bibnamefont{Schubank}},
  \bibinfo{author}{\bibfnamefont{J.~A.} \bibnamefont{Cameron}},
  \bibnamefont{and} \bibinfo{author}{\bibfnamefont{V.~P.}
  \bibnamefont{Janzen}}, \bibinfo{journal}{Phys. Rev. C}
  \textbf{\bibinfo{volume}{40}}, \bibinfo{pages}{2310 } (\bibinfo{year}{1989}),
  \urlprefix\url{https://link.aps.org/doi/10.1103/PhysRevC.40.2310}.

\bibitem[{\citenamefont{Pougheon et~al.}(1972)\citenamefont{Pougheon, Roussel,
  Colombani, Doubre, and Roynette}}]{Pougheon1972}
\bibinfo{author}{\bibfnamefont{F.}~\bibnamefont{Pougheon}},
  \bibinfo{author}{\bibfnamefont{P.}~\bibnamefont{Roussel}},
  \bibinfo{author}{\bibfnamefont{P.}~\bibnamefont{Colombani}},
  \bibinfo{author}{\bibfnamefont{H.}~\bibnamefont{Doubre}}, \bibnamefont{and}
  \bibinfo{author}{\bibfnamefont{J.}~\bibnamefont{Roynette}},
  \bibinfo{journal}{Nuclear Physics A} \textbf{\bibinfo{volume}{193}},
  \bibinfo{pages}{305 } (\bibinfo{year}{1972}), ISSN \bibinfo{issn}{0375-9474},
  \urlprefix\url{https://www.sciencedirect.com/science/article/pii/0375947472902552}.

\bibitem[{\citenamefont{Kondev et~al.}(2021)\citenamefont{Kondev, Wang, Huang,
  Naimi, and Audi}}]{Kondev2021}
\bibinfo{author}{\bibfnamefont{F.}~\bibnamefont{Kondev}},
  \bibinfo{author}{\bibfnamefont{M.}~\bibnamefont{Wang}},
  \bibinfo{author}{\bibfnamefont{W.}~\bibnamefont{Huang}},
  \bibinfo{author}{\bibfnamefont{S.}~\bibnamefont{Naimi}}, \bibnamefont{and}
  \bibinfo{author}{\bibfnamefont{G.}~\bibnamefont{Audi}},
  \bibinfo{journal}{Chinese Physics C} \textbf{\bibinfo{volume}{45}},
  \bibinfo{pages}{030001} (\bibinfo{year}{2021}),
  \urlprefix\url{https://doi.org/10.1088/1674-1137/abddae}.

\bibitem[{\citenamefont{Brown and Rae}(2014)}]{Brown2014}
\bibinfo{author}{\bibfnamefont{B.}~\bibnamefont{Brown}} \bibnamefont{and}
  \bibinfo{author}{\bibfnamefont{W.}~\bibnamefont{Rae}},
  \bibinfo{journal}{Nuclear Data Sheets} \textbf{\bibinfo{volume}{120}},
  \bibinfo{pages}{115 } (\bibinfo{year}{2014}), ISSN \bibinfo{issn}{0090-3752},
  \urlprefix\url{https://www.sciencedirect.com/science/article/pii/S0090375214004748}.

\bibitem[{\citenamefont{Smirnova et~al.}(2017)\citenamefont{Smirnova, Blank,
  Brown, Richter, Benouaret, and Lam}}]{Smirnova2017}
\bibinfo{author}{\bibfnamefont{N.~A.} \bibnamefont{Smirnova}},
  \bibinfo{author}{\bibfnamefont{B.}~\bibnamefont{Blank}},
  \bibinfo{author}{\bibfnamefont{B.~A.} \bibnamefont{Brown}},
  \bibinfo{author}{\bibfnamefont{W.~A.} \bibnamefont{Richter}},
  \bibinfo{author}{\bibfnamefont{N.}~\bibnamefont{Benouaret}},
  \bibnamefont{and} \bibinfo{author}{\bibfnamefont{Y.~H.} \bibnamefont{Lam}},
  \bibinfo{journal}{Phys. Rev. C} \textbf{\bibinfo{volume}{95}},
  \bibinfo{pages}{054301} (\bibinfo{year}{2017}),
  \urlprefix\url{https://link.aps.org/doi/10.1103/PhysRevC.95.054301}.

\bibitem[{\citenamefont{Honma et~al.}(2004)\citenamefont{Honma, Otsuka, Brown,
  and Mizusaki}}]{Honma2004}
\bibinfo{author}{\bibfnamefont{M.}~\bibnamefont{Honma}},
  \bibinfo{author}{\bibfnamefont{T.}~\bibnamefont{Otsuka}},
  \bibinfo{author}{\bibfnamefont{B.~A.} \bibnamefont{Brown}}, \bibnamefont{and}
  \bibinfo{author}{\bibfnamefont{T.}~\bibnamefont{Mizusaki}},
  \bibinfo{journal}{Phys. Rev. C} \textbf{\bibinfo{volume}{69}},
  \bibinfo{pages}{034335} (\bibinfo{year}{2004}),
  \urlprefix\url{https://link.aps.org/doi/10.1103/PhysRevC.69.034335}.

\bibitem[{\citenamefont{Ormand and Brown}(1995)}]{Ormand1995}
\bibinfo{author}{\bibfnamefont{W.~E.} \bibnamefont{Ormand}} \bibnamefont{and}
  \bibinfo{author}{\bibfnamefont{B.~A.} \bibnamefont{Brown}},
  \bibinfo{journal}{Phys. Rev. C} \textbf{\bibinfo{volume}{52}},
  \bibinfo{pages}{2455 } (\bibinfo{year}{1995}),
  \urlprefix\url{https://link.aps.org/doi/10.1103/PhysRevC.52.2455}.

\bibitem[{\citenamefont{J\"anecke}(1966)}]{Jaenecke1966}
\bibinfo{author}{\bibfnamefont{J.}~\bibnamefont{J\"anecke}},
  \bibinfo{journal}{Phys. Rev.} \textbf{\bibinfo{volume}{147}},
  \bibinfo{pages}{735 } (\bibinfo{year}{1966}),
  \urlprefix\url{https://link.aps.org/doi/10.1103/PhysRev.147.735}.

\bibitem[{\citenamefont{Seth et~al.}(1986)\citenamefont{Seth, Iversen, Kaletka,
  Barlow, Saha, and Soundranayagam}}]{Seth1986}
\bibinfo{author}{\bibfnamefont{K.~K.} \bibnamefont{Seth}},
  \bibinfo{author}{\bibfnamefont{S.}~\bibnamefont{Iversen}},
  \bibinfo{author}{\bibfnamefont{M.}~\bibnamefont{Kaletka}},
  \bibinfo{author}{\bibfnamefont{D.}~\bibnamefont{Barlow}},
  \bibinfo{author}{\bibfnamefont{A.}~\bibnamefont{Saha}}, \bibnamefont{and}
  \bibinfo{author}{\bibfnamefont{R.}~\bibnamefont{Soundranayagam}},
  \bibinfo{journal}{Physics Letters B} \textbf{\bibinfo{volume}{173}},
  \bibinfo{pages}{397 } (\bibinfo{year}{1986}), ISSN \bibinfo{issn}{0370-2693},
  \urlprefix\url{https://www.sciencedirect.com/science/article/pii/0370269386904028}.

\bibitem[{\citenamefont{Okuma et~al.}(1985)\citenamefont{Okuma, Motobayashi,
  Takimoto, Shimoura, Fukada, Suehiro, and Yanabu}}]{Okuma1985}
\bibinfo{author}{\bibfnamefont{Y.}~\bibnamefont{Okuma}},
  \bibinfo{author}{\bibfnamefont{T.}~\bibnamefont{Motobayashi}},
  \bibinfo{author}{\bibfnamefont{K.}~\bibnamefont{Takimoto}},
  \bibinfo{author}{\bibfnamefont{S.}~\bibnamefont{Shimoura}},
  \bibinfo{author}{\bibfnamefont{M.}~\bibnamefont{Fukada}},
  \bibinfo{author}{\bibfnamefont{T.}~\bibnamefont{Suehiro}}, \bibnamefont{and}
  \bibinfo{author}{\bibfnamefont{T.}~\bibnamefont{Yanabu}},
  \emph{\bibinfo{title}{Ann. rept. 1984}}, \bibinfo{howpublished}{RCNP (Osaka)}
  (\bibinfo{year}{1985}).

\bibitem[{\citenamefont{Weber et~al.}(1979)\citenamefont{Weber, Crawley,
  Benenson, Kashy, and Nann}}]{Weber1979}
\bibinfo{author}{\bibfnamefont{D.}~\bibnamefont{Weber}},
  \bibinfo{author}{\bibfnamefont{G.}~\bibnamefont{Crawley}},
  \bibinfo{author}{\bibfnamefont{W.}~\bibnamefont{Benenson}},
  \bibinfo{author}{\bibfnamefont{E.}~\bibnamefont{Kashy}}, \bibnamefont{and}
  \bibinfo{author}{\bibfnamefont{H.}~\bibnamefont{Nann}},
  \bibinfo{journal}{Nuclear Physics A} \textbf{\bibinfo{volume}{313}},
  \bibinfo{pages}{385 } (\bibinfo{year}{1979}), ISSN \bibinfo{issn}{0375-9474},
  \urlprefix\url{https://www.sciencedirect.com/science/article/pii/0375947479905086}.

\bibitem[{\citenamefont{Koning and Rochman}(2012)}]{Koning2012}
\bibinfo{author}{\bibfnamefont{A.}~\bibnamefont{Koning}} \bibnamefont{and}
  \bibinfo{author}{\bibfnamefont{D.}~\bibnamefont{Rochman}},
  \bibinfo{journal}{Nuclear Data Sheets} \textbf{\bibinfo{volume}{113}},
  \bibinfo{pages}{2841 } (\bibinfo{year}{2012}), ISSN
  \bibinfo{issn}{0090-3752}, \bibinfo{note}{special Issue on Nuclear Reaction
  Data},
  \urlprefix\url{https://www.sciencedirect.com/science/article/pii/S0090375212000889}.

\bibitem[{\citenamefont{Mardor et~al.}(2021)\citenamefont{Mardor, Andr\'es,
  Dickel, Amanbayev, Beck, Bergmann, Geissel, Gr\"of, Haettner, Hornung
  et~al.}}]{Mador2021}
\bibinfo{author}{\bibfnamefont{I.}~\bibnamefont{Mardor}},
  \bibinfo{author}{\bibfnamefont{S.~A.~S.} \bibnamefont{Andr\'es}},
  \bibinfo{author}{\bibfnamefont{T.}~\bibnamefont{Dickel}},
  \bibinfo{author}{\bibfnamefont{D.}~\bibnamefont{Amanbayev}},
  \bibinfo{author}{\bibfnamefont{S.}~\bibnamefont{Beck}},
  \bibinfo{author}{\bibfnamefont{J.}~\bibnamefont{Bergmann}},
  \bibinfo{author}{\bibfnamefont{H.}~\bibnamefont{Geissel}},
  \bibinfo{author}{\bibfnamefont{L.}~\bibnamefont{Gr\"of}},
  \bibinfo{author}{\bibfnamefont{E.}~\bibnamefont{Haettner}},
  \bibinfo{author}{\bibfnamefont{C.}~\bibnamefont{Hornung}},
  \bibnamefont{et~al.}, \bibinfo{journal}{Phys. Rev. C}
  \textbf{\bibinfo{volume}{103}}, \bibinfo{pages}{034319}
  (\bibinfo{year}{2021}),
  \urlprefix\url{https://link.aps.org/doi/10.1103/PhysRevC.103.034319}.

\end{thebibliography}

\end{document}